\documentclass[journal=jpcafh,manuscript=article,layout=twocolumn]{achemso}
\setkeys{acs}{articletitle=true}

\usepackage[T1]{fontenc}       
\usepackage{verbatim}
\usepackage{graphicx}
\usepackage{amsmath}
\usepackage{xcolor}
\usepackage{amssymb}
\usepackage{textcomp}
\usepackage{multirow}
\usepackage{siunitx}
\newcommand{\cm}{\ensuremath{\textrm{cm}^{-1}}}

\newcommand{\K}{K$^{-1}$}

\author{Shubhadip Chakraborty}
\author{Giacomo Mulas}
\altaffiliation{Istituto Nazionale di Astrofisica (INAF), Osservatorio Astronomico di Cagliari, 09047 Selargius (CA), Italy}
\author{Karine Demyk}
\author{Christine Joblin}
\email{christine.joblin@irap.omp.eu}
\affiliation
{Institut de Recherche en Astrophysique et Plan\'etologie, Universit\'e de Toulouse, CNRS, CNES, 9 Av. du Colonel Roche, 31028 Toulouse Cedex 4, France}
\title
  {Experimental Approach to the Study of Anharmonicity in the Infrared Spectrum of Pyrene from 14 to 723\,K}
\abbreviations{IR Spectroscopy,Anharmonicity,Pyrene}

\begin{document}

%%%%%%%%%%%%%%%%%%%%%%%%%%%%%%%%%%%%%%%%%%%%%%%%%%%%%%%%%%%%%%%%%%%%%

%%%%%%%%%%%%%%%%%%%%%%%%%%%%%%%%%%%%%%%%%%%%%%%%%%%%%%%%%%%%%%%%%%%%%
%\begin{tocentry}
%\begin{center}
%\includegraphics[width=0.75\columnwidth]{pyrene_expt_paper_GA.pdf} \\
%end{center}
%\end{tocentry}

%%%%%%%%%%%%%%%%%%%%%%%%%%%%%%%%%%%%%%%%%%%%%%%%%%%%%%%%%%%%%%%%%%%%%

%%%%%%%%%%%%%%%%%%%%%%%%%%%%%%%%%%%%%%%%%%%%%%%%%%%%%%%%%%%%%%%%%%%%%
\begin{abstract}
Quantifying the effect of anharmonicity on the infrared spectrum of large molecules such as polycyclic aromatic hydrocarbons (PAHs) at high temperature is the focus of a number of theoretical and experimental studies, many of them motivated by astrophysical applications. We recorded the infrared spectrum of pyrene C$_{16}$H$_{10}$ microcrystals embedded in a KBr pellet over a wide range of temperature (14 -723 \,K) and studied the evolution of band positions, widths and integrated intensities with temperature. We identified jumps for some of the spectral characteristics of some bands in the [423-473]\,K range. These were attributed to a change of phase from crystal to molten in condensed pyrene, which appears to affect more strongly bands involving large CH motions. Empirical anharmonic factors that describe the evolution of band positions and widths with temperature were retrieved from both phases over an unprecedented temperature range. The derived values were found to be consistent with available gas-phase data. We provide recommended values for anharmonic factors and conclude about the interest of the methodology to provide data of interest for comparison with theoretical models and as inputs of models that simulate the infrared emission of astro-PAHs.

\end{abstract}

%%%%%%%%%%%%%%%%%%%%%%%%%%%%%%%%%%%%%%%%%%%%%%%%%%%%%%%%%%%%%%%%%%%%%
%% Start the main part of the manuscript here.
%%%%%%%%%%%%%%%%%%%%%%%%%%%%%%%%%%%%%%%%%%%%%%%%%%%%%%%%%%%%%%%%%%%%%
\section{Introduction}
Polycyclic aromatic hydrocarbons (PAHs) are a class of organic compounds containing more than one fused benzene ring in a honeycomb fashion with peripheral hydrogen atoms. They are ubiquitous in astrophysical environments including the diffuse interstellar medium and regions associated with massive star formation and evolution.\cite{peeters2002,Tielens2008,vermeij2002} Because of their low heat capacity and extreme isolation in space, the vibrational temperature of these molecules can transiently rise up to several thousands of Kelvin. The excitation mechanism involves the absorption of a single UV photon from nearby stars via an electronic transition and a sequence of radiationless transitions converting most of the absorbed energy into vibrational excitation in the electronic ground state. The hot  molecule then relaxes by emitting IR photons in a cascade of vibrational transitions, thus contributing to the so called Aromatic Infrared Bands (AIBs) at $\sim$3.3, 6.7, 7.7, 8.6, 11.3, 12.7 and 16.4 $\mu$m. \cite{Leger1984,Allamandola1985} Each of these bands results from the superposition of a large number of hot bands, all slightly shifted with respect to the corresponding 1$\rightarrow$0 fundamentals due to anharmonicity.\cite{barker1987,Joblin1995}
Several related weaker bands are also observed in astronomical spectra, for instance at $\sim$ 3.4, 3.5, 5.25, 5.75, 6.0, 6.9, 11.0, 13.5, 14.2, 16.4 and 17.4.\cite{Allamandola1989,peeters2011} Some of them could also be induced by anharmonic effects. Modeling the emission features of the ``astronomical mixture'' of PAHs in space is a long standing goal of the astronomers, and this cannot be achieved without taking into account the impact of anharmonicity on the spectra, in particular on band positions and widths.\cite{pech2002}

Several experimental studies have been performed to study the effect on  anharmonicity on the infrared (IR) spectra of PAHs. The effect of temperature was first reported in the solid phase (KBr and CsI pellets) for temperatures up to 500\,K.\cite{bernard1989,blanco1990,colangeli1992,mennella1992} 
Most of these studies evidenced band shifts towards lower wavenumbers (towards the red) with increasing temperature. Joblin et al. \cite{Joblin1995} derived that the band positions and widths follow a near-linear dependence with temperature, based on high-temperature gas-phase data obtained for pyrene (C$_{16}$H$_{10}$), coronene (C$_{24}$H$_{12}$), and ovalene (C$_{32}$H$_{14}$) up to $\sim$900~K. Empirical effective anharmonicity factors were derived to quantify these slopes for the most intense bands. These values were used by Pech et al.\cite{pech2002} to model the IR emission of PAHs in astrophysical environments, assuming that these empirical values are suitable to describe the evolution with temperature of the IR bands of ''astronomical'' PAHs (the astro-PAHs). Robinson et al. \cite{robinson1995} recorded the IR spectrum of naphthalene in vapor phase using a long-path cell from 300 to 1000\,K to investigate the variation of the band position and intensity with temperature of the CH stretch band at 3066\,$\cm$ and the CH out-of-plane bend at 782\,$\cm$. 
Oomens et al.\cite{oomens2003} determined global anharmonicity shifts for several bands of several PAH cations as the difference in band position observed between rare-gas matrices and IR multiphoton dissociation experiments.

In parallel with experiments, several theoretical calculations were performed to model the evolution of the infrared spectrum of neutral or ionic PAHs with temperature. 
The first type of calculations allow to derive the global evolution of the band positions with temperatures and derive empirical anharmonicity factors similarly to experiments. Calculations at finite temperature were performed using a quantum statistical approach for naphthalene \cite{basire2009} and pyrene-based PAHs. \cite{calvo2011} 
Except for the 482 $\cm$ band of naphthalene which shifts towards the blue up to 600\,K, all other bands are found to shift towards the red with a linear trend with temperature.
Classical and quantum molecular dynamics trajectories were also used to model the anharmonic IR spectra of neutral and ionic PAHs and their derivatives. \cite{joalland2010,calvo2010, simon2011}  
Other type of codes aim at modeling in details the band structures by describing the connection between states with explicit consideration of the resonances. \cite{pirali2009,mackie2015,mackie2016,mulas2018}
For these theoretical studies, the best way to test and validate the method is by comparison between calculated IR anharmonic spectra and experimental spectra for a range of temperatures. Unfortunately, experimental data remain scarce.

In this work, we report detailed experimental data on the evolution of the pyrene spectrum with temperatures between 14 and 723\,K. This is the first time that the vibrational spectrum of a PAH molecule is uniformly studied over such a wide temperature range; this was made possible by embedding pyrene micro-crystals in solid KBr. The discussion will therefore focus on the relevance of these data to quantify anharmonicity in isolated PAHs.
This work is laid out as follows: in Section II, we provide a detailed description of the experimental set up; in Section III we report the procedure for the data processing and our results; in Section IV, we discuss our results and compare them with previous measurements; finally, we wrap up in Section V with our conclusions.

\section{Experiment}
The infrared spectrum of solid pyrene in KBr pellets was recorded in transmission mode from 14 to 723\,K using the following equipment and procedures.
\subsection{Sample preparation}
Pyrene (98\%) was purchased from Sigma Aldrich and was used without further purification. 
Sample pellets were prepared by mixing $\sim$ 0.30, 0.33, 0.83 and 0.86 mg of solid pyrene with  $\sim$ 200 mg of potassium bromide (KBr, FTIR grade, 98\%) in an agate mortar.  The mixture was then put in a 13 mm diameter pellet die and pressed under 10 tons for 5 minutes. The recording of the different spectral sets is summarized in Table\,\ref{Tab:py_conc}. The lowest concentration pellets were used to study the strongest bands at 709.7, 749.4, 839.9, and 1184.8\,\cm (see Figure\,\ref{tot_spec}).

\begin{table*}
\begin{center}
\caption{Bands analyzed with different concentrations of pyrene}
\hspace{2cm}
\begin{tabular}{c c l}
\hline
\multicolumn{1}{c}{Temperature range}  & \multicolumn{1}{c}{Concentration}& \multicolumn{1}{c}{Bands analyzed}\\
\multicolumn{1}{c}{(K)}  & \multicolumn{1}{c}{(mg)}& \multicolumn{1}{c}{(\cm)}\\
\hline
\vspace{2mm}
\emph{Low temperature} & &\\
14-300 & 0.86 & 820.3, 964.7, 1000.9, 1096.3, 1241.7\\ 
       &      & 1313.2, 1433.6, 1486.5, 1599.8$^{\#}$, 3048.3$^{\#}$\\
14-300 & 0.33 & 709.7$^{*}$, 749.4$^{*}$, 839.9$^{*}$, 1184.8$^{*}$ \\ 
\emph{High temperature} & & \\
300-723 & 0.83 & 820.3, 964.7, 1000.9, 1096.3, 1184.3, 1241.7\\ 
       &      & 1313.2, 1433.6, 1486.5, 1599.8$^{\#}$, 3048.3$^{\#}$\\
300-723 & 0.30 & 709.7$^{*}$, 749.4$^{*}$, 839.9$^{*}$, 1184.8$^{*}$\\
\hline
\label{Tab:py_conc}
\end{tabular}
\end{center}
\begin{flushleft}
\footnotesize{$^{\#}$ Bands with complex structure for which the peak maxima of the highest intensity band at 300~K are reported. For the other bands, weighted average band positions at 300~K obtained from multicomponent fitting are reported.\\
$^{*}$ Bands with high oscillator strengths, which have been analyzed using the low concentration pellets.} 
\end{flushleft}
\end{table*}

The IR spectrum of a coarsely ground pyrene mixture with KBr suffers from the deleterious Christiansen effect \cite{christiansen1884,christiansen1885}: small transparent particles suspended in a non-absorbing medium can enhance the transmission when the refractive index of the sample and the matrix are accidentally (almost) equal. 
This causes severe distortion of the experimental IR bands and the analysis becomes very difficult. To reduce this effect and to obtain high quality spectra in the mid-infrared region with good spectral contrast, it is essential to have grain sizes as small as possible. A grinding time of 15 min. was found to be enough to remove the Christiansen effect (cf. Figure~S1 of the Supporting Information), which ensures that the grains are (sub-)micronic in size. Additional evidence for the necessity of these small sizes is provided by the low contribution of scattering in the recorded transmission spectra.

\subsection{Spectral acquisition}
The experiment was performed using the ESPOIRS setup at IRAP in Toulouse,\cite{karine2016} which consists in a Vertex 70V Fourier Transform vacuum infrared spectrometer from Bruker Optics equipped with two sample compartments. 
All spectra were recorded with a room temperature deuterated triglycine sulfate (DLaTGS) detector and a KBr beamsplitter. The spectral resolution was kept at 0.2 $\cm$ and the acquired interferograms were averaged over 256 scans.  
Both a background and a sample spectrum were recorded at each temperature. The transmission spectra of pyrene were obtained by dividing the single channel spectrum of the sample pellet by that of the blank pellet, both measured at the same temperature.
They were converted in absorbance spectra with the following formula: A = \textminus Log(T). The continuum, which remained from scattering, was subtracted for each spectrum using the OPUS 7.0 software. 

\subsubsection{Low temperature experiment}
The spectra at low temperatures (14 to 300\,K) were recorded using a pulsed tube cooler cryogenic cooling system from QMC instruments equipped with a sample wheel. The blank KBr and the sample pellets were placed inside the sample wheel. A thermal probe was attached close to the wheel for reading the temperature of the wheel. In order to achieve a good thermal contact, an indium o-ring was installed between the wheel and the pellets.

The cryocooling system was placed inside the sample compartment of the spectrometer and was pumped with a turbomolecular pump. The average pressure inside the sample compartment was $\sim$ 3.7 $\times$ 10$^{-7}$ mbar at 14 K. The temperature of the sample wheel was controlled by a LakeShore temperature controller (Model 335). Background and sample spectra of pyrene were recorded from 14\,K (the lowest temperature we can reach in our set up) up to 300\,K in steps of 50\,K. At each temperature we used a stabilization time of about 10 min.  prior to the spectral acquisition, in order to avoid thermal fluctuations. 

\subsubsection{High temperature experiment}
The spectra at high temperatures (300 to 723\,K) were recorded using a high pressure-high temperature cell from Specac (Model number P/N GS05850) equipped with ZnSe windows. The temperature of the cell was controlled by a temperature controller provided by Specac. A temperature step of 50\,K was used, except in the 423 to 473 \,K range over which a finer sampling by 10\,K step was performed to better sample spectral features which have been observed in this range, as will be presented below.
The cell body is cooled by water at $\sim$\ang{10}C.
The average pressure inside the sample compartment varies from $\sim$10$^{-6}$ (at 300\,K) to 10$^{-4}$ (at 723\,K) mbar. 
We performed the measurements up to a maximum temperature of 723\,K since the transparency of the pellet drastically decreases beyond this value. In order to test the precision of our measurements over the full temperature range (14\,K to 723\,K), we compared the band positions which were recorded at 300\,K in the two experiments at low and high temperatures. We found that the positions match at the 1\% level.

\section{Results}

\subsection{Global spectral evolution with temperature}

\begin{figure*}
\centering
\includegraphics[width=0.8\hsize]{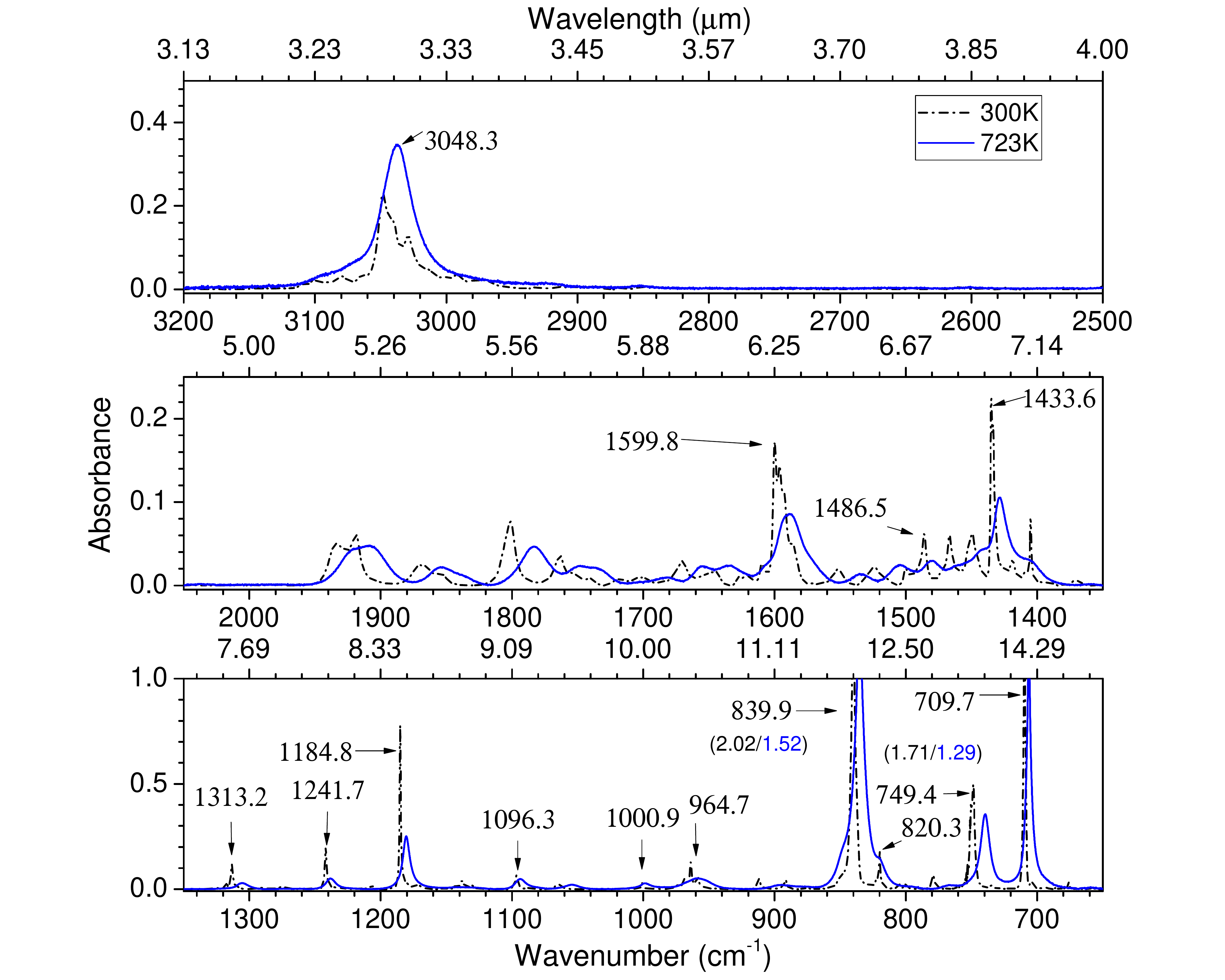} \\
\caption{Continuum subtracted infrared spectrum of the (0.83\,mg) pyrene pellet at 300\,K and 723\,K from 3200 to 650 $\cm$. Band positions marked in the figures were derived at 300~K from the multi-component fitting except for the bands at 3048.3 and 1599.8~\cm\ for which the peak maxima of the highest intensity band at 300 K are reported. }
\label{tot_spec}
\end{figure*}

Figure~\ref{tot_spec} displays two spectra of pyrene in KBr pellet, which were recorded in the 3200-650~$\cm$ spectral range at 300 and 723~K.
The top, middle and lowest panels of Figure \ref{tot_spec} focus on the C-H stretching region, in-plane deformations (C-C stretch, C-H in-plane bending and C-C-C in-plane angle deformations) and the out-of-plane deformations (C-H out-of-plane bend, C-C torsion), respectively. It is evident from Figure \ref{tot_spec} that, at room temperature (300\,K), the bands have resolved structures while with increasing temperature (723\,K), the band structure becomes blurred. The structures are resolved even better at lower temperatures (Figures\,\ref{band_pro_T} and S2). In the following, we study the detailed band properties as a function of temperature.

\begin{figure*}
\centering
\begin{tabular}{c c}
\includegraphics[width=0.5\hsize]{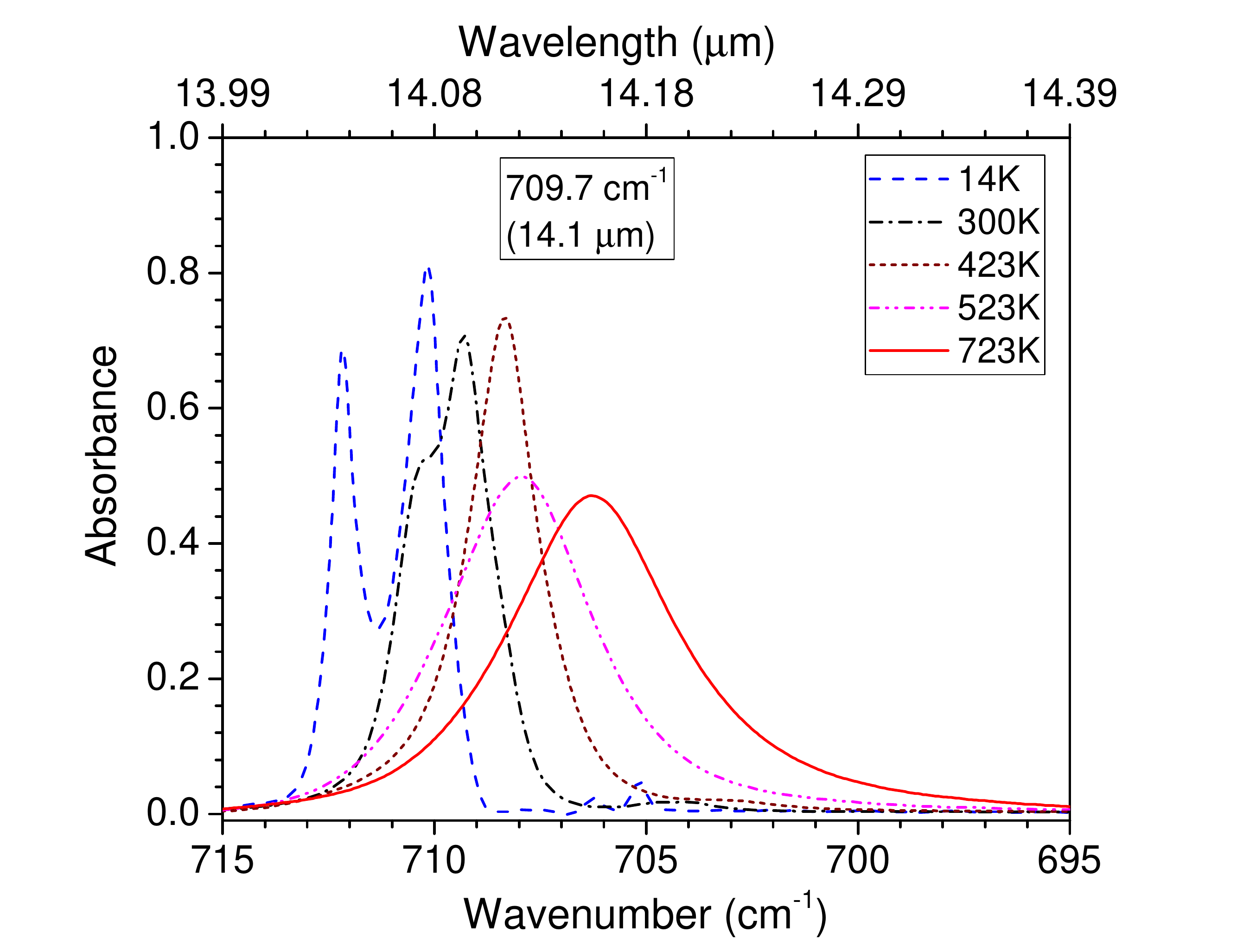}& \includegraphics[width=0.5\hsize]{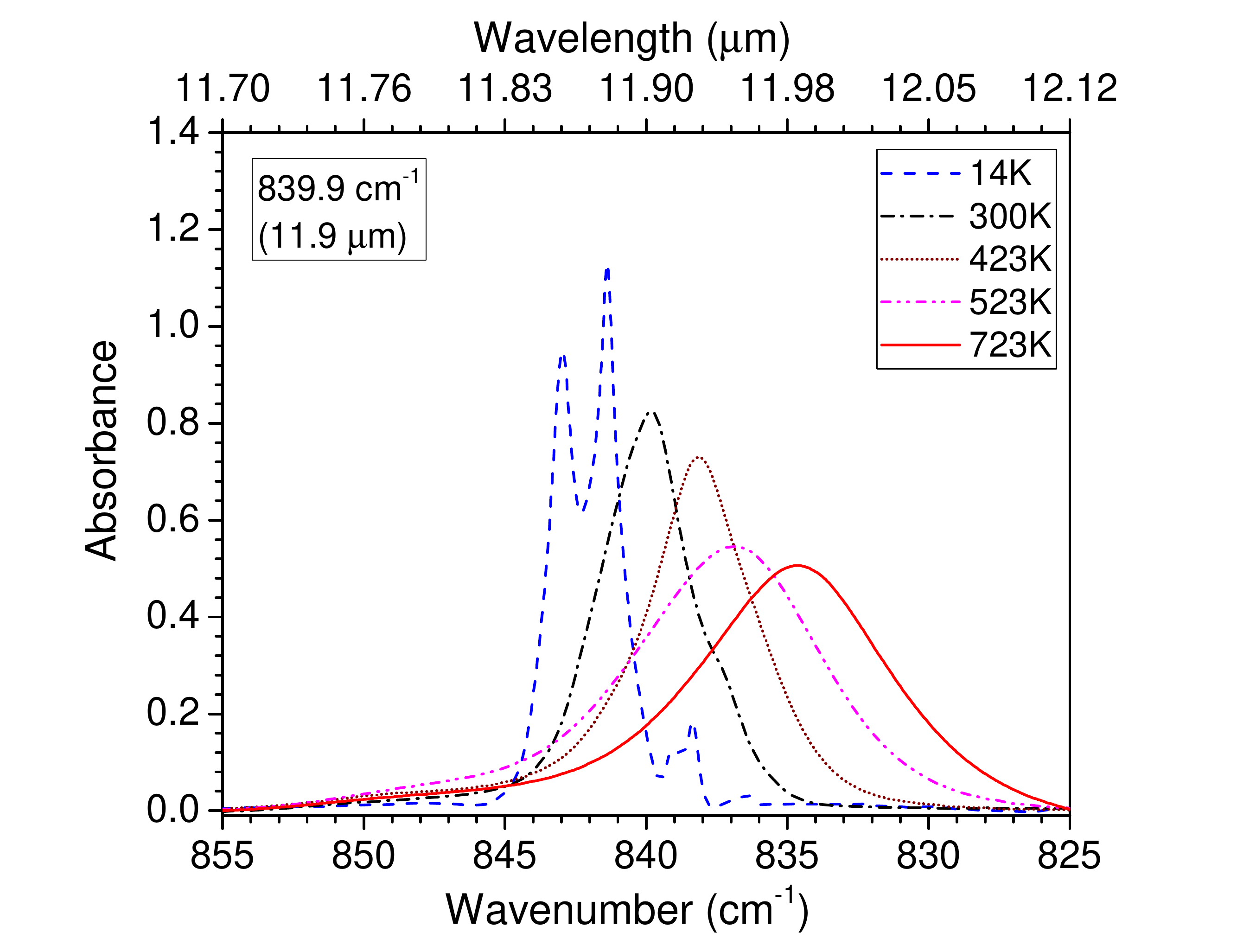} \\
(a) & (b)\\
\includegraphics[width=0.5\hsize]{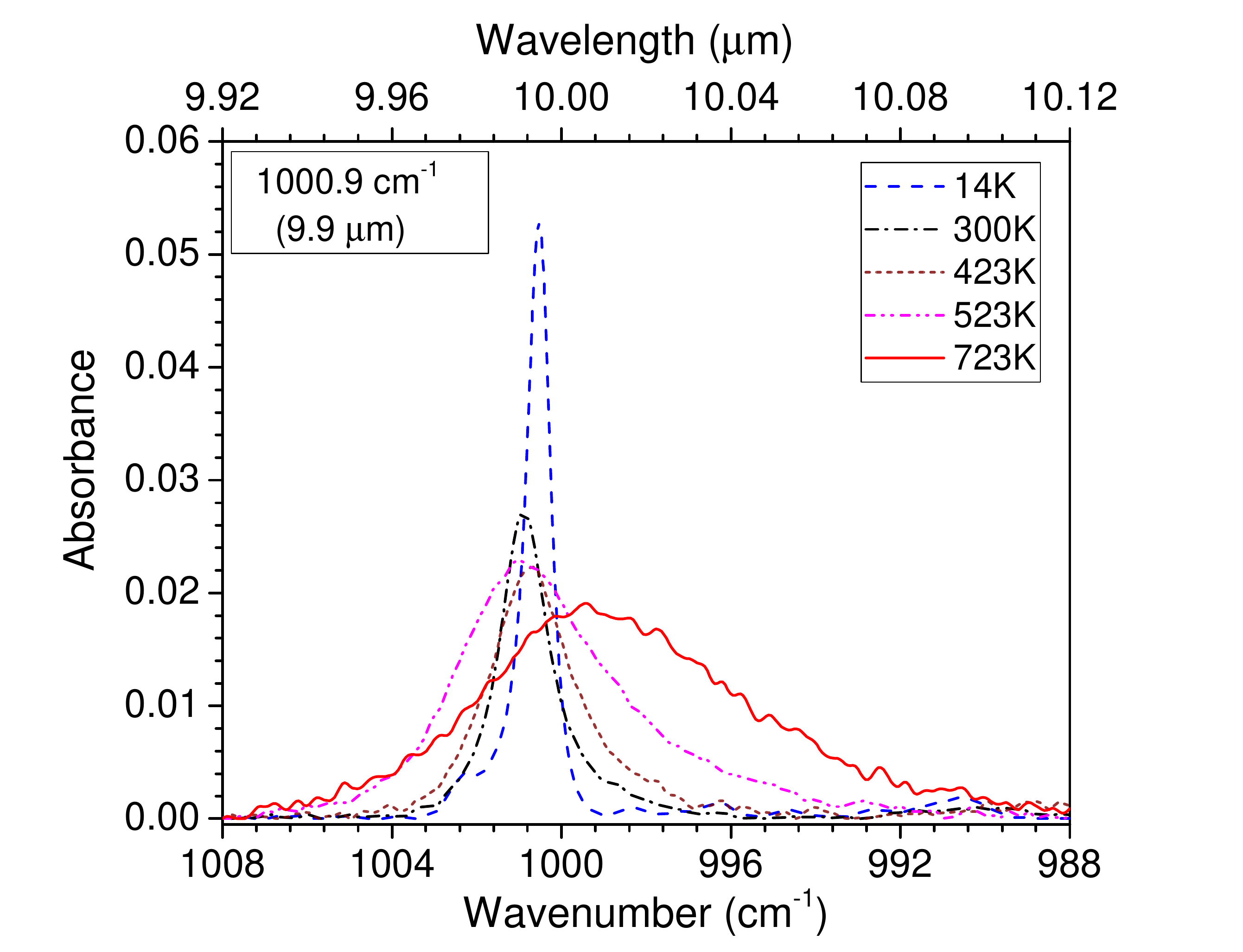} & \includegraphics[width=0.5\hsize]{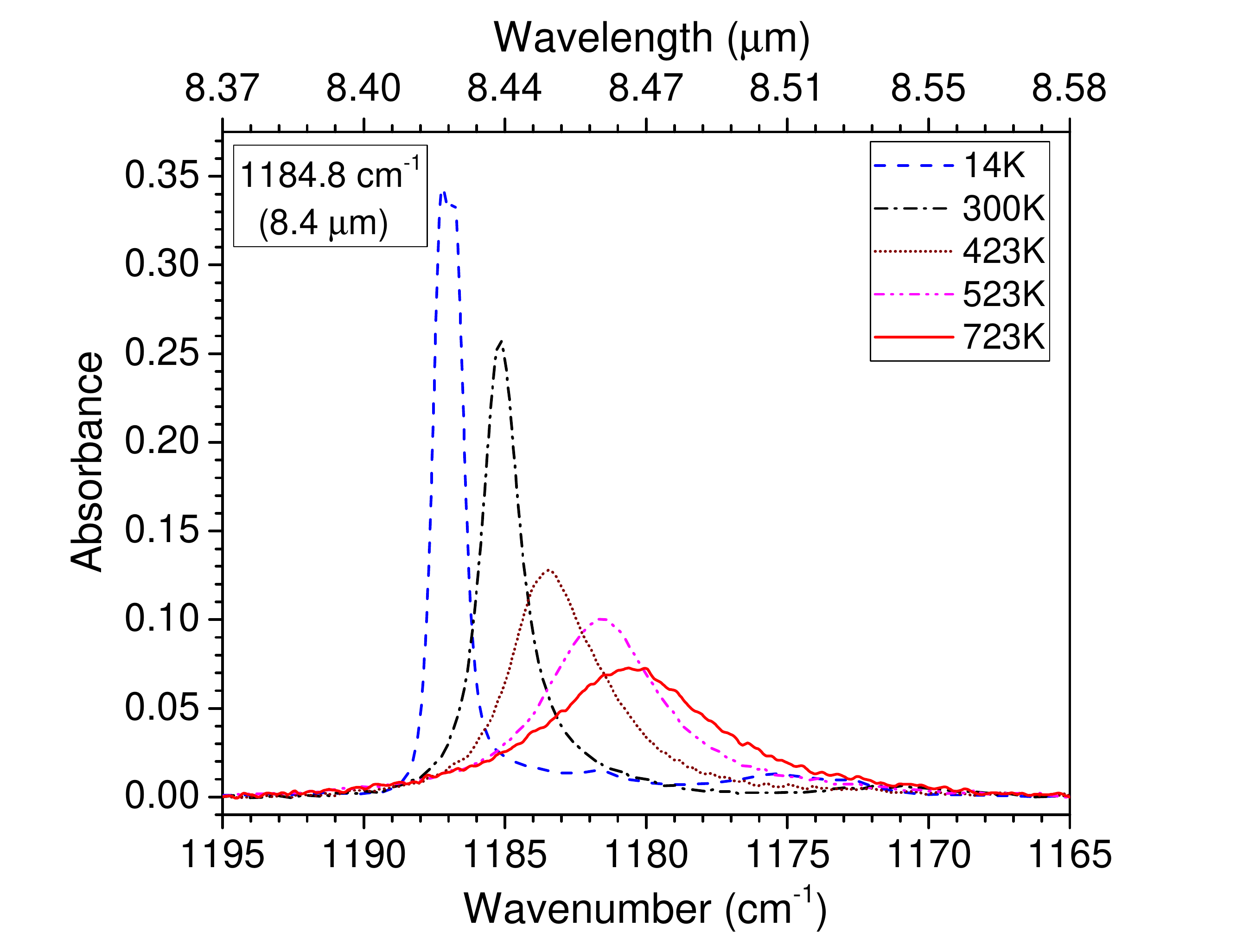}\\
(c) & (d)\\
\includegraphics[width=0.5\hsize]{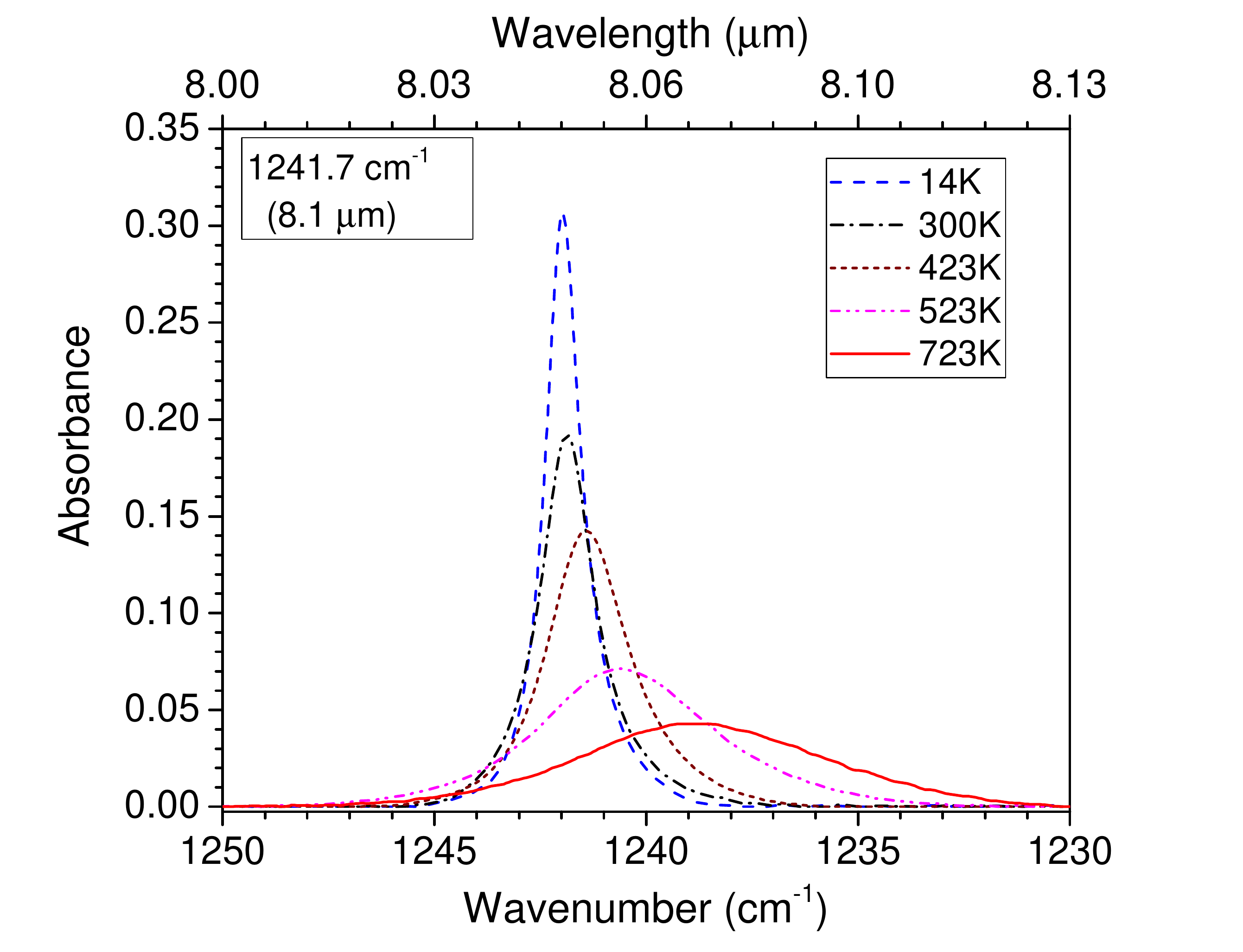} & \includegraphics[width=0.5\hsize]{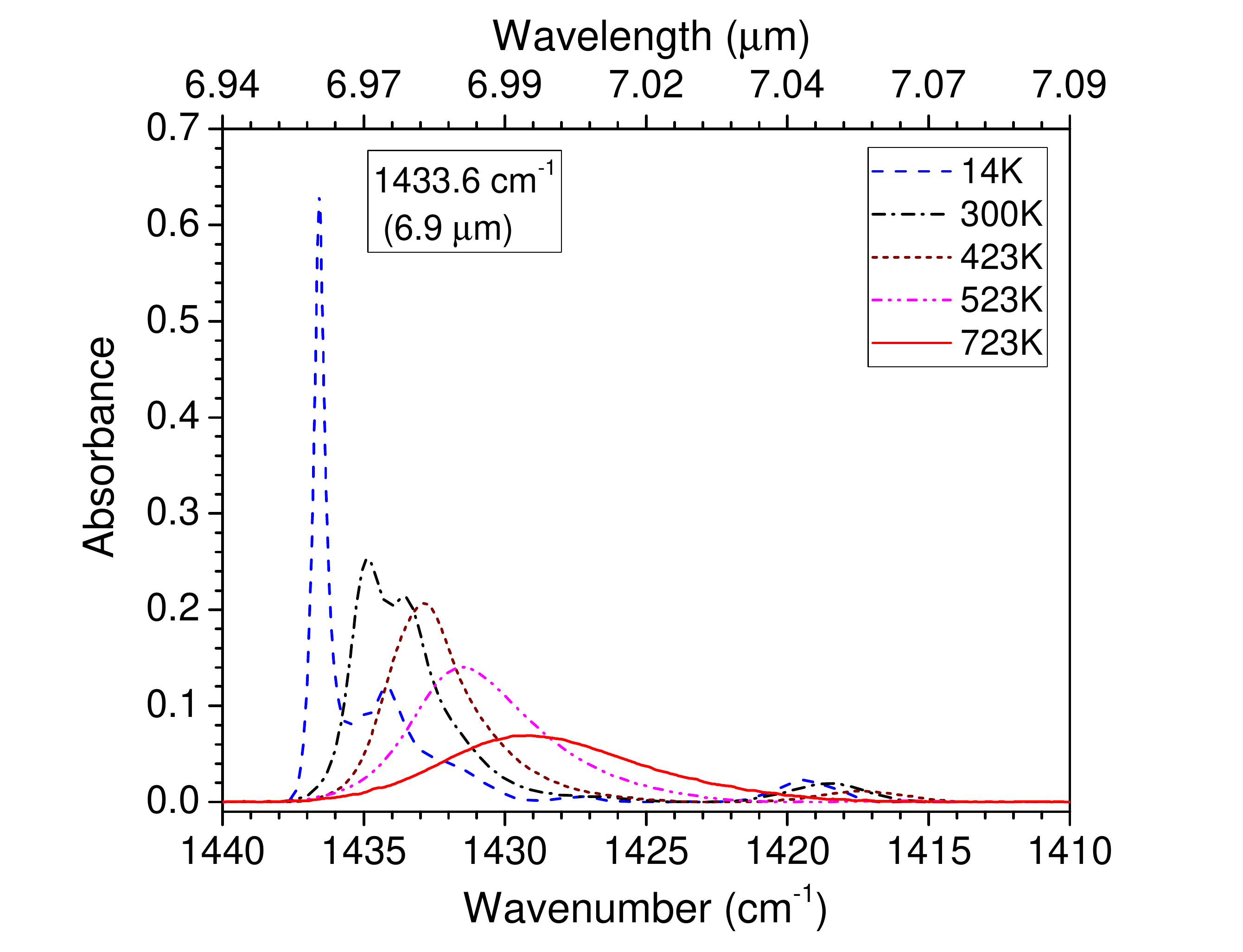}\\
(e) & (f)\\
\end{tabular}
\caption{Evolution of the IR band profiles with temperature for some fundamental transitions of pyrene in KBr pellets. Weighted average band positions from the multi-component fitting at 300~K are listed in Figures \ref{band_pro_T} (a) to (f). Multi-component analysis was not possible due to complex band profiles for the bands shown in Figures \ref{band_pro_T} (g) and (h). Therefore, we report the maxima of these bands at 300~K.}
\end{figure*}
\begin{figure*}
\ContinuedFloat
\centering
\begin{tabular}{c c}
\includegraphics[width=0.5\hsize]{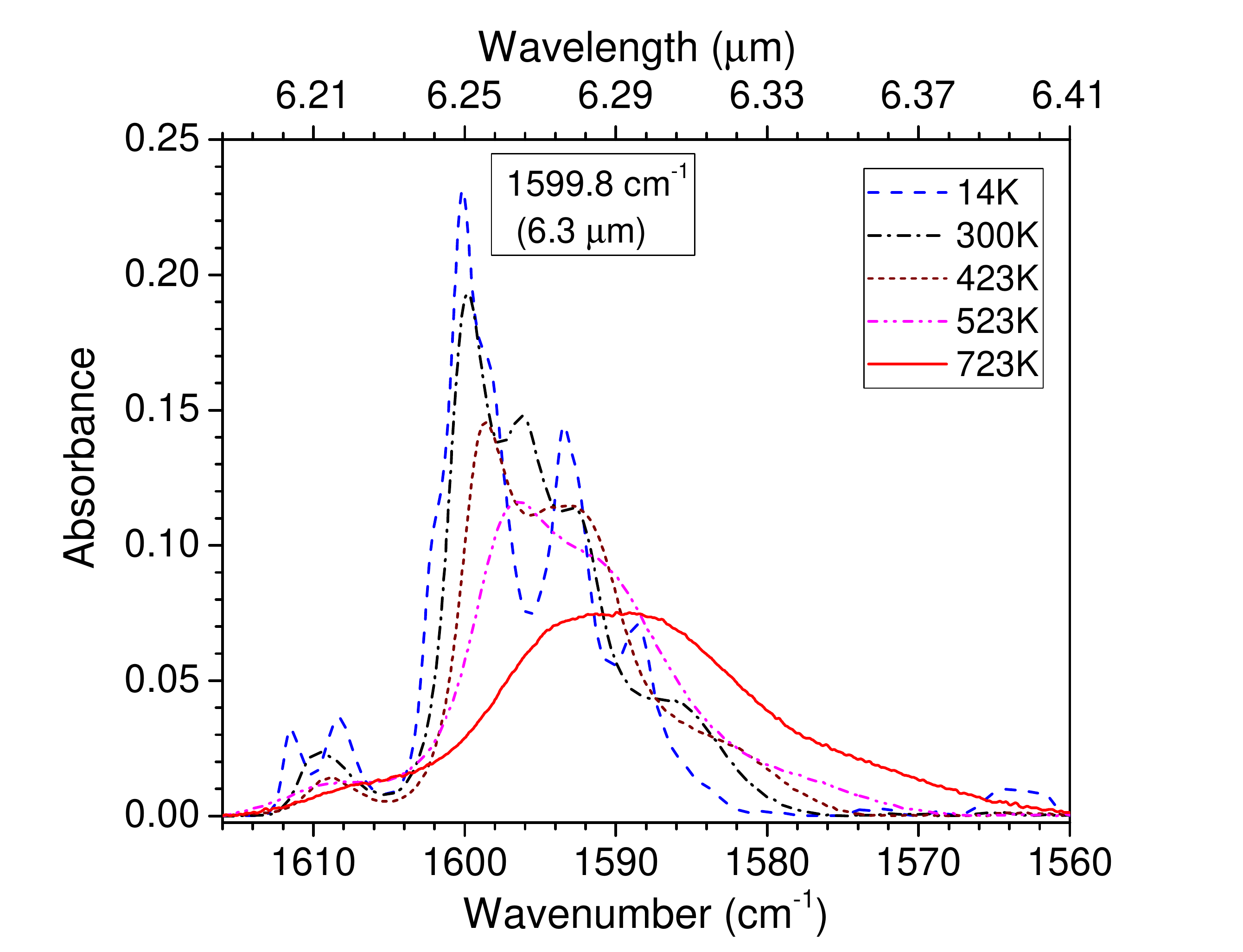} & \includegraphics[width=0.5\hsize]{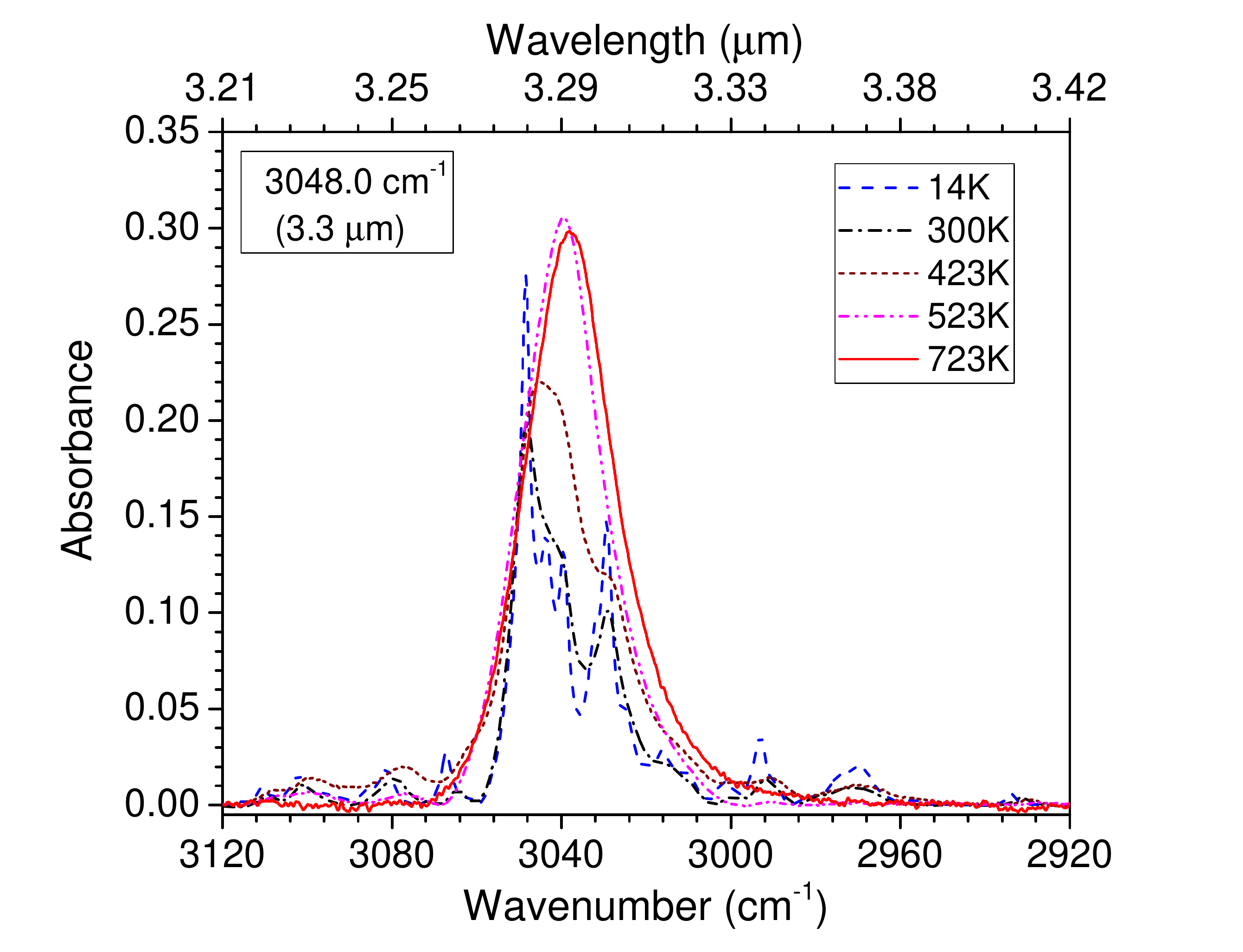}\\
(g) & (h)\\
\end{tabular}
\caption{Evolution of the IR band profiles with temperature for some fundamental transitions of pyrene in KBr pellets. Weighted average band positions from the multi-component fitting at 300~K are listed in Figures \ref{band_pro_T} (a) to (f). Multi-component analysis was not possible due to complex band profiles for the bands shown in Figures \ref{band_pro_T} (g) and (h). Therefore, we report the maxima of these bands at 300~K. (continued from the previous page)}
\label{band_pro_T}
\end{figure*}

\subsection{Multi-component fitting}
Pyrene has 72 normal modes among which 29 are IR active. In our experiment we observed 14 bands
which result from fundamental transitions (several can be clustered within the same band) together with combination bands and overtones. In addition, effects due to solid state are expected. In particular, anharmonic calculations performed at 0\,K show that there is no substructure for the bands at 709.7, 749.4, 820.3 and 839.9~cm$^{-1}$, whereras a clear splitting is observed in our low-temperature measurements (Figure\,\ref{band_pro_T} and Figures S2, S3 and S4 from the Supporting Information).
In this context, we determined the band positions at the various temperatures by performing a multi-component fit. We excluded the two most complex bands at 1599.8 and 3048.2 cm$^{-1}$, which exhibit a lot of substructures at low temperatures  (see Figure\,\ref{band_pro_T} (g) and (h)). Other bands were classified into three categories namely, isolated (Type 1), low mixed (Type 2) and highly mixed (Type 3) as shown in Table\,\ref{Tab:bp_T}.
Before analysis, a local linear continuum was subtracted for all the studied bands with the same anchor points for all spectra. The fits were performed with as many Gaussians as necessary to reproduce the overall profile. The band positions were derived by averaging individual Gaussian positions after weighting them by their integrated intensity. Examples of the multi-component fitting are provided in Figures S3, S4 and S5 of the Supporting Information. Over the manuscript, we decided to refer to the bands by their position at 300\,K as determined by the multi-component fitting except for the 1599.8 and 3048.2 cm$^{-1}$ bands for which we took the position of the strongest peak.
For Type 1 isolated bands, the band width was determined by taking the Full Width at Half Maximum (FWHM) of the overall  fit of the band. 
Finally, the integrated band intensities were determined both for the full spectral range and for all 14 bands reported in Figures \ref{band_pro_T} and S2 in the Supporting Information.

\subsection{Evolution of band positions with temperature}

\begin{figure}
\includegraphics[width=1.05\columnwidth]{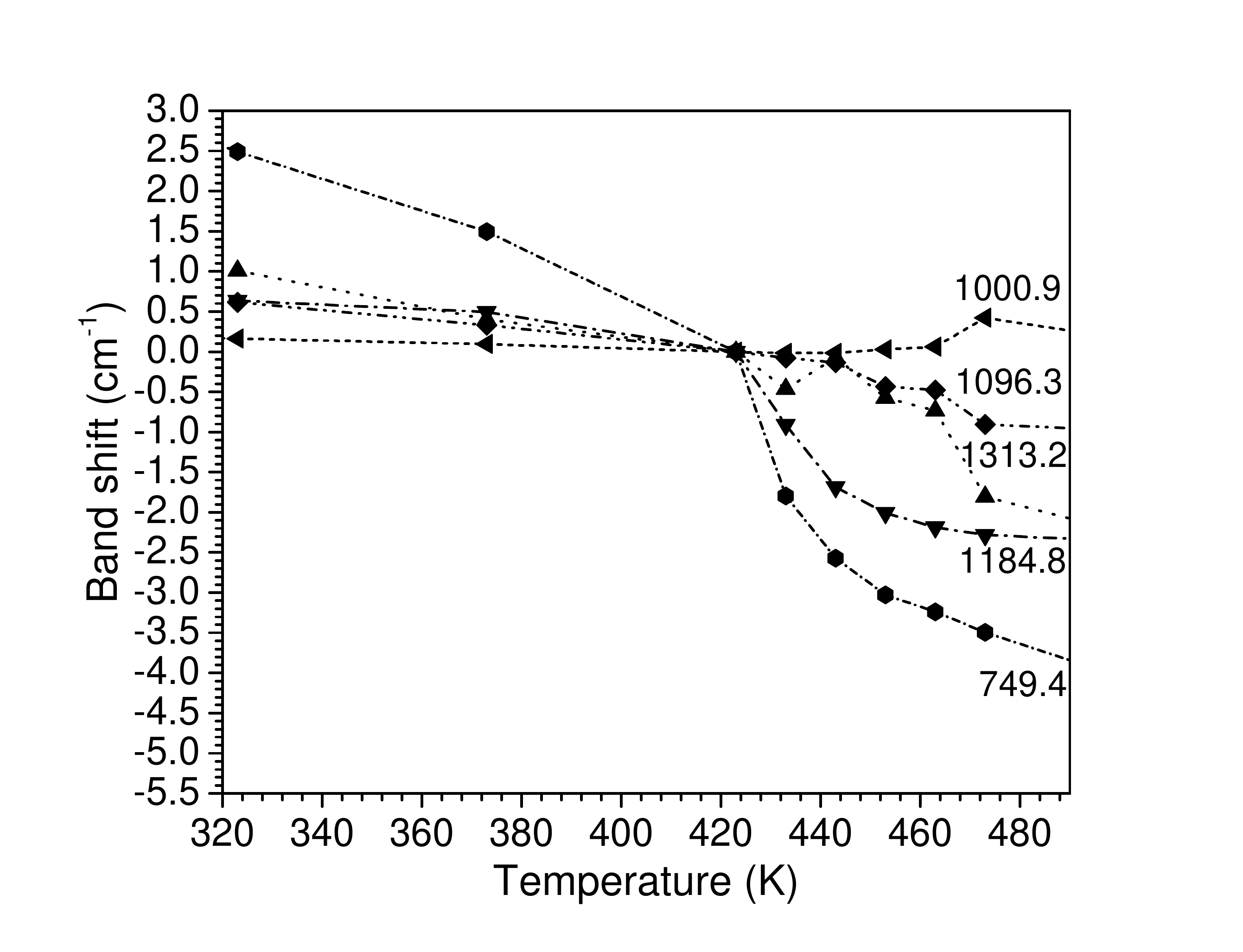}\\
\caption{Band shifts with respect to the positions at 423\,K. In the range [423-473]\,K a discontinuity is observed in the position of the displayed bands. Each curve is labelled with the corresponding band position in \cm\ and at 300 \,K.}
\label{phase_transition}
\end{figure}

\begin{figure*}
\centering
\begin{tabular}{c c}
\includegraphics[width=0.5\hsize]{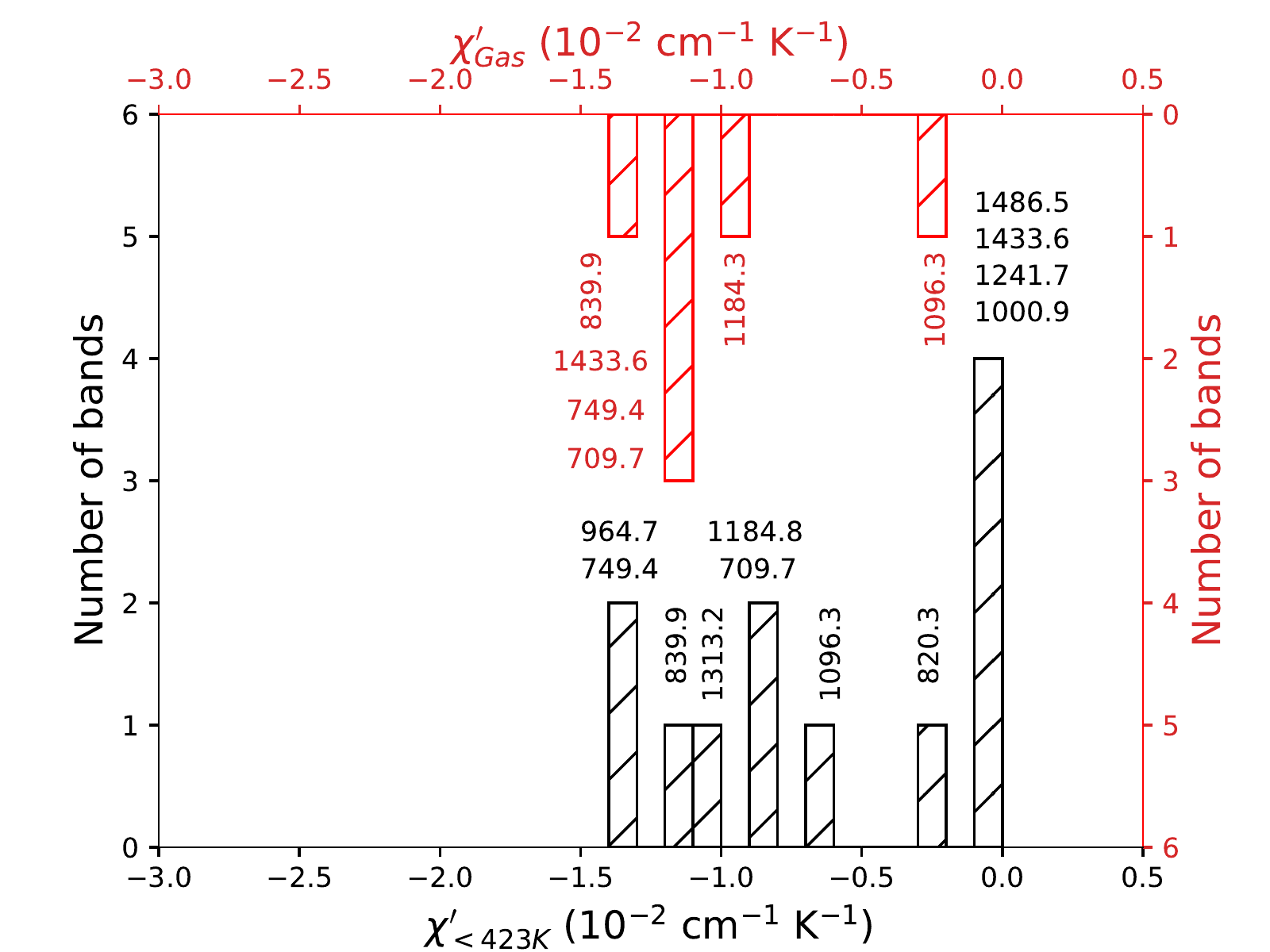}&\includegraphics[width=0.5\hsize]{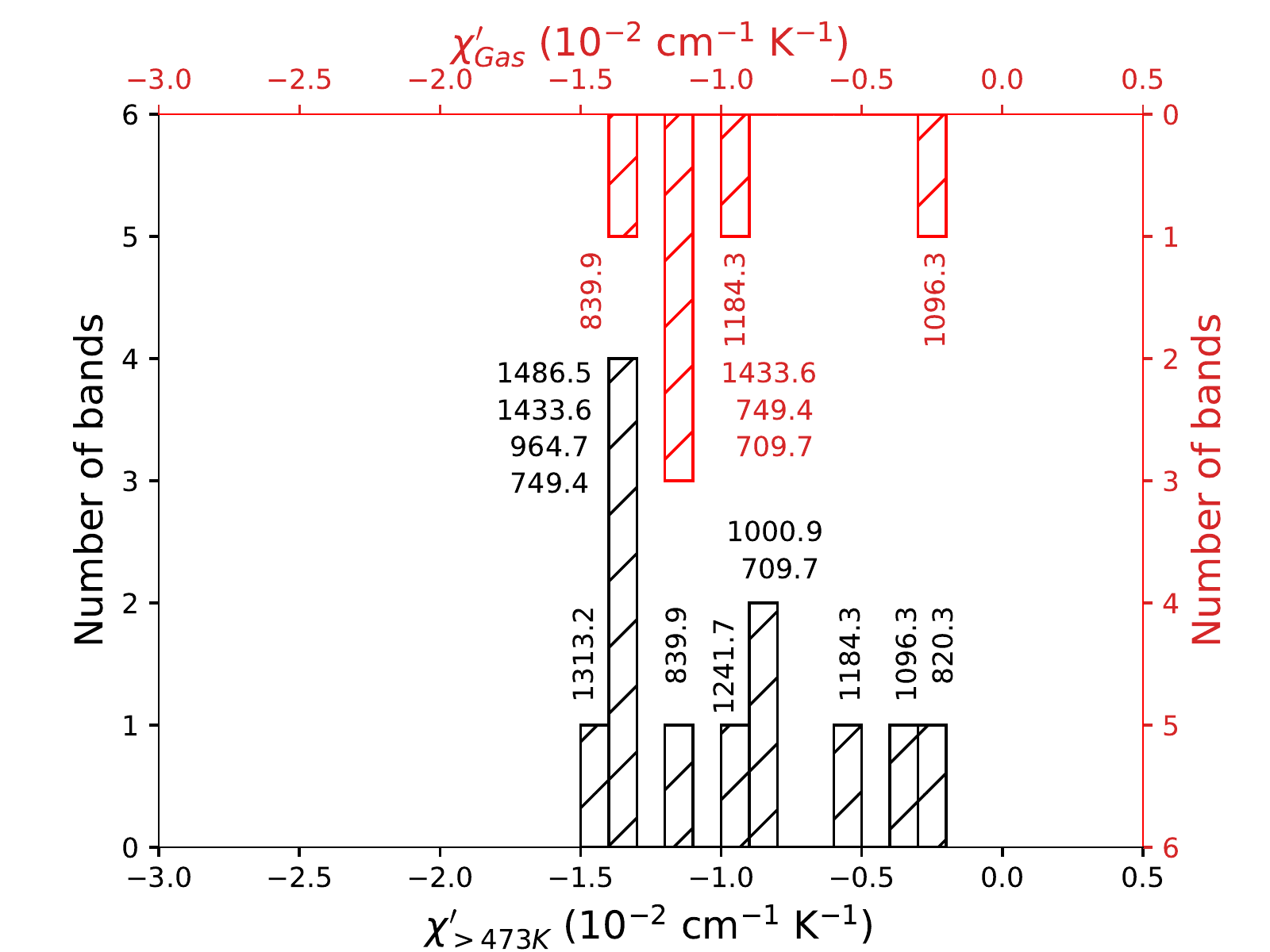}\\
(a) & (b) \\
\end{tabular}
\caption{Statistical analysis of the empirical anharmonicity factors of pyrene in different conditions. The lower histograms represent the values obtained for pyrene in KBr pellets at T<423\,K and at T> 473\,K, in (a) and (b), respectively. Values in gas phase \cite{Joblin1992,Joblin1995} are also shown by the top histograms.}
\label{anh_fact_microcrystal_melted_gas}
\end{figure*}

The evolution of band positions with temperature is shown in Figures\,\ref{posn_T} (a)-(f)) and Figure~S6 of the Supporting Information. For comparison and discussion, we also report in these figures the band positions which have been previously measured in gas-phase at high temperatures (573 - 873\,K) \cite{Joblin1995,Joblin1992} and in matrices at very low temperatures ($\sim$4\,K) \cite{Joblin1994, bahou2013, maltseva2016}.
Globally, the evolution of band positions with temperature can be described by two slopes: a low-temperature and a high-temperature one, with a gradual change between them. The temperature at which this change occurs, depends on the specific bands; it is typically around 150~K but can be as high as 250~K for the band at 1486.5~$\cm$.
An exception is the band at 1241.7~$\cm$ which does not seem to require a different low-temperature slope.
In addition, in the high-temperature range, the positions of the bands at 749.4, 1000.9, 1096.3, 1184.8, 1241.7 and 1313.2~\cm\ are found to exhibit a jump between 423 and 473\,K (see Figure\,\ref{phase_transition}), with a subsequent change of slope in the evolution of the band position with temperature.
\\
From the evolution of the band position with temperature, we can derive empirical anharmonicity factors. These consist in the slopes described above, and quantified by the $\chi'$ parameter as defined in Joblin et al.\cite{Joblin1995}.
That led us to report in Table~\ref{Tab:bp_T} two or three values for $\chi'$ over the studied temperature range, depending on the presence or not of a jump in the 423-473\,K range.
The change in slope after the 423-473~K range leads to variations by up to 50\%. The case of the 1000.9\,\cm\ band is peculiar since it is not clear if the  absence of band shift before 423~K is a temperature or a solid phase effect. 
We note that, on average, the anharmonicity coefficients are larger below 423~K and smaller at higher temperatures. The band at 1313.2\,\cm\ shows the opposite trend though.
These values are also compared in Table~\ref{Tab:bp_T} with gas-phase measurements either coming from Joblin et al.\cite{Joblin1995} or derived using the original gas-phase spectra of the authors\cite{Joblin1992} (bands at 709, 750, 1096 and 1435\,\cm).
 Both data sets (before and after the 423-473~K range) are found to show a reasonable agreement compared to the gas-phase values (cf. Figure\,\ref{anh_fact_microcrystal_melted_gas}).

\begin{figure*}
\centering
\begin{tabular}{c c}
\includegraphics[width=0.5\hsize]{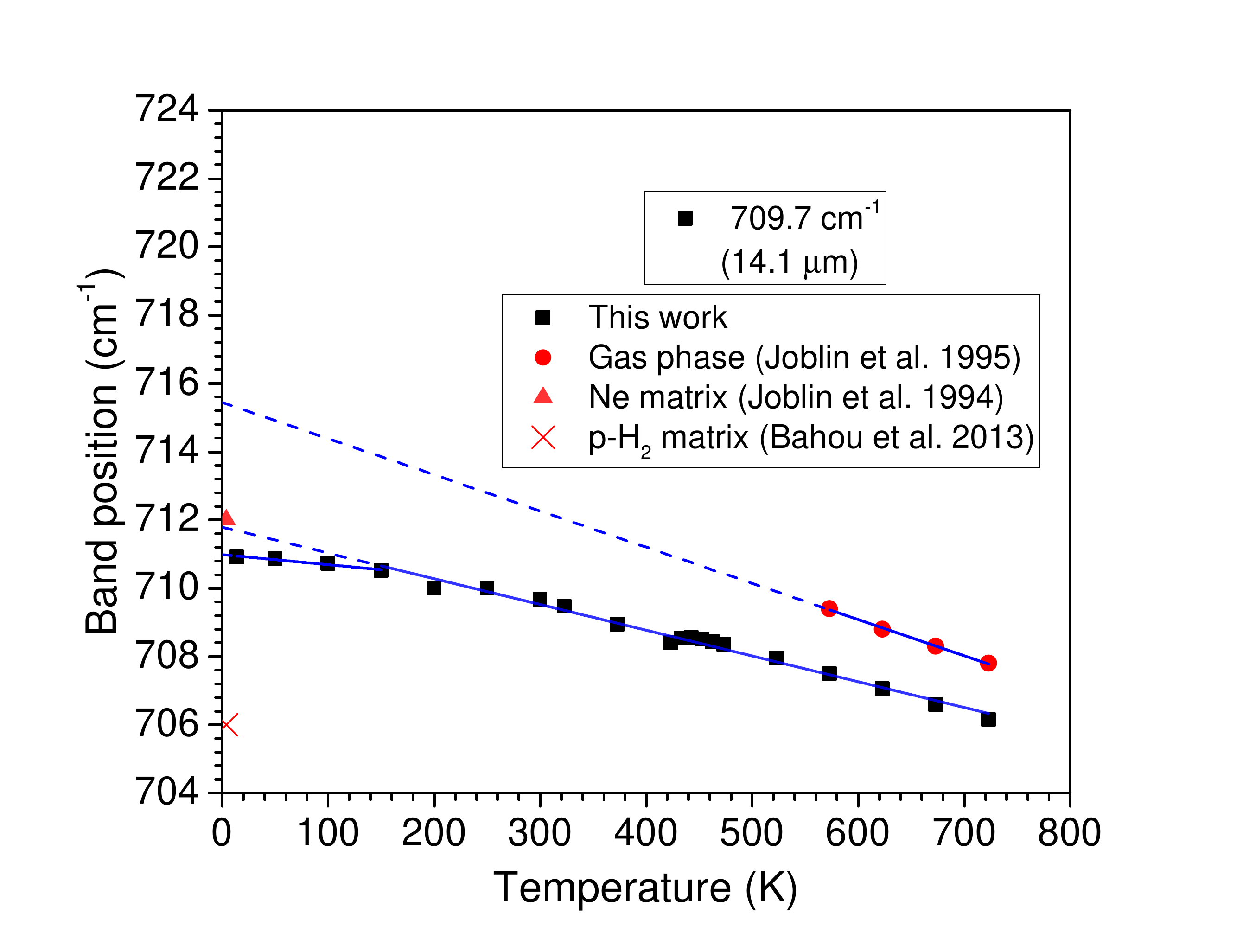}& \includegraphics[width=0.5\hsize]{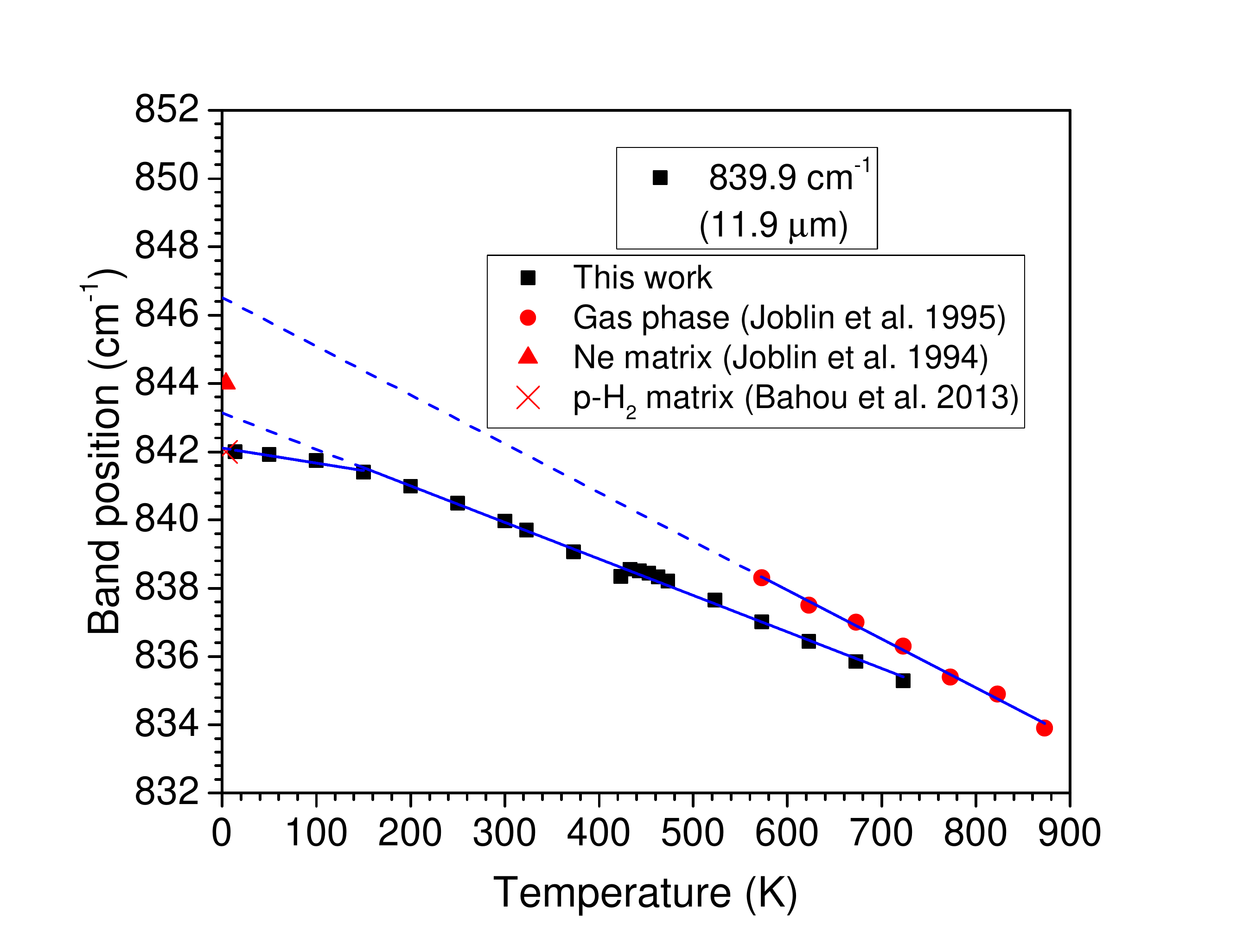}\\
(a) & (b)\\
\end{tabular}
\caption{Evolution of the band positions with temperature for fundamental transitions of pyrene in KBr pellets (black squares). 
Band positions (in \cm\ and $\mu$m) at 300\,K are mentioned in the top right corner of each panel.
The calculated linear fits in the low and high temperature ranges are shown by blue lines. Also shown with red marks are published data both at high temperature in the gas-phase \cite{Joblin1995} and at very low temperature (3-4\,K) either in the gas-phase \cite{maltseva2016}, Ne matrix\cite{Joblin1994} or p-H$_{2}$ matrix\cite{bahou2013}.}

\end{figure*}
\begin{figure*}
\ContinuedFloat
\centering
\begin{tabular}{c c}
\includegraphics[width=0.5\hsize]{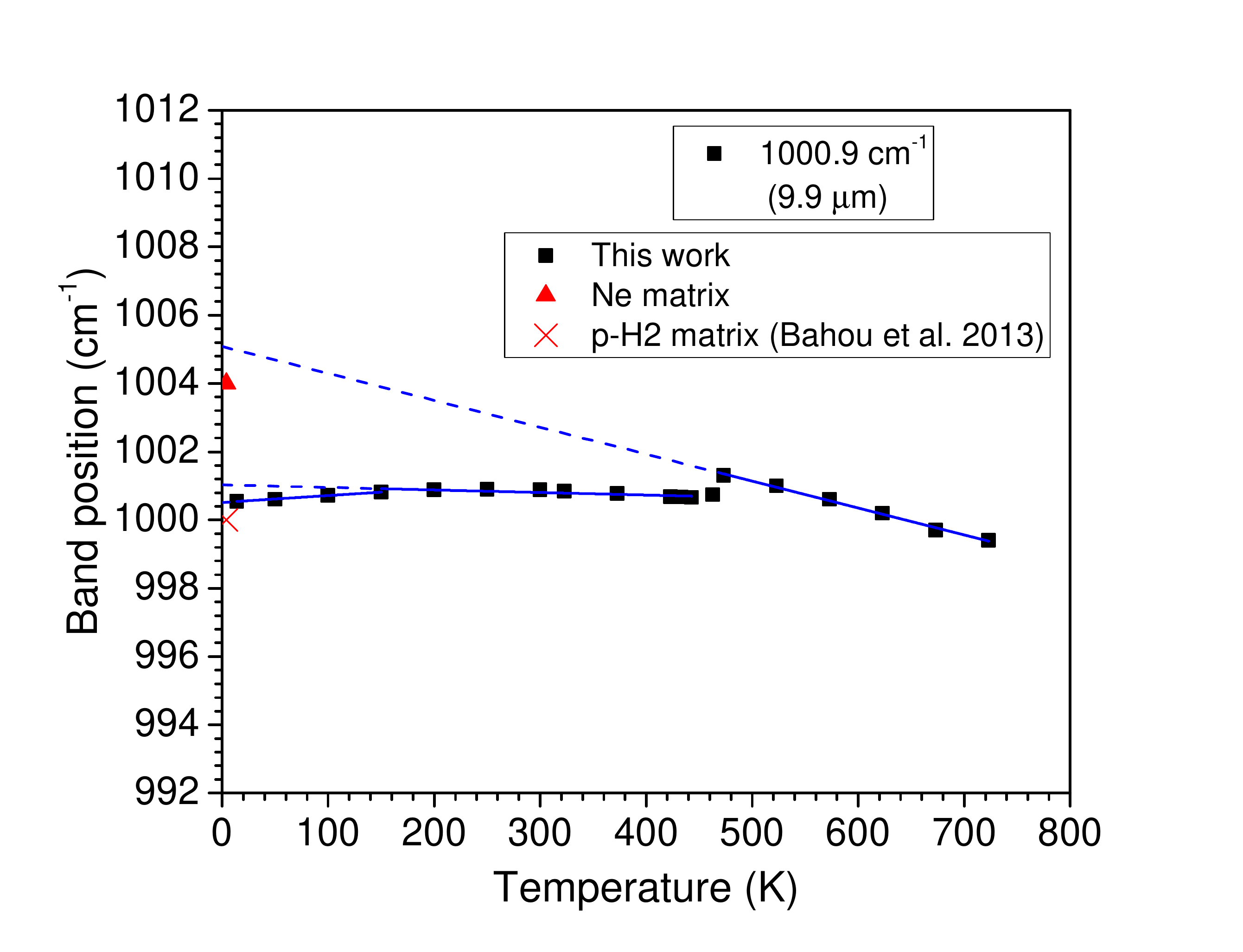} &\includegraphics[width=0.5\hsize]{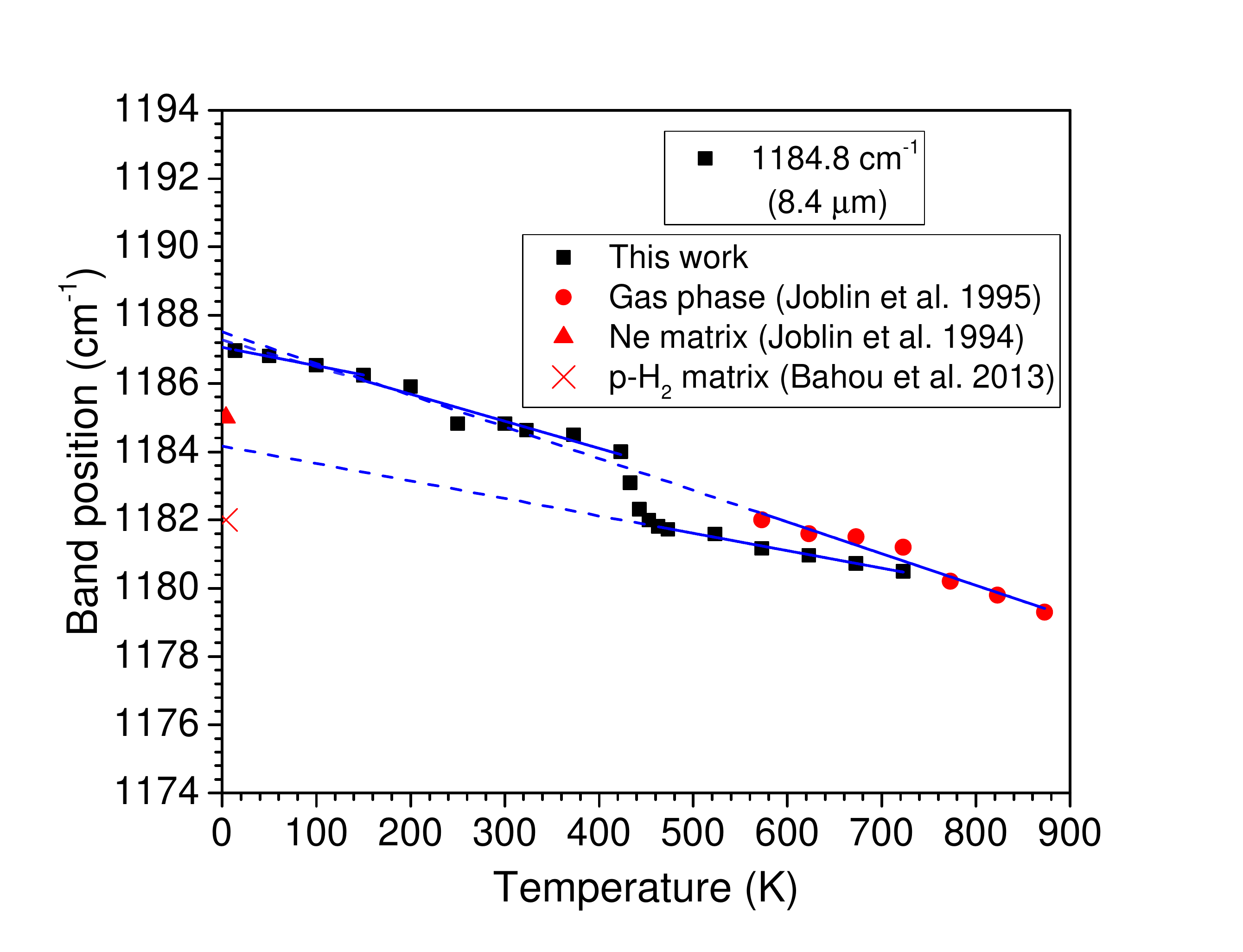}\\
(c) & (d)\\
\includegraphics[width=0.5\hsize]{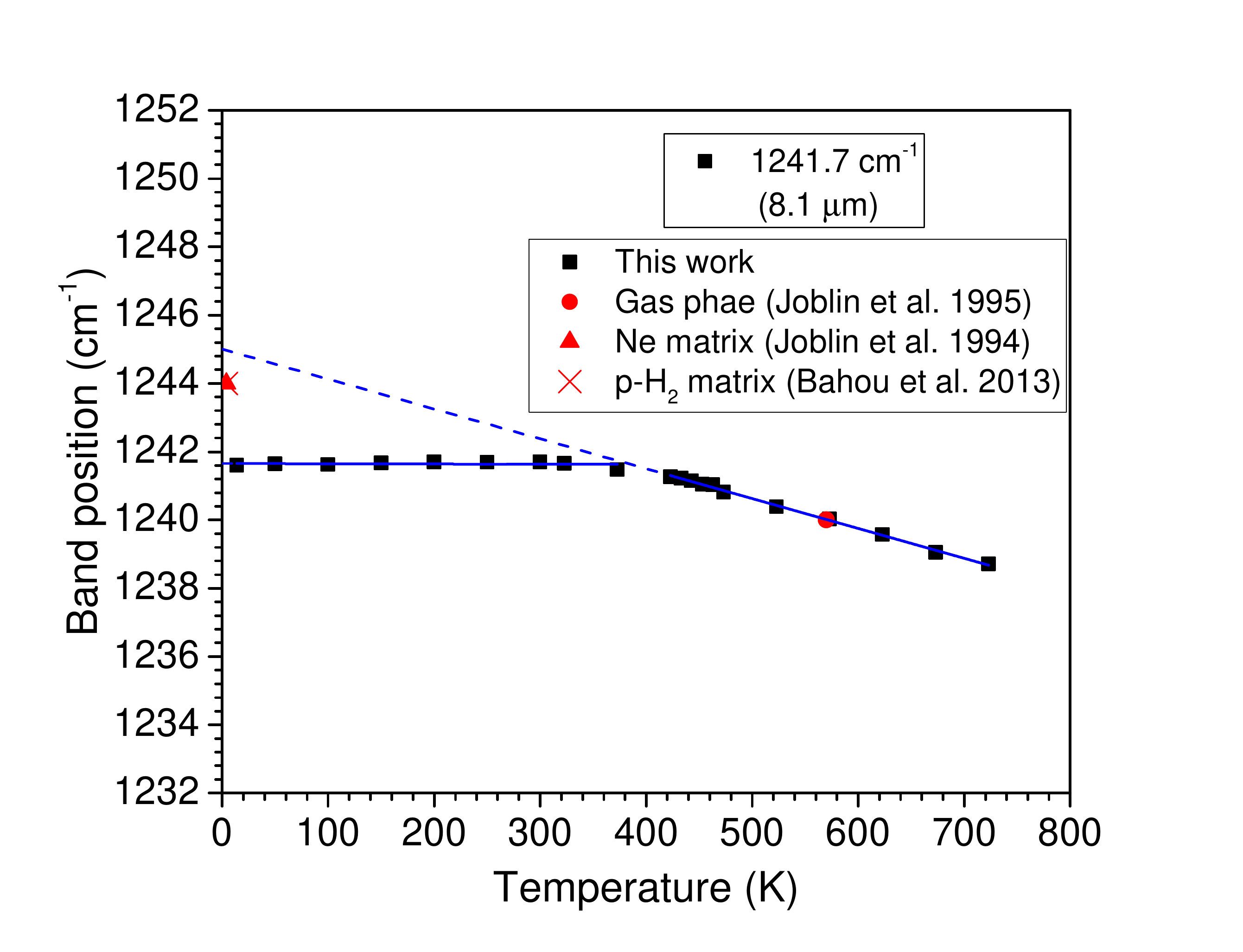} & \includegraphics[width=0.5\hsize]{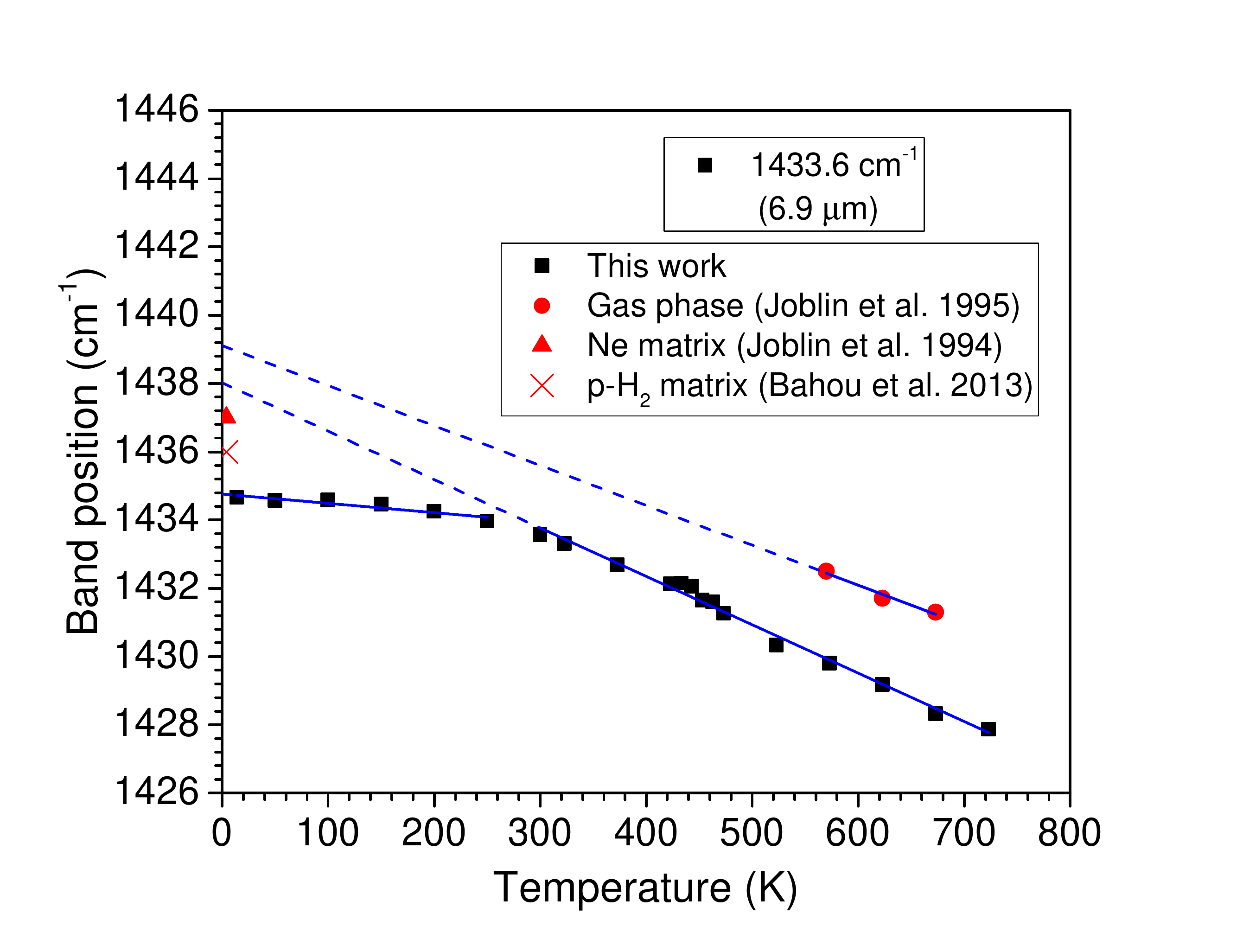}\\
(e) & (f)\\
\end{tabular}
\caption{Evolution of the band positions with temperature for fundamental transitions of pyrene in KBr pellets (black squares). 
Band positions (in \cm\ and $\mu$m) at 300\,K are mentioned in the top right corner of each panel.
The calculated linear fits are shown by blue lines. Also shown with red marks are published data both at high temperature in the gas-phase \cite{Joblin1995} and at very low temperature (3-4\,K) either in the gas-phase \cite{maltseva2016}, Ne matrix\cite{Joblin1994} or p-H$_{2}$ matrix\cite{bahou2013}. (continued from the previous page)}
\label{posn_T}
\end{figure*}

\begin{table*}
\centering
\footnotesize
\caption{Empirical anharmonicity factors ($\chi'$) of the main IR bands of pyrene derived from the linear fitting of the band position over different temperature ranges
(Figures\,\ref{posn_T} and S3 of the Supporting Information). 
(a) are values from this work and (b) from gas-phase measurements\cite{Joblin1995,Joblin1992}.
$\chi'_\mathrm{rec}$ are the recommended values for the anharmonicity coefficients, which have been obtained from either the value of $\chi'$ at high temperature in case of no jump in band position, or the average value of $\chi'$ values before and after the jump.  The error only takes into account the change of slope before and after the jump since the fitting error is below the precision limit reported here. Type 1, 2 and 3 denotes the isolated, low mixed and highly mixed band profiles, respectively.}

\hspace{2cm}
\begin{tabular}{c c c c c c c c}
\hline
\multicolumn{2}{c}{Position} & \multicolumn{1}{c}{Type} & \multicolumn{1}{c}{Absorbance}& \multicolumn{1}{c}{Fit Range}& \multicolumn{3}{c} {Empirical anharmonicity factors}\\
\multicolumn{2}{c}{(300\,K)} & & \multicolumn{1}{c}{(300\,K)} &  & \multicolumn{1}{c}{$\chi'^{(a)}$} & \multicolumn{1}{c}{$\chi'_\mathrm{rec}$} & \multicolumn{1}{c}{$\chi'^{(b)}$} \vspace{1mm}\\
\multicolumn{1}{c}{$\mu$m} & \multicolumn{1}{c}{$\cm$} &  &  & \multicolumn{1}{c}{K} & \multicolumn{3}{c} {10$^{-2}$\cm\K} \vspace{1mm}\\
\hline \\
\multirow{2}{*}{14.1}& \multirow{2}{*}{709.7} & \multirow{2}{*}{2} & \multirow{2}{*}{0.71} &14 - 150 & $-0.3$ & & \\
%                              &  & & & & & -0.6 &  -1.1\\
                & & & & 150 - 723&$-0.8$ & $-0.8$ & $-1.1$ \vspace{2mm}\\ 
\multirow{3}{*}{13.3}& \multirow{3}{*}{749.4} & \multirow{3}{*}{2} &  \multirow{3}{*}{0.19} &14 - 150& $-0.7$ & & \\
                   & & & &150 - 423& $-1.6$ & \multirow{2}{*}{$-1.5 \pm 0.15$}&\multirow{2}{*}{$-1.1$}\\
                  &    & & &473 - 723& $-1.3$ & &\vspace{2mm}\\
\multirow{2}{*}{12.2}& \multirow{2}{*}{820.3} & \multirow{2}{*}{2} & \multirow{2}{*}{0.19} & 14 - 150& $-0.3$ & & \\
                  &    & & &150 - 723& $-0.2$ & $-0.2$ \vspace{2mm}\\
\multirow{2}{*}{11.9} & \multirow{2}{*}{839.9} & \multirow{2}{*}{2} & \multirow{2}{*}{0.83} &14 - 150& $-0.4$ & & \\
                & & & &150 - 723& $-1.1$ & $-1.1$ & $-1.4$  \vspace{2mm}\\
\multirow{2}{*}{10.4}& \multirow{2}{*}{964.7} & \multirow{2}{*}{3} & \multirow{2}{*}{0.16} &14 - 150& 0.3 & & \\
                  & &  & &150 - 723& $-1.4$ & $-1.4$ \vspace{2mm}\\
\multirow{3}{*}{9.9} & \multirow{3}{*}{1000.9} & \multirow{3}{*}{1} & \multirow{3}{*}{0.03} & 14 - 150& 0.2 &
& \multirow{3}{*}{}\\
                    & & &   &150 - 423& 0.0 &\multirow{2}{*}{$-0.4\pm 0.4$}\\
   &       &     & & 473 - 723 & $-0.8$ & \vspace{2mm}\\
\multirow{3}{*}{9.1}& \multirow{3}{*}{1096.3} & \multirow{3}{*}{1}& \multirow{3}{*}{0.07} & 14 - 150 & $-0.3$ &
& \\
                & & &  & 150 - 423 & $-0.6$ & \multirow{2}{*}{$-0.5$ $\pm$ 0.15} & \multirow{2}{*}{$-0.2$}\\
               &   & & & 473 - 723 & $-0.3$ &\vspace{2mm}\\
\multirow{3}{*}{8.4} & \multirow{3}{*}{1184.8} & \multirow{3}{*}{1} & \multirow{3}{*}{0.29} & 14 - 150 & $-0.5$ &
& \\
                & & &  & 150 - 423 & $-0.8$ &\multirow{2}{*}{$-0.7$ $\pm$ 0.15} & \multirow{2}{*}{$-0.9$}\\
                &  & & & 453 - 723 & $-0.5$ & \vspace{2mm}\\
\multirow{2}{*}{8.1} & \multirow{2}{*}{1241.7} & \multirow{2}{*}{1} & \multirow{2}{*}{0.22} & 14 - 423 & 0.0 &
\multirow{2}{*}{$-0.4\pm0.4$} & \multirow{2}{*}{}\\
    &       &     & & 423 - 723  & $-0.8$ &\vspace{2mm}\\ 
\multirow{3}{*}{7.6} & \multirow{3}{*}{1313.2} & \multirow{3}{*}{2} & \multirow{3}{*}{0.14} & 14 - 200 & $-0.3$ &
& \multirow{3}{*}{}\\
                    & & & & 200 - 423 & $-1.0$ &\multirow{2}{*}{$-1.3$ $\pm$ 0.25}\\
                &  & & & 473 - 723 & $-1.5$ & \vspace{2mm}\\
\multirow{2}{*}{6.9} & \multirow{2}{*}{1433.6} & \multirow{2}{*}{3} & \multirow{2}{*}{0.26} & 14 - 250 & $-0.3$ & & \\
                 &  & & & 300 - 723 & $-1.4$ & $-1.4$ & $-1.1$ \vspace{2mm}\\
\multirow{2}{*}{6.7} & \multirow{2}{*}{1486.5} & \multirow{2}{*}{1} & \multirow{2}{*}{0.06} & 14 - 250 & $-0.3$  & & \multirow{2}{*}{}\\
                 & & & & 300 - 723 & $-1.3$& $-1.3$ \vspace{2mm}\\
\hline
\label{Tab:bp_T}
\end{tabular}
\end{table*}

\subsection{Evolution of the band widths and intensities with temperature}
\begin{comment}
\begin{figure*}
\centering
\begin{tabular}{c c}
\includegraphics[width=0.5\hsize]{Bandwidth_1241_1096_1001_Karine_new_data.pdf} & \includegraphics[width=0.5\hsize]{Bandwidth_1486_1186_Karine_new_data.pdf}\\
(a) & (b)\\
\end{tabular}
\caption{Evolution of the bandwidth with temperature for five Type 1 bands (c.f. Table \ref{Tab:bp_T}) of pyrene. 
Band positions (in \cm) at 300\,K are mentioned in the top left corner of each figure.}
\label{bw_T}
\end{figure*}
\end{comment}
\begin{figure}
\centering
\includegraphics[width=1.1\columnwidth]{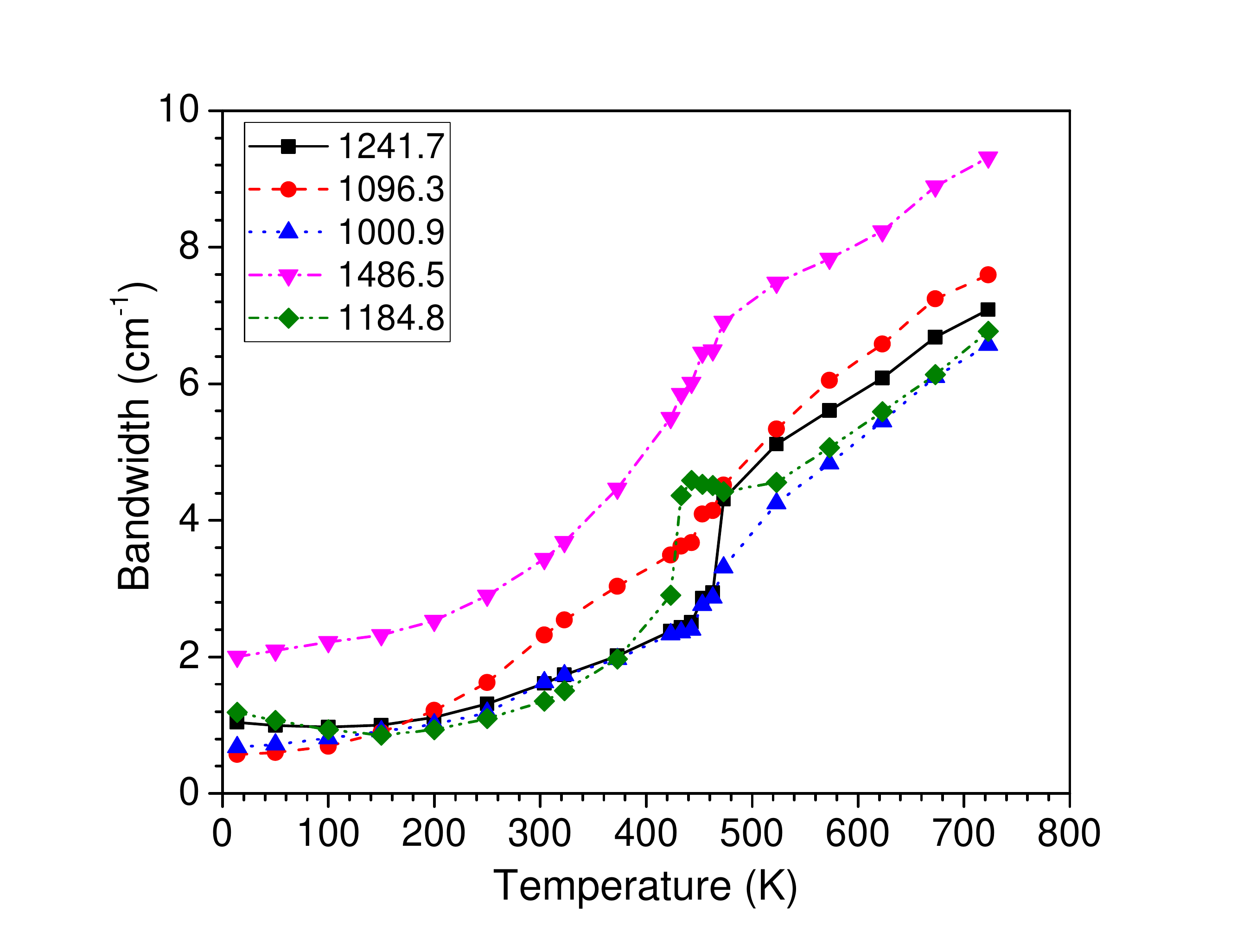}\\
\caption{Evolution of the bandwidth with temperature for five Type 1 bands (c.f. Table \ref{Tab:bp_T}) of pyrene. 
Band positions (in \cm) at 300\,K are mentioned in the top left corner of each figure.}
\label{bw_T}
\end{figure}

In Figure~\ref{bw_T} we present the variation of the bandwidth (FWHM) with temperature for several bands which are rather well isolated and do not exhibit significant structure. Globally the band width increases with temperature but with different regimes: it varies little at very low temperatures, then it steepens gradually above $\sim$200~K. Similarly to the case of some band positions, a change of slope is observed below and above the [443-473]\,K range. We note that the discontinuity in the [443-473]\,K range appears clearly in the band width of some bands, especially those at 1184.3 and 1241.7\,\cm.
\\
The band width includes many effects, among which mode coupling (anharmonicity) of interest for this study. In previous gas-phase measurements at thermal equilibrium there was in addition a strong contribution due to rotation, which had to be removed in astrophysical models.\cite{Joblin1995,pech2002}
In condensed phases, band broadening is expected to be dominated by interaction with the environment with minor or no contribution from rotational broadening. Only the band at 1184.8 cm$^{-1}$ could be compared between our new data and previous gas-phase data. The agreement is within 50\% (see Table \ref{Tab:bw_T}).
\begin{table*}
\centering
\footnotesize
\caption{Empirical anharmonicity factors ($\chi''$) derived from fitting the bandwidths over different temperature ranges for the five Type 1 (c.f. Table \ref{Tab:bp_T} for the definition of Type 1 bands) bands of pyrene. (a) are values from this work and (b) from gas-phase measurements\cite{Joblin1995,Joblin1992}.
$\chi''_\mathrm{rec}$ are the recommended values for the anharmonicity coefficients, which have been obtained from the average value of $\chi''$ values derived below 423~K and above 473~K.}

\hspace{2cm}
\begin{tabular}{c c c c c c c c}
\hline
\multicolumn{2}{c}{Position} & \multicolumn{1}{c}{Type} & \multicolumn{1}{c}{Absorbance}& \multicolumn{1}{c}{Fit Range}& \multicolumn{3}{c} {Empirical anharmonicity factors}\\
\multicolumn{2}{c}{(300\,K)} &  & \multicolumn{1}{c}{(300\,K)} &  & \multicolumn{1}{c}{$\chi''^{(a)}$} & \multicolumn{1}{c}{$\chi''^{(a)}_{rec}$} & \multicolumn{1}{c}{$\chi''^{(b)}$} \vspace{1mm}\\
\multicolumn{1}{c}{$\mu$m} & \multicolumn{1}{c}{$\cm$} &  &  & \multicolumn{1}{c}{K}& \multicolumn{3}{c} {10$^{-2}$\cm\K} \vspace{1mm}\\
\hline \\
\multirow{2}{*}{9.9} & \multirow{2}{*}{1000.9} & \multirow{2}{*}{1}  & \multirow{2}{*}{0.03} & 250 - 443 & 0.6 & \multirow{2}{*}{$1.0 \pm 0.4$}\\
    &       &     &      & 473 - 723 & 1.3 & \vspace{2mm}\\
\multirow{2}{*}{9.1} & \multirow{2}{*}{1096.3} & \multirow{2}{*}{1}  &  \multirow{2}{*}{0.07} & 200 - 443 & 1.0& \multirow{2}{*}{$1.1\pm 0.1$}\\
                  &   & & & 473 - 723 & 1.2& \vspace{2mm}\\
8.4 & 1184.8      & 1 & 0.29  & 523 - 723 & 1.1 & 1.1 & 1.6 \vspace{2mm}\\
\multirow{2}{*}{8.1} & \multirow{2}{*}{1241.7} & \multirow{2}{*}{1}   & \multirow{2}{*}{0.22} & 200 - 443 & 0.6& \multirow{2}{*}{$0.9 \pm 0.3$}\\
    &        &     & & 473 - 723  & 1.1& \vspace{2mm} \\ 
\multirow{2}{*}{6.7} & \multirow{2}{*}{1486.5} & \multirow{2}{*}{1} &\multirow{2}{*}{ 0.06} & 250 - 443 & 1.8& \multirow{2}{*}{$1.5\pm 0.4$}\\
                 & & & & 473 - 723 & 1.1& \vspace{2mm}\\
\hline
\label{Tab:bw_T}
\end{tabular}
\end{table*}

Finally, we present in Figures S6 and S7 (see Supporting Information), the variation of the integrated band intensities with temperature for the complete spectrum (3200-650~\cm) and for the individual bands, respectively. The values were normalized to their value at 300\,K. For the complete spectrum the intensity is rather constant up to $\sim$400~K. Then there is a jump up to another nearly constant value 20\% higher from $\sim$500~K to 723~K.
From the variation of the integrated band intensity of individual bands (cf. Figure S7 in Supporting Information) with temperature we can identify three categories. (i) constant value within 10\% for the 749.4, 964.7, 1184.8 and 1599.8~\cm~bands, (ii) increase by up to 85\% after a jump in the 423-473~K range for the 709.7, 839.9, 1000.9, 1096.3 and 3048.3 ~\cm~bands, and (iii) decrease for the 820.3, 1241.7, 1313.2, 1433.6 and 1486.5~\cm~bands. This decrease reaches typically 50\% but is especially strong for the 820.3~\cm~band with a drop of the intensity by a factor of 5. A decrease of the integrated band intensity with temperature is expected when the intensity is leaking in various hot bands which fall outside the integration range. However, in general, both the increase and decrease of the intensity appear correlated with a change in the 423-473~K range, whose origin is discussed in the next section.

\section{Discussion}
\label{discussion}

The main objective of this work is to provide experimental data that allow us to quantify the effect of anharmonicity on the IR spectra of hot PAHs. In particular the empirical anharmonicity factors, which can be derived from the evolution of the band positions and widths with temperature can be used in astronomical models to simulate the emission of PAHs following the absorption of UV photons.\citep{pech2002} In addition, they can be valuable data to test the reliability and accuracy of codes that calculate anharmonic spectra.\cite{mackie2016,mulas2018,chen2018} The biggest advancement of this work is to report uniformly measured data over the full 14 to 723\,K temperature range, whereas previous studies were either limited to low temperature ($\sim$4\,K)\cite{maltseva2016,Joblin1994,bahou2013} or to high temperatures [573-873]\,K.\cite{Joblin1995}  The main drawback of the used experimental methodology, though, is that the data are recorded in condensed phase, consisting of solid grains embedded in KBr pellets.

The stable configuration of solid pyrene in normal conditions is referred to as Phase~I (cf. Figure~S9 of the Supporting Information). It has been initially imaged in X rays by Robertson and White, \cite{Robertson1947} and later classified as a sandwich herringbone structure based on geometrical considerations. \cite{Desiraju1989} It is composed of dimers and involves both van der Waals forces associated with the PAH stacking and \mbox{C-H$\cdots \pi$} interaction between dimers arranged in the herringbone pattern.
The evolution of the spectroscopic signatures associated with the phase transitions of pyrene has been studied by several authors, most of them focusing on the low temperature transition. \cite{wincke1970,brehat1976, zallen1976} Brehat et al.\cite{brehat1976} reported evidence for a contraction of the crystal lattice at low temperature (below ~93 K) by observing the evolution of the low-frequency IR bands with temperature. Also, Zallen et al.\cite{zallen1976} recorded the Raman spectrum of solid pyrene crystal as a function of temperature and pressure. They observed a discontinuity in the band position of pyrene for the modes below 130\,\cm. They attributed it to the phase transition from the Phase I form to the more compact Phase II form, \cite{frampton2000} which can be obtained both at low temperature (110~K, 0~kbar) and at high pressure (4~kbar, 300~K).
Evidence for a Phase III form with a different packing structure compared to the others is reported by some authors.\cite{fabbiani2006} 
\\
In our data, we do not see evidence for a phase transition at low temperature ($\sim$ 110\,K). 
We however observe a clear jump in the characteristics (position, width and intensity) of some bands at higher temperatures, in the [423-473]\,K range. These values are close to the melting temperature of solid pyrene, which was measured at 424\,K.\cite{wong1971}
The fact that this transition happens on an extended range might be related to the time it requires to achieve the full melting of the solid grains, which are of different sizes.  This scenario is however difficult to reconcile with the timing of the experiments. Our measurements are performed with typically 45 min. interval between two spectral acquisitions, which corresponds, including the time to record the spectra, to a couple of hours spent on the transition region. Another possibility would be that the grain size distribution leads to a range of melting temperatures. Indeed, molecular dynamics simulations show that the melting temperature of PAH clusters decreases with the size of the clusters and is expected to reach the value of the solid at large sizes.\cite{chen2014}.  Still, this could explain the width of the transition but not the overall shift towards higher temperatures compared with the melting temperature of solid pyrene.
Several studies have reported an elevation of the melting point of submicron sized particles embedded in higher melting matrices. In particular, Allen et al. reported an increase of 12$^\circ$C in the case of microcrystals of tin in amorphous carbon.\cite{allen1980}. A similar effect was reported in other systems\cite{saka1988,malhotra1991} and is likely due to a strain energy effect. We propose that a similar scenario is involved for pyrene grains embedded in solid KBr, which melts at a temperature of 1007\,K.

All the data we gathered correspond therefore to the crystal Phase I below 423~K and to the molten phase above $\sim$473~K.
The features associated with our IR spectra contain then information on both intramolecular anharmonicity (which we are interested in) and intermolecular anharmonicity related to the two involved condensed phases. The latter can impact band positions, widths and intensities \cite{lucazeau2003}. In both cases, the intermolecular forces are dominated by the interactions between pyrene molecules. In the Phase I crystal (see Figure~S9 of the Supporting Information), one side of each molecule interacts in a $\pi-\pi$ stacking configuration whereas the other side involves mainly \mbox{C-H$\cdots\pi$} interactions.\cite{mcKinnon2004} How intermolecular forces are going to be affected by melting will depend on the possibility of volume expansion within the KBr pellet. If this can be fully achieved then intermolecular distances are expected to increase and molecules to have a larger mobility both in rotation and vibration. \cite{wong1971, dunitz1999} However, as we discussed above, the molten phase is not free to fully expand and is somehow confined by the KBr matrix.

\begin{figure}
\centering
\includegraphics[width=1.1\columnwidth]{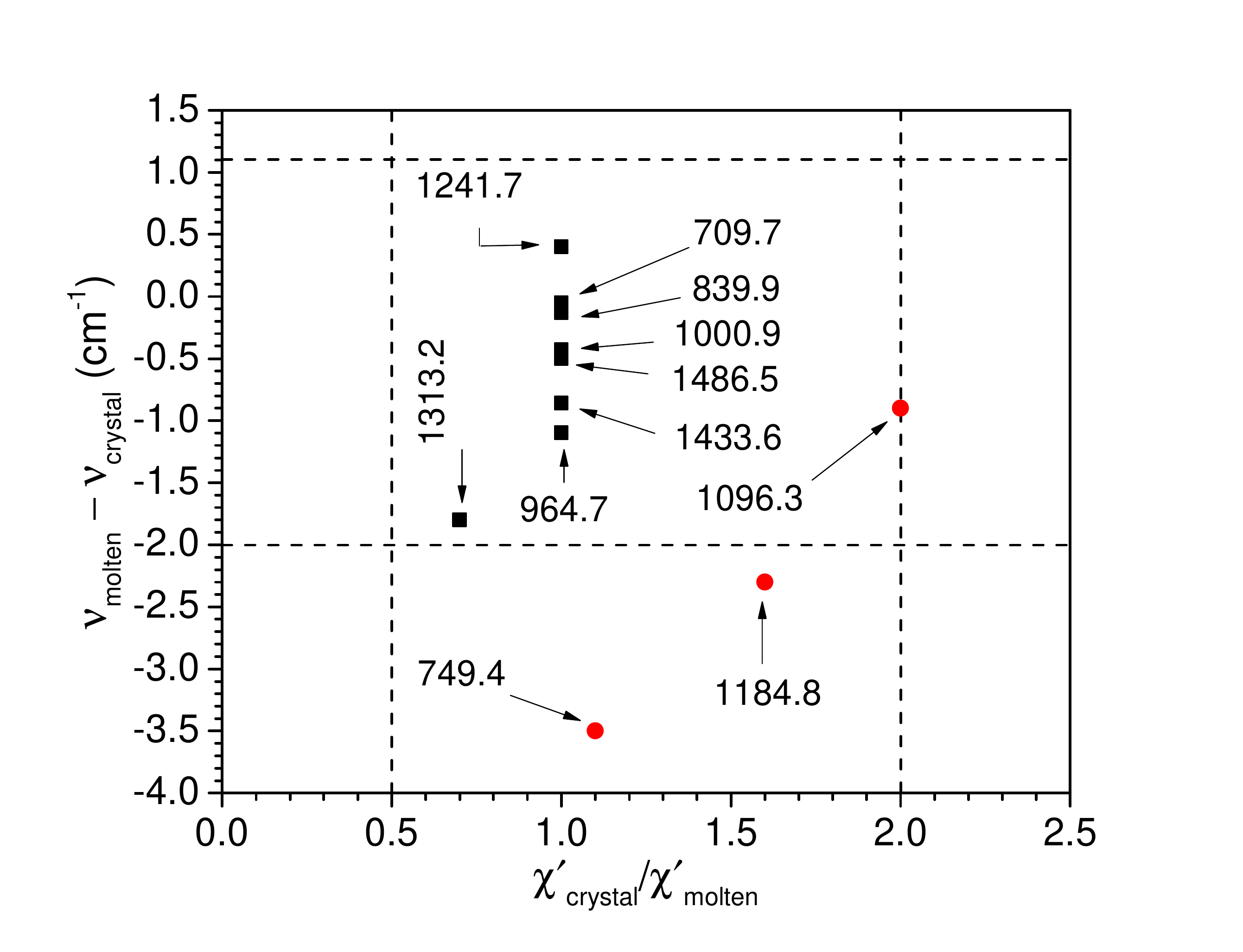}\\
\caption{Band shift between the molten and crystal phases of pyrene (cf. Figure~\ref{phase_transition}) versus the ratio of the anharmonicity factors between both phases (Table\,\ref{Tab:bp_T}). The band shift corresponds to the difference in band position between 473 and 423~K. The ratio of anharmonicity factors is one for bands having the same anharmonic factor over the studied temperature range. The points, which significantly deviate from average values, are identified by red dots.}
\label{anh_bs}
\end{figure}

Both phases, microcrystal and molten, are condensed and involve intermolecular interactions which are not present in the gas-phase. We have seen that the transition between both phases can induce jumps in some of the characteristics (position, width and/or intensity) of some specific bands. Although our analysis of band widths is limited to a smaller set of bands, we can see that a jump in band position is not systematically associated with a jump in band width or integrated intensity. In fact, we show in Figure~S5 of the Supporting Information the case of the 1184.8~\cm\ band, which is the only case for which we have identified a jump in both band position and width (and not intensity). Our data suggest that there are several processes that affect in a different way the band characteristics. 
Figure\,\ref{anh_bs} summarizes the variations of the band positions and anharmonic factors induced by the change of phases of pyrene. We identified the bands that are more affected by intermolecular forces, leading either to jump in band position which is over 2\,\cm\ or to a change in the anharmonicity factor by more than a factor of 2. From our selected band set, the first criteria concern the two bands at 749.4 and 1184.8~\cm\ and only the band at 1096.3 cm$^{-1}$ meets at the limit the second criteria. The band at 749.4~\cm\ is a C-H out-of-plane bend and, similarly to the 3048.3\,\cm\ band (the complex CH stretch motion range that we excluded from our analysis), involves large amplitude motions of the H nuclei. This induces a significant volume change (cf. (b), (n) and (o) in Figure\,S10 from Supporting Information) and therefore an expected stronger interaction with the surrounding. This is less the case for the C-H in-plane bend, although we can see that the band at 1184.8~\cm\ somehow interacts with the environment. On average, we can conclude that intermolecular effects are expected to be a concern in studies of intramolecular anharmonicity for only a couple of bands (those involving large CH motions).

From this study, we can conclude that the derived empirical anharmonicity factors are on the average consistent with available gas-phase data despite the phase change experienced by pyrene (Table\,\ref{Tab:bp_T} and Figure\,\ref{anh_fact_microcrystal_melted_gas}). This is also true for the two bands which have been identified above as more responsive to the environment. 
It is difficult to further elaborate on which condensed phases is more appropriate to retrieve anharmonic factors which would be consistent with gas-phase values and therefore relevant for astrophysical models. Such comparison would require to process all the data in a consistent way. The determination of the band positions is indeed somehow dependent on the analysis method. We have used here a weighted average, which differs from what was done in earlier gas-phase experiments. In addition, gas-phase data contain additional rotational broadening. In Tables~\ref{Tab:bp_T} and \ref{Tab:bw_T}, we report recommended values of the anharmonicity factors, which were retrieved from the measurements in both condensed phases. We consider these values as the best values we can provide for inputs in astronomical models and comparison with molecular dynamics simulations.

\section{Conclusions}
The IR spectrum of condensed pyrene is reported for the first time over a wide range of temperature (14 -723 \,K). The sample was prepared by embedding pyrene microcrystals in a KBr matrix. Over the studied temperature range, pyrene experienced two phases: Phase I microcrystal and molten.
We found spectral fingerprints of this phase transition given by jumps in the positions, widths and/or integrated intensities of several bands. We could not however identify systematic effects, except that the response to the environment seems more important for bands that involve 
large CH motions.
\\
This study allows us to retrieve values for the empirical anharmonic factors that describe the evolution of the band positions over an unprecedented temperature range. As far as the comparison can be performed, the values obtained in condensed phase are consistent with those published from gas-phase data. The experimental methodology used in this work is much easier than gas-phase studies and this opens perspectives to generalize these measurements on other species, in particular large PAHs (50-100 C atoms) which are better analogs of the astro-PAHs.

%%%%%%%%%%%%%%%%%%%%%%%%%%%%%%%%%%%%%%%%%%%%%%%%%%%%%%%%%%%%%%%%%%%%%

%%%%%%%%%%%%%%%%%%%%%%%%%%%%%%%%%%%%%%%%%%%%%%%%%%%%%%%%%%%%%%%%%%%%%
\begin{acknowledgement}
The research leading to these results was supported by the funding received from European Research Council under the European Union's seventh framework program (FP/2007-2013) ERC-2013-SyG, Grant agreement n. 610256 NANOCOSMOS. We thank Dr. Anthony Bonnamy and Lo\"ic Nogu\`es for their technical assistance during the experiment. We sincerely thank the reviewers for their insightful comments on the manuscript.
\end{acknowledgement}

%%%%%%%%%%%%%%%%%%%%%%%%%%%%%%%%%%%%%%%%%%%%%%%%%%%%%%%%%%%%%%%%%%%%%
\section{Dataset}

The dataset associated with this work can be found under 10.5281/zenodo.2605449

%%%%%%%%%%%%%%%%%%%%%%%%%%%%%%%%%%%%%%%%%%%%%%%%%%%%%%%%%%%%%%%%%%%%%
\begin{suppinfo}
\begin{itemize}
    \item IR spectra of pyrene showing Christiansen effect 
    \item IR band profiles of six fundamental transitions of pyrene at various temperatures
    \item Evolution with temperature and multi-component fitting of the 749.4~\cm~band 
    \item Evolution with temperature and multi-component fitting of the 839.9~\cm~band
    \item Evolution with temperature and multi-component fitting of the 1184.8~\cm~band
    \item Evolution of band positions with temperature 
    \item Variation with temperature of band intensity, integrated over the complete spectral range 3200-650\,\cm
    \item Variation of integrated band intensity with temperature zoomed in on each fundamental transition 
    \item Schematic representation of phase 1 structure of pyrene 
    \item Figures showing harmonic vibrational modes of pyrene.
\end{itemize}
\end{suppinfo}

%%%%%%%%%%%%%%%%%%%%%%%%%%%%%%%%%%%%%%%%%%%%%%%%%%%%%%%%%%%%%%%%%%%%%

%%%%%%%%%%%%%%%%%%%%%%%%%%%%%%%%%%%%%%%%%%%%%%%%%%%%%%%%%%%%%%%%%%%%%
\bibliography{pyrene_expt}

\end{document}

\begin{abstract}
This Supporting Information contains:
\begin{itemize}
    \item Figure~\ref{christiansen}: IR spectra of pyrene showing Christiansen effect 
    \item Figure~\ref{spectra}: IR band profiles of six fundamental transitions of pyrene at various temperatures 
    \item Figure~\ref{749band}: Evolution with temperature and multi-component fitting of 
    the 749.4~\cm~band 
    \item Figure~\ref{839band}: Evolution with temperature and multi-component fitting of the 839.9~\cm~band
    \item Figure~\ref{1185band}: Evolution with temperature and multi-component fitting of the 1184.8~\cm~band
    \item Figure~\ref{posn}: Evolution of band positions with temperature 
    \item Figure~\ref{tot_int}: Variation with temperature of band intensity, integrated over the complete spectral range 3200-650\,\cm
    \item Figure~\ref{band_int}: Variation of integrated band intensity with temperature zoomed in on each fundamental transition
    \item Figure~\ref{pyr_str}: Schematic representation of phase 1 structure of pyrene 
    \item Figure~\ref{vib_str}: Figures showing harmonic vibrational modes of pyrene.
\end{itemize}
\end{abstract}

\maketitle

\begin{figure}
\centering
\includegraphics[width=1.15\columnwidth]{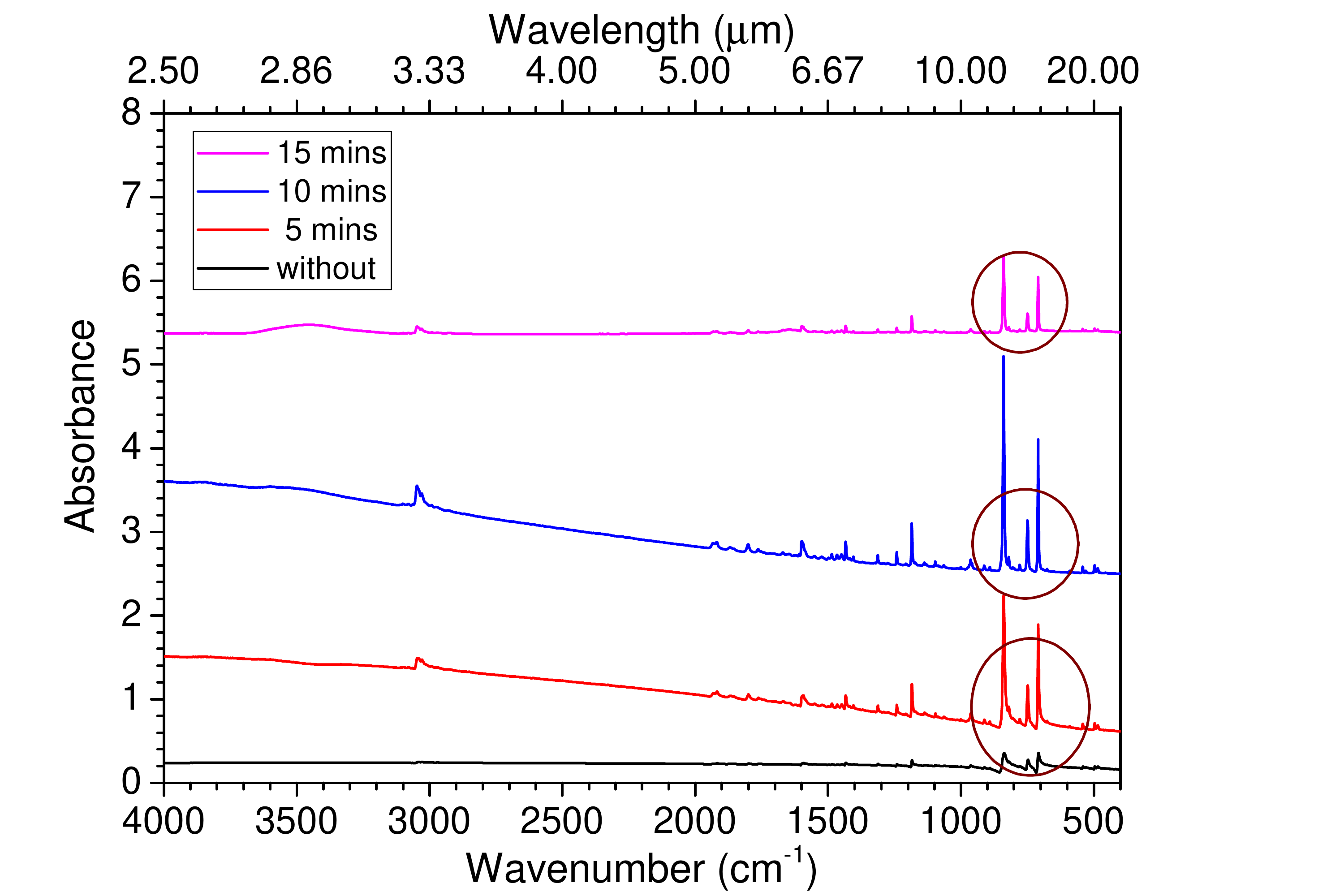} \\
\caption{Infrared spectrum of solid pyrene in KBr pellet at 300\,K with different grinding time. Christiansen effect below 1000~\cm\ is quite evident when the spectrum was taken without and after 5 minutes of grinding the mixture of KBr and pyrene. Best spectrum was obtained after grinding the mixture of KBr and pyrene for 15 minutes.}
\label{christiansen}
\end{figure}

\begin{figure*}
\centering
\begin{tabular}{c c}
\includegraphics[width=0.5\hsize]{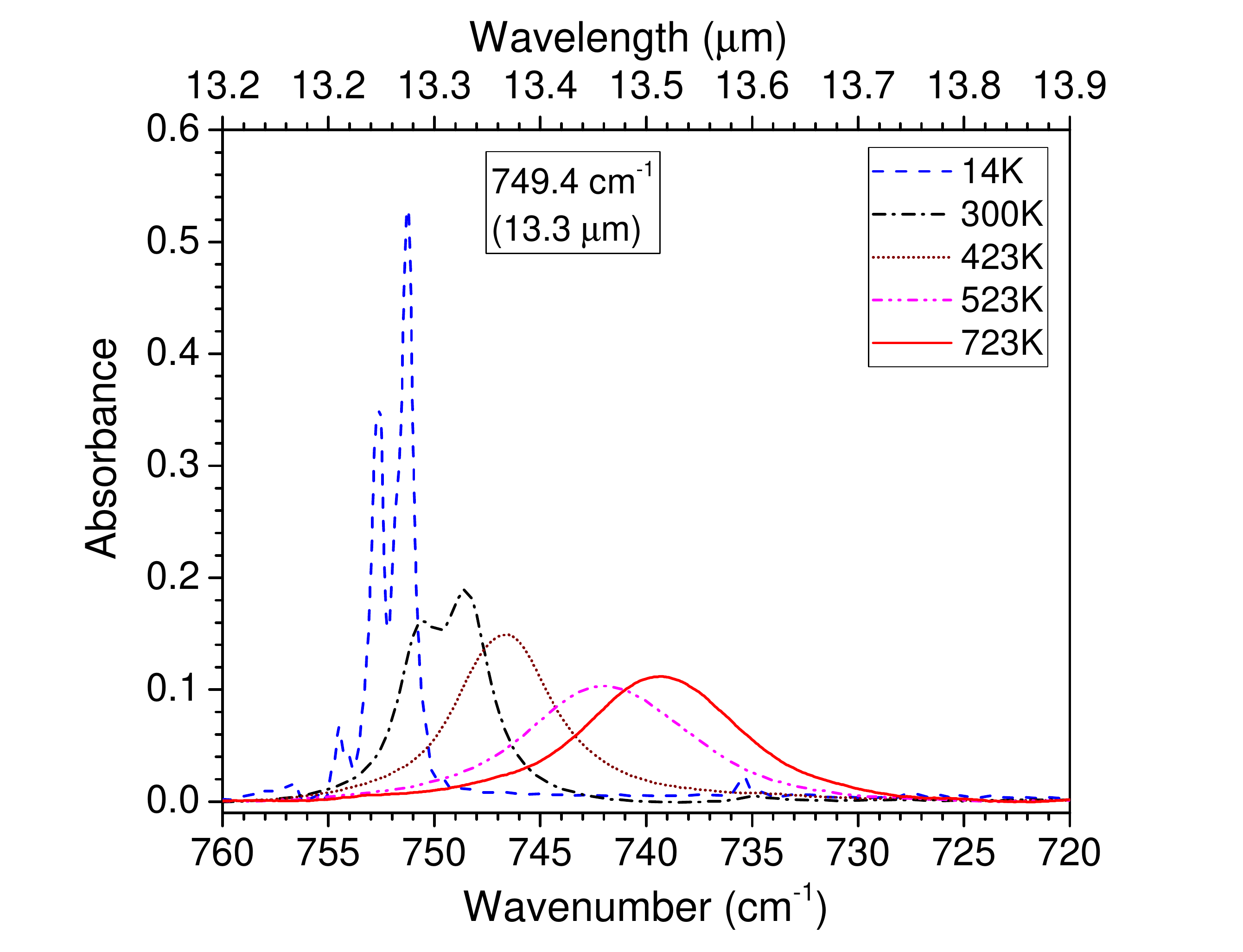} & \includegraphics[width=0.5\hsize]{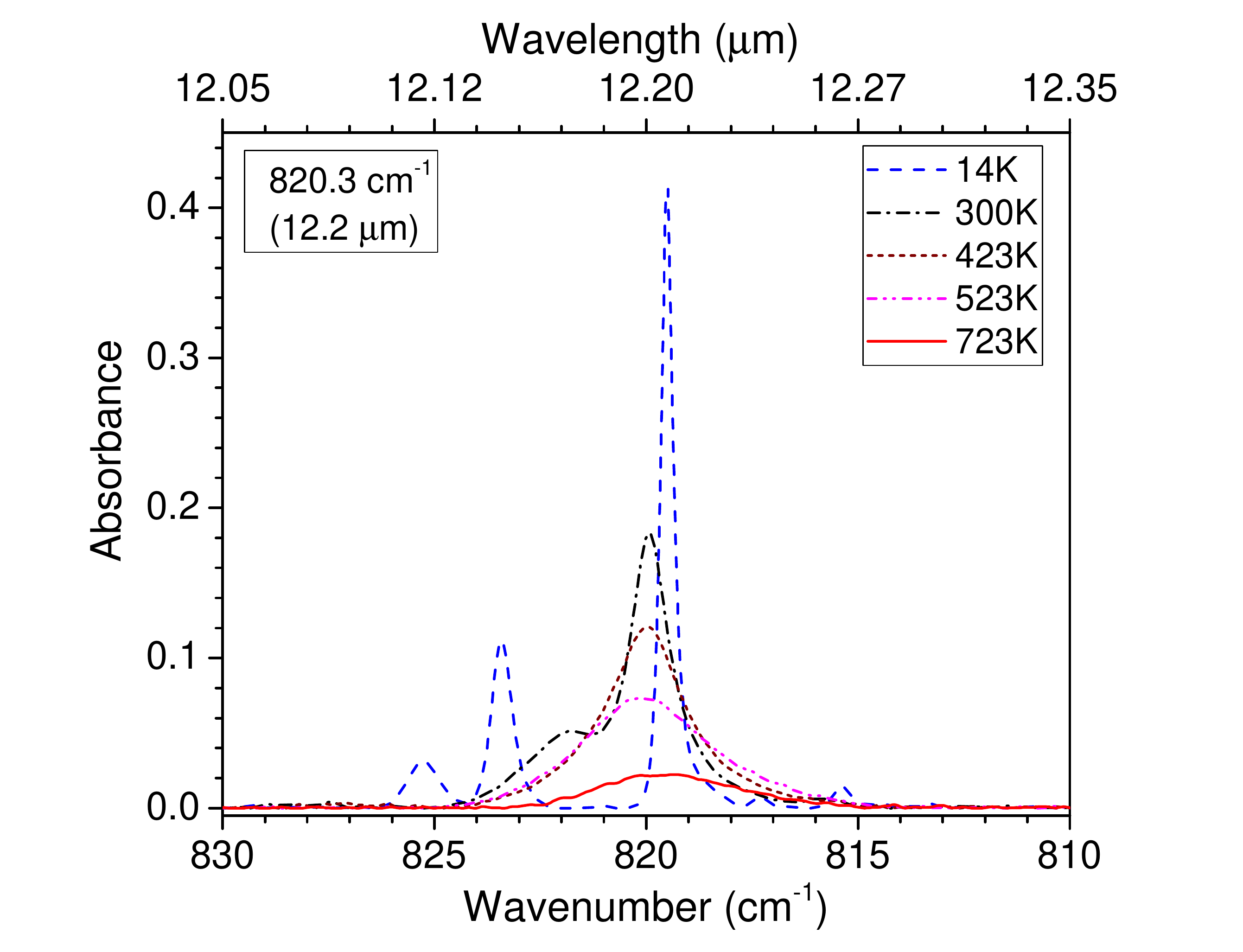}\\
(a) & (b)\\
 \includegraphics[width=0.5\hsize]{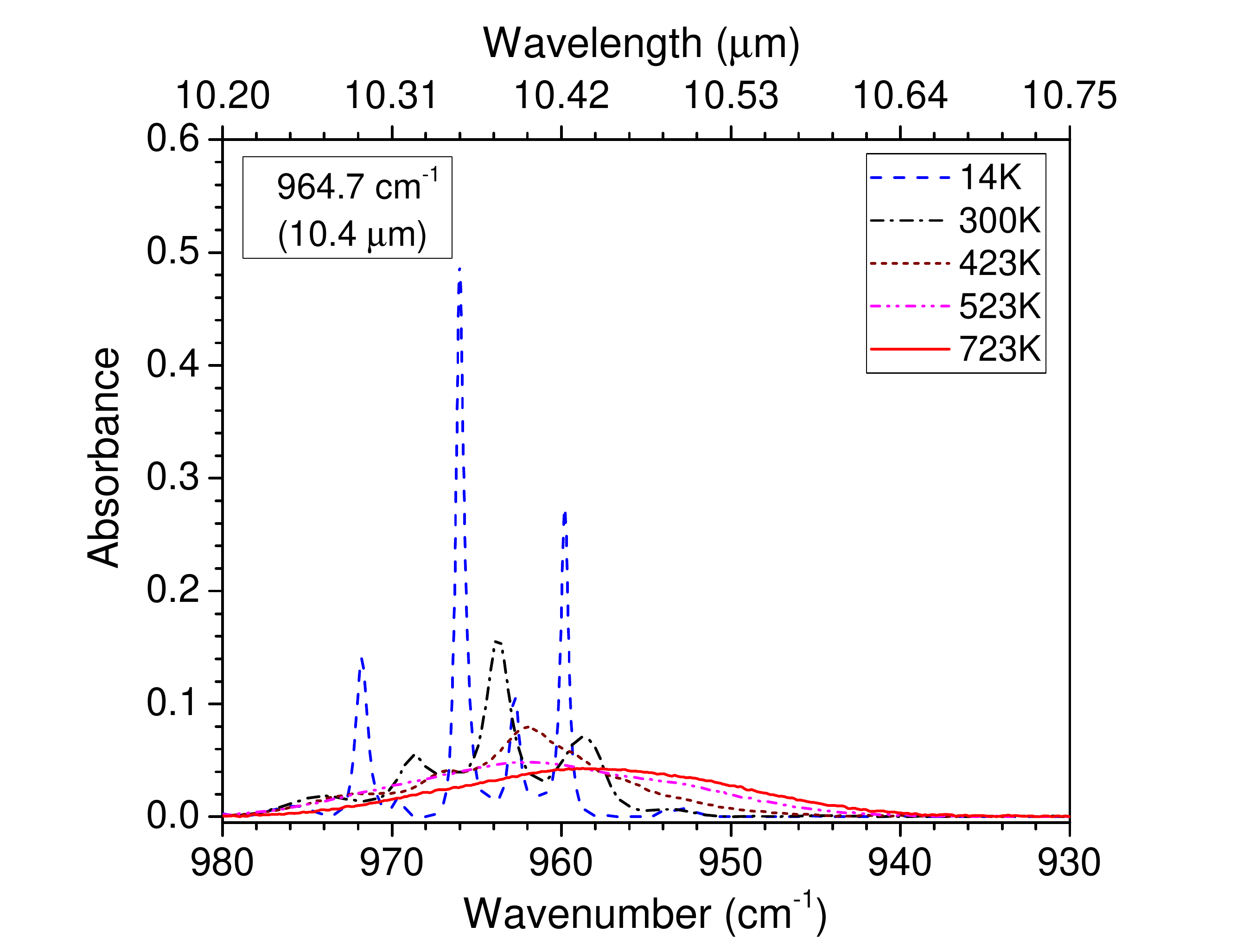}& \includegraphics[width=0.5\hsize]{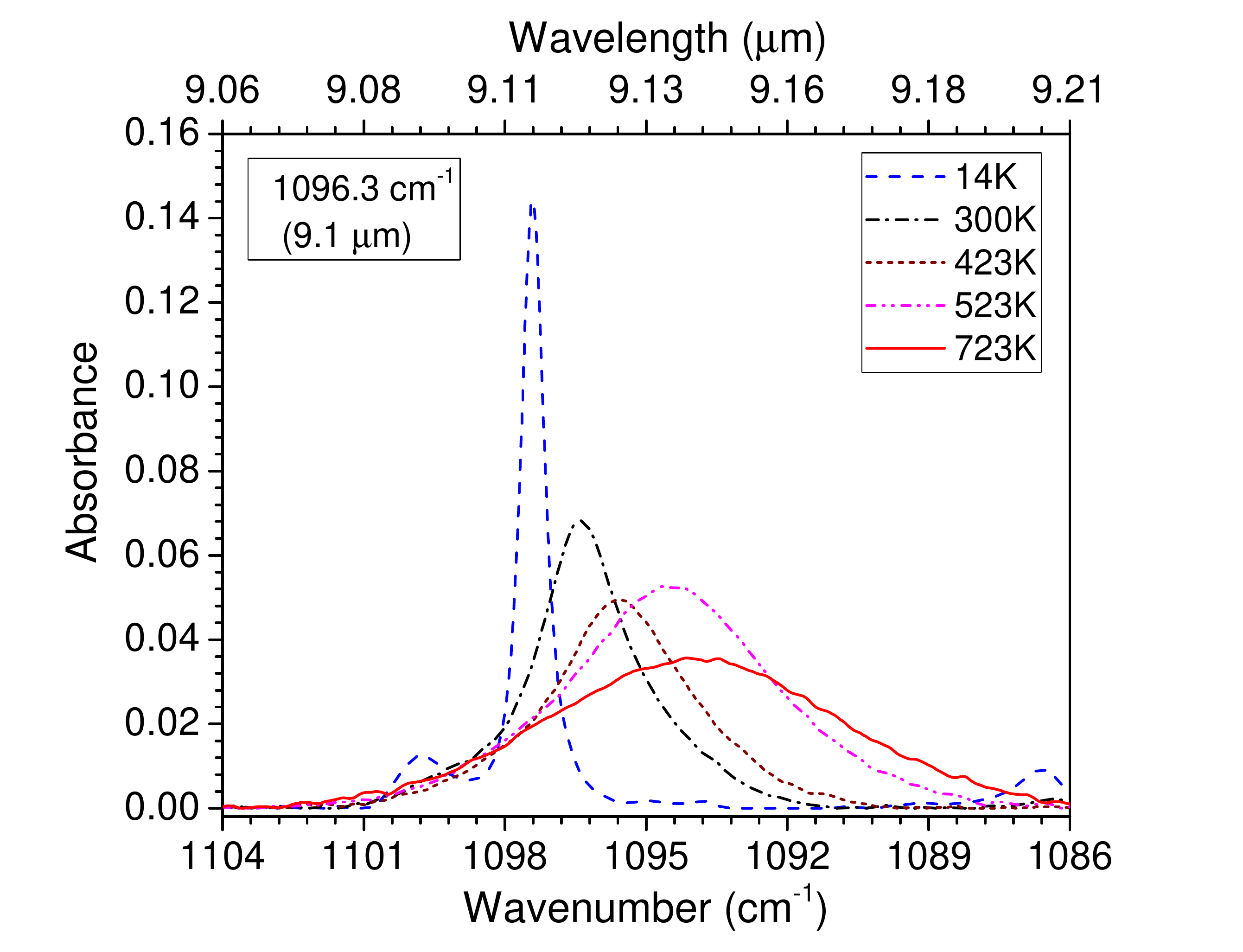}\\
(c) & (d)\\
 \includegraphics[width=0.5\hsize]{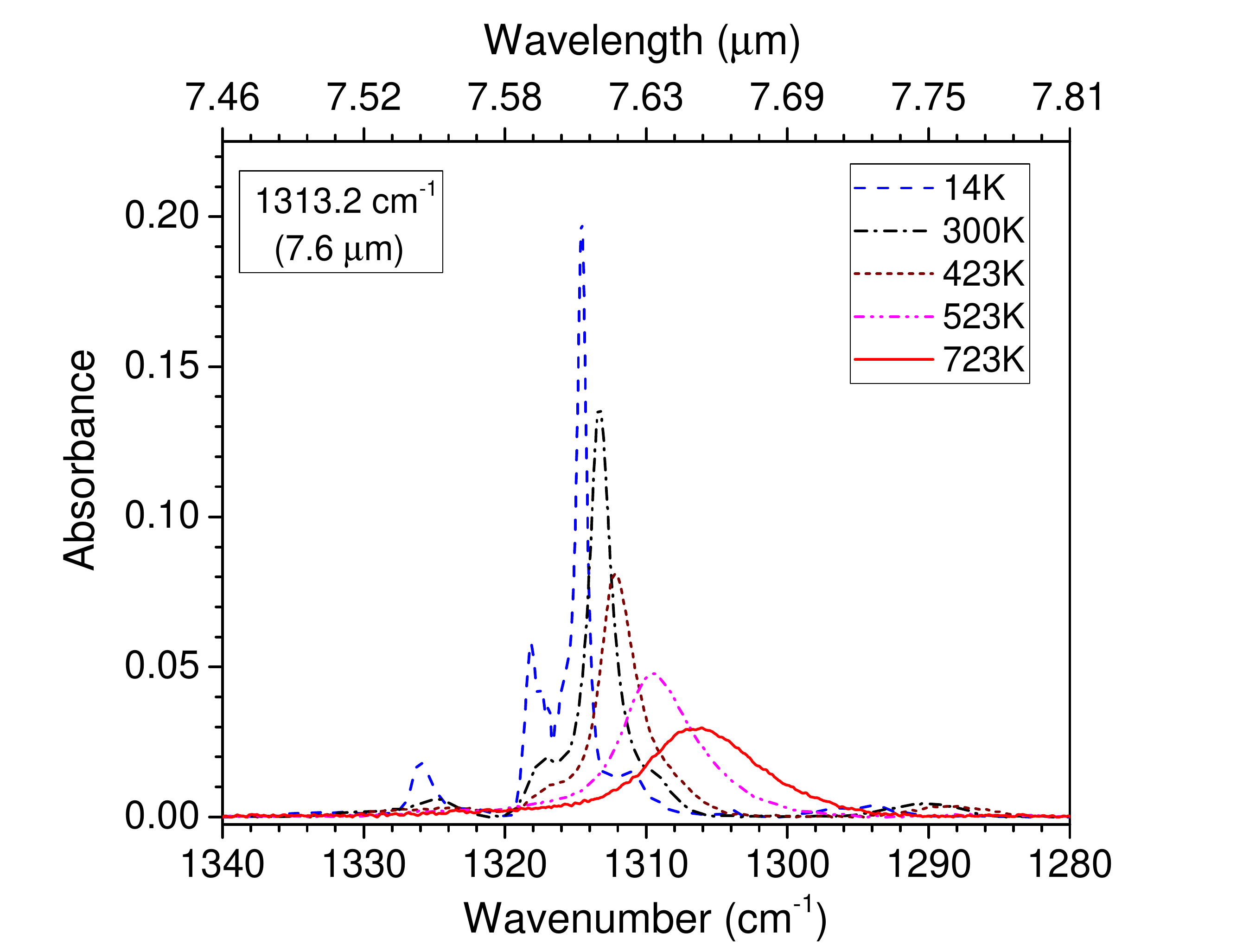}& \includegraphics[width=0.5\hsize]{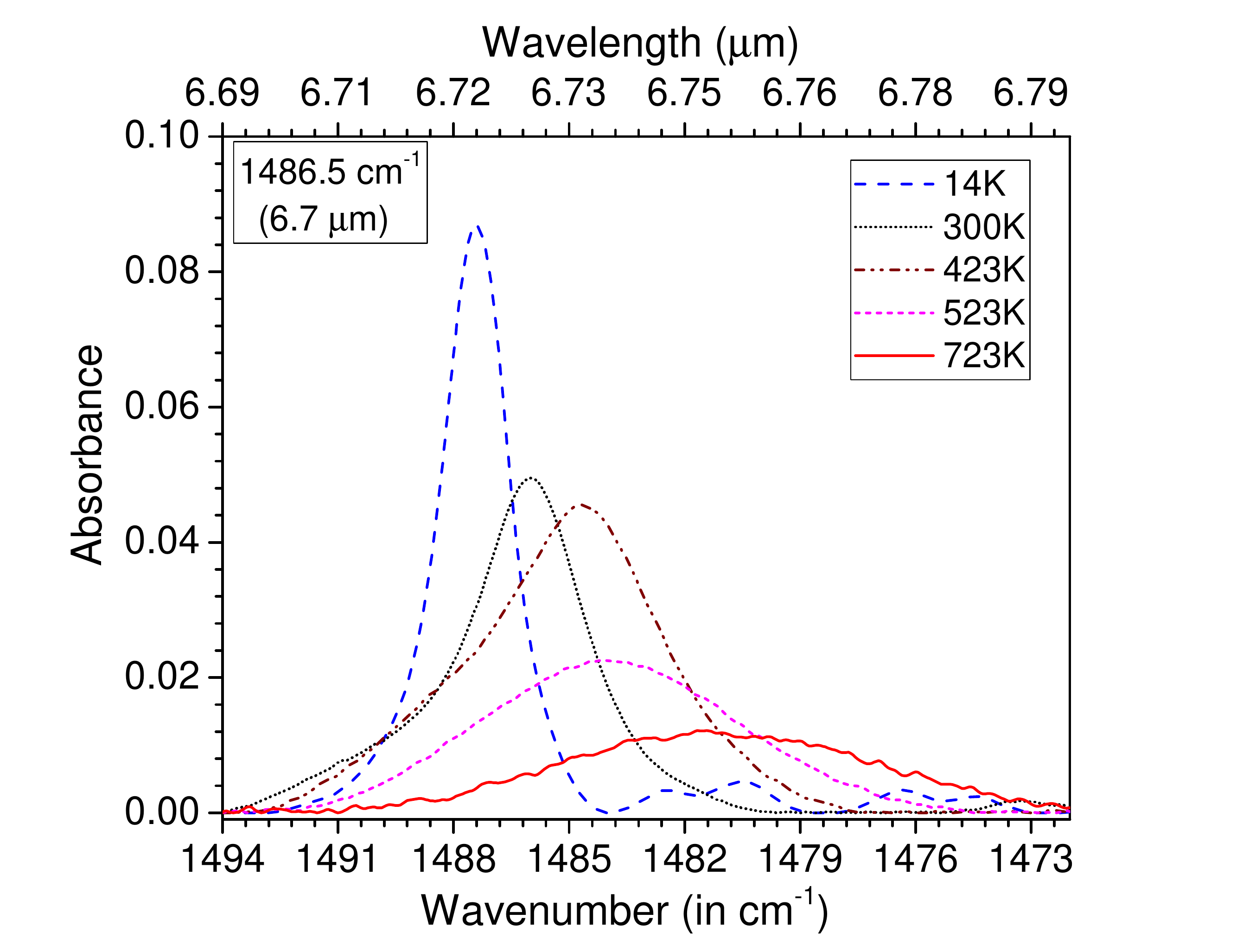}\\
(e) & (f)\\
\end{tabular}
\caption{Evolution of the IR band profile with temperature for additional fundamental transitions of pyrene in KBr pellets. Band positions (in \cm and $\mu$m ) at 300\,K are given in the top left corner of each panels.}
\label{spectra}
\end{figure*}

\begin{figure*}
\centering
\includegraphics[width=1.0\hsize]{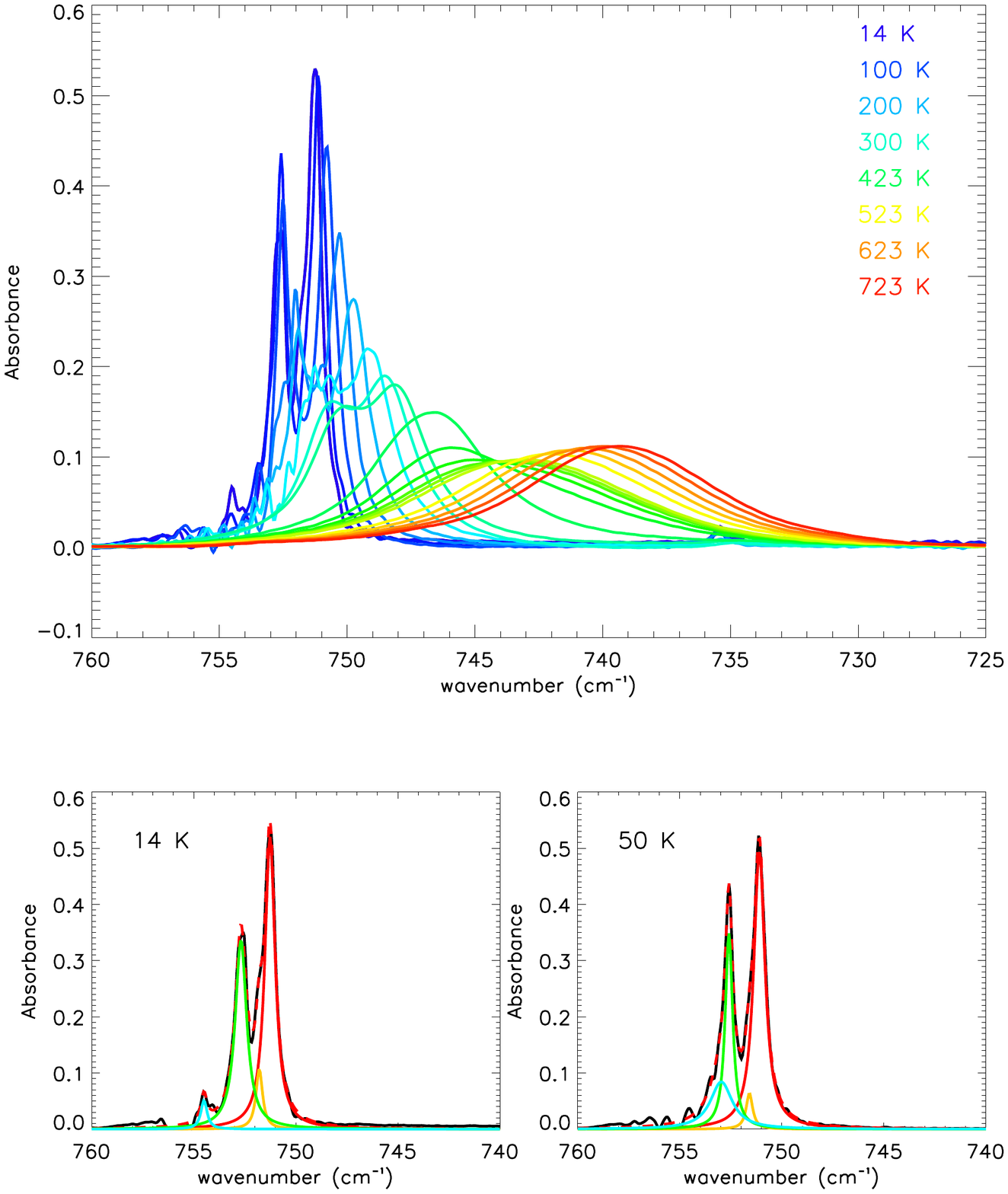} \\
\caption{Top: Evolution with temperature of the 749.4~\cm~band of pyrene in KBr pellets. Bottom: Multi-component fit of the 749.4~\cm~band at 14 and 50~K}
\label{749band}
\end{figure*}

\begin{figure*}
\ContinuedFloat
\centering
\includegraphics[width=1.0\hsize]{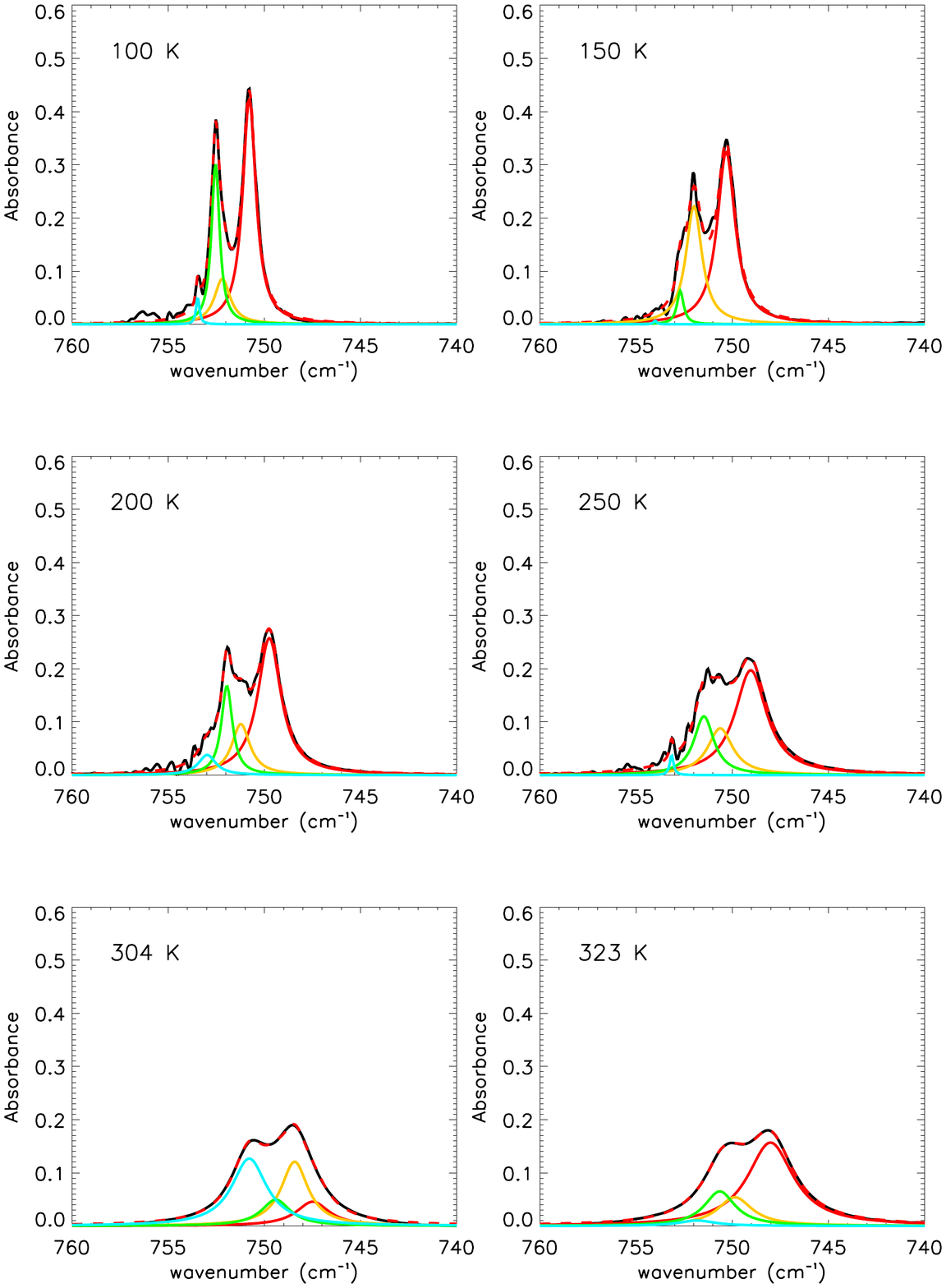} \\
\caption{Multi-component fit of the 749.4~\cm~band at 100, 150, 200, 250, 304, and 323~K (continued from the previous page)}
%\label{749band}
\end{figure*}

\begin{figure*}
\ContinuedFloat
\centering
\includegraphics[width=1.0\hsize]{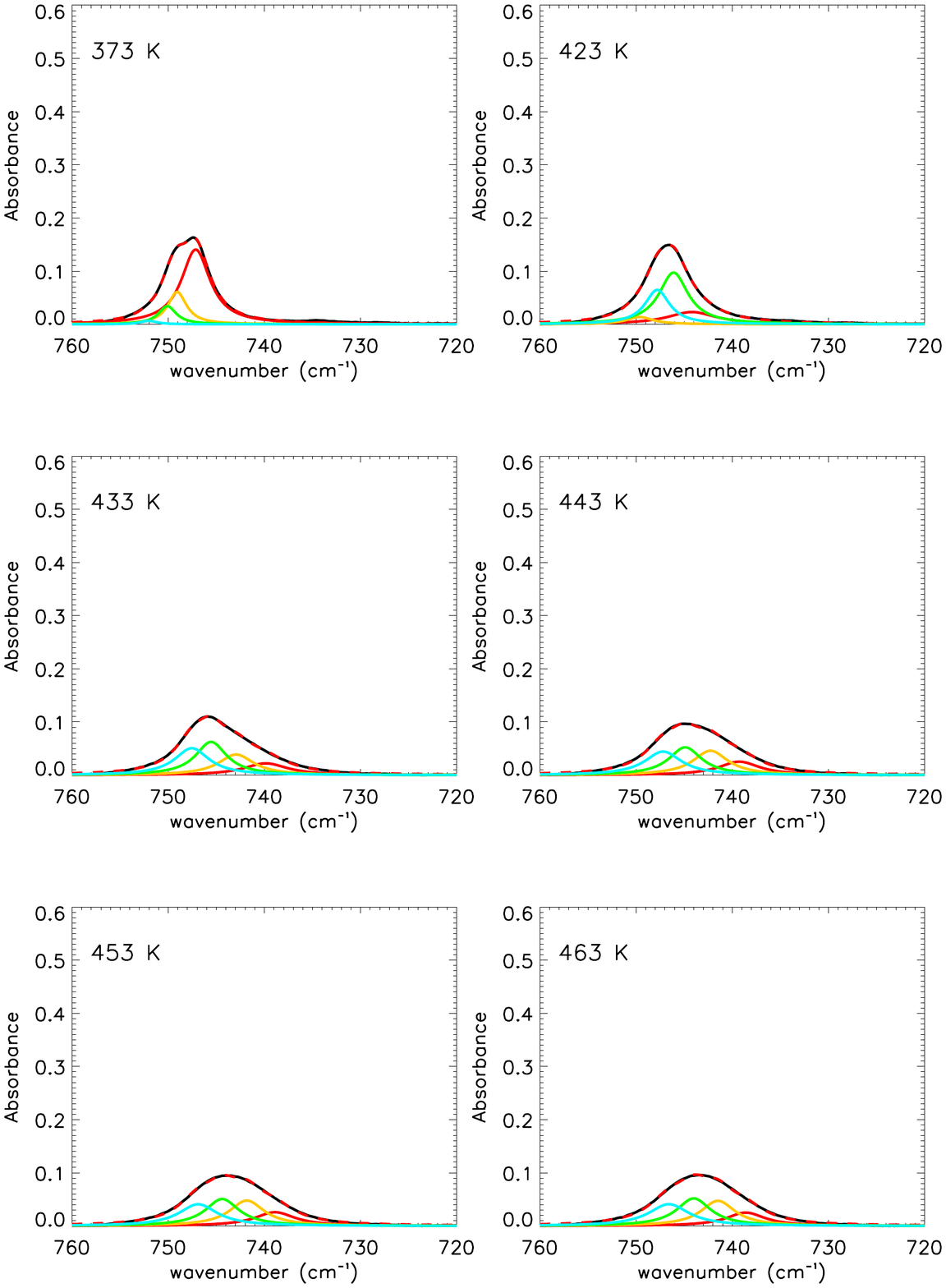} \\
\caption{Multi-component fit of the 749.4~\cm~band at 373, 423, 433, 443, 453, and 463~K (continued from the previous page)}
%\label{749band}
\end{figure*}

\begin{figure*}
\ContinuedFloat
\centering
\includegraphics[width=1.0\hsize]{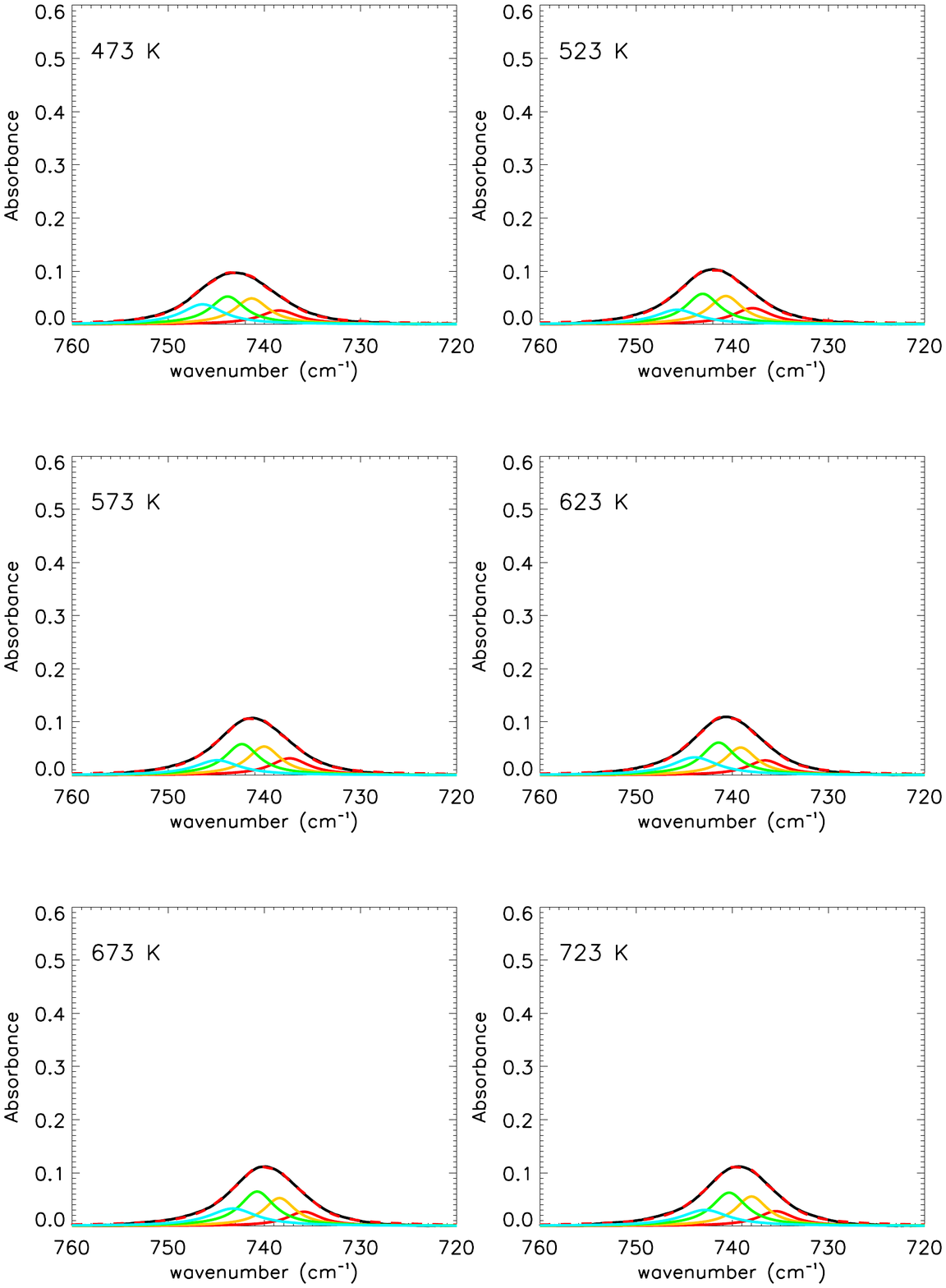} \\
\caption{Multi-component fit of the 749.4~\cm~band at 473, 523, 573, 623, 673, and 723~K (continued from the previous page)}
%\label{749band}
\end{figure*}

\begin{figure*}
\centering
\includegraphics[width=1.0\hsize]{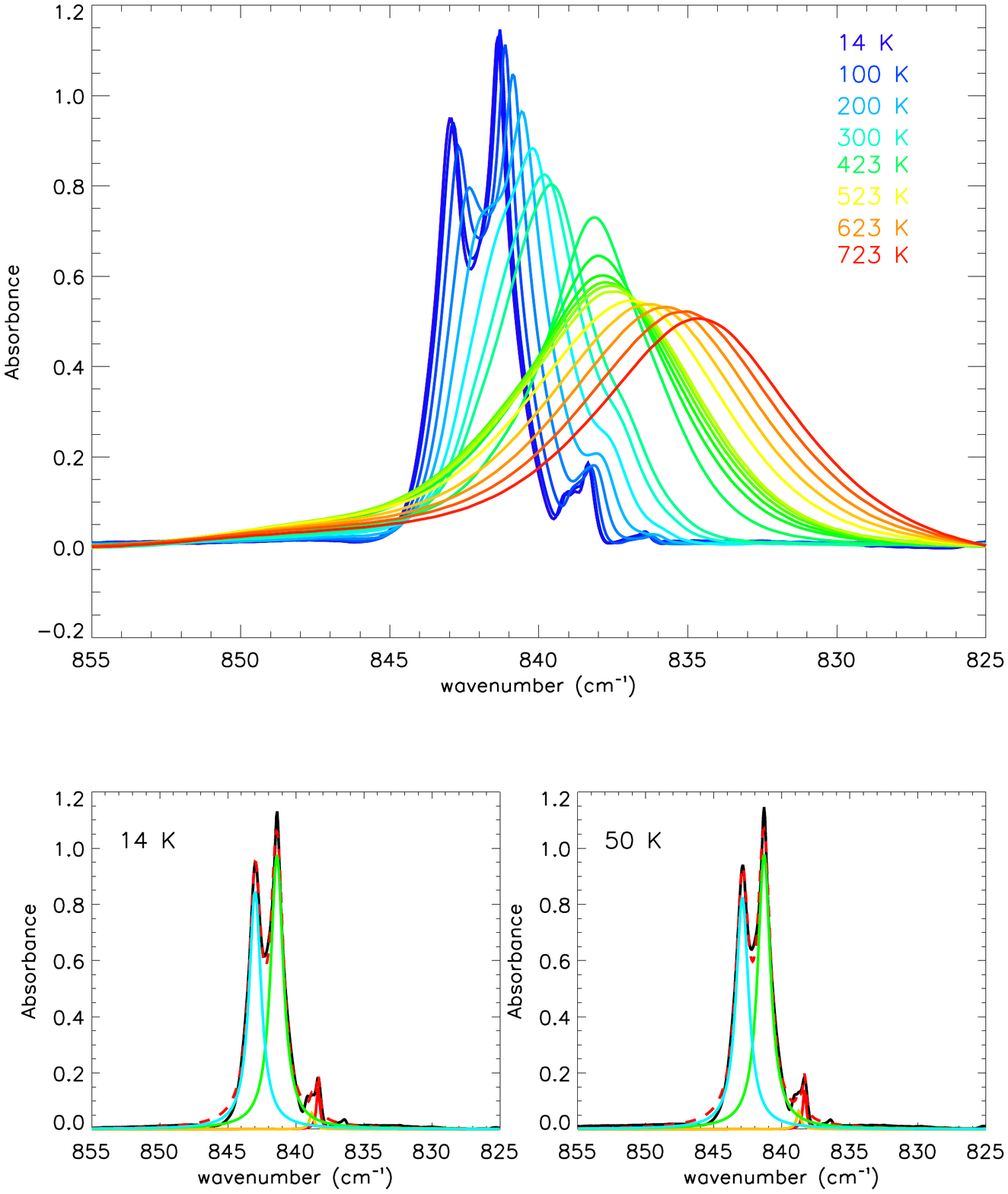} \\
\caption{Top: Evolution with temperature of the 839.9~\cm~band of pyrene in KBr pellets. Bottom: Multi-component fit of the 839.9~\cm~band at 14 and 50~K}
\label{839band}
\end{figure*}

\begin{figure*}
\ContinuedFloat
\centering
\includegraphics[width=1.0\hsize]{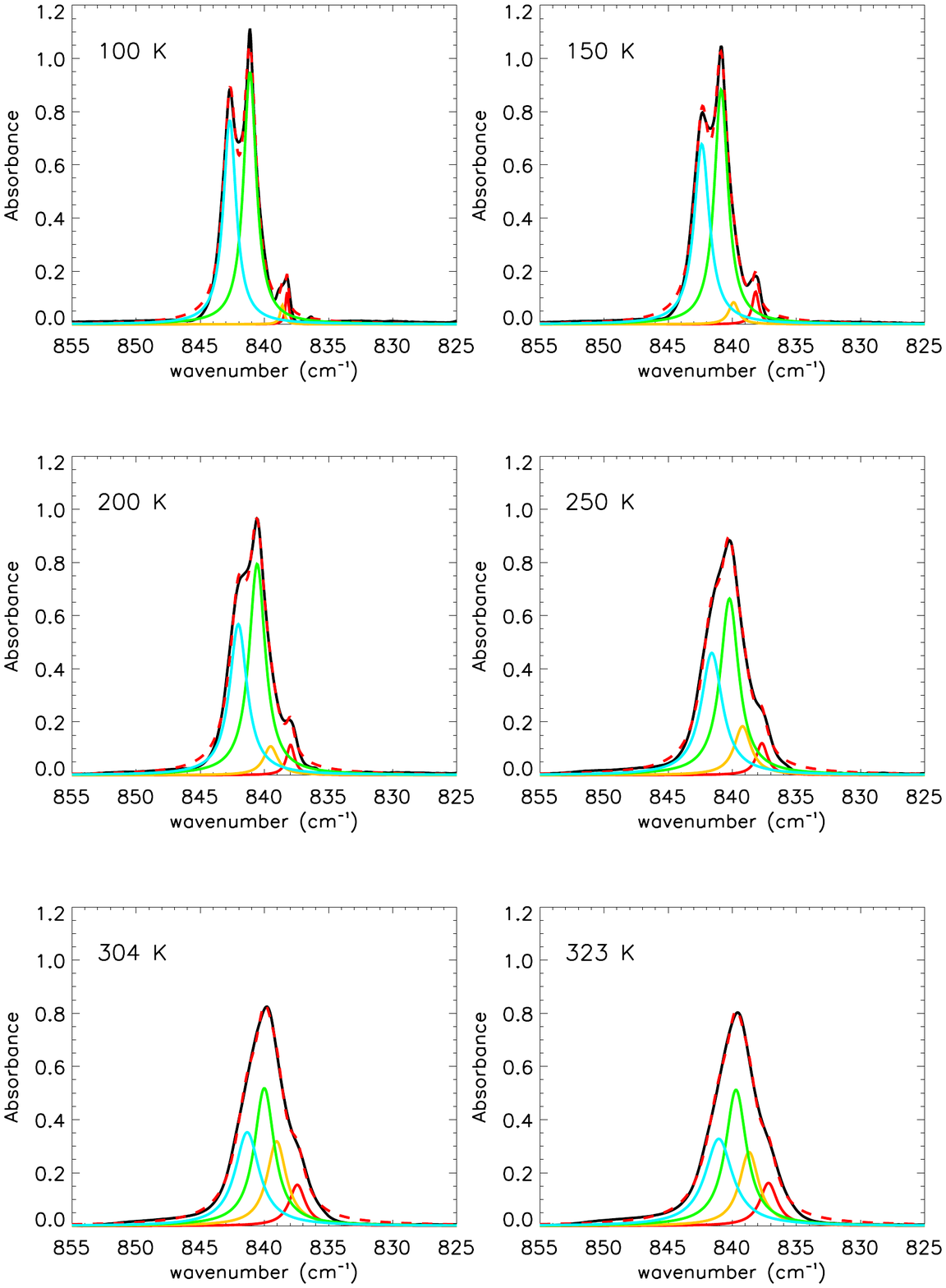} \\
\caption{Multi-component fit of the 839.9~\cm~band at 100, 150, 200, 250, 304, and 323~K (continued from the previous page)}
%\label{839band}
\end{figure*}

\begin{figure*}
\ContinuedFloat
\centering
\includegraphics[width=1.0\hsize]{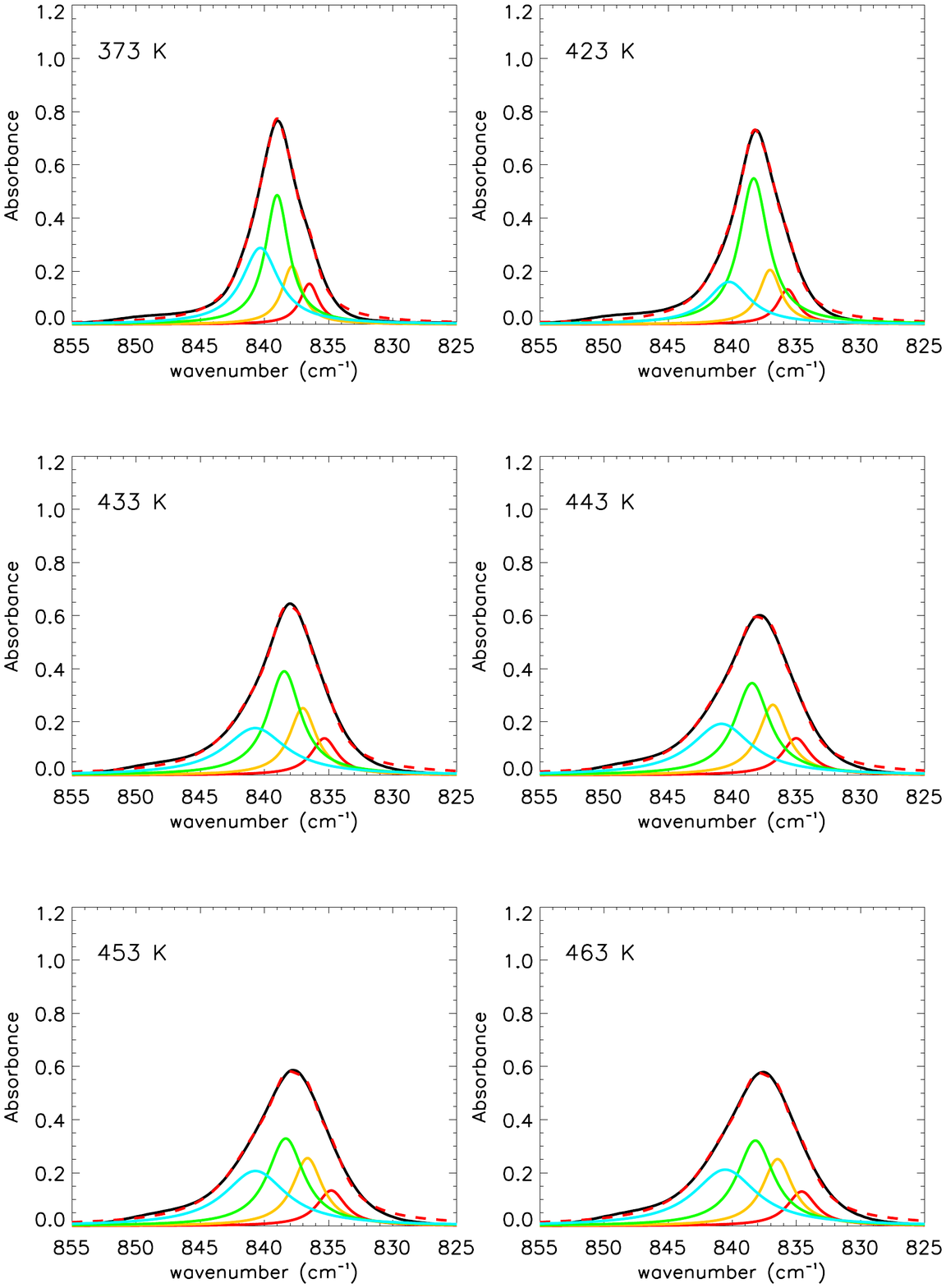} \\
\caption{Multi-component fit of the 839.9~\cm~band at 373, 423, 433, 443, 453, and 463~K (continued from the previous page)}
%\label{839band}
\end{figure*}

\begin{figure*}
\ContinuedFloat
\centering
\includegraphics[width=1.0\hsize]{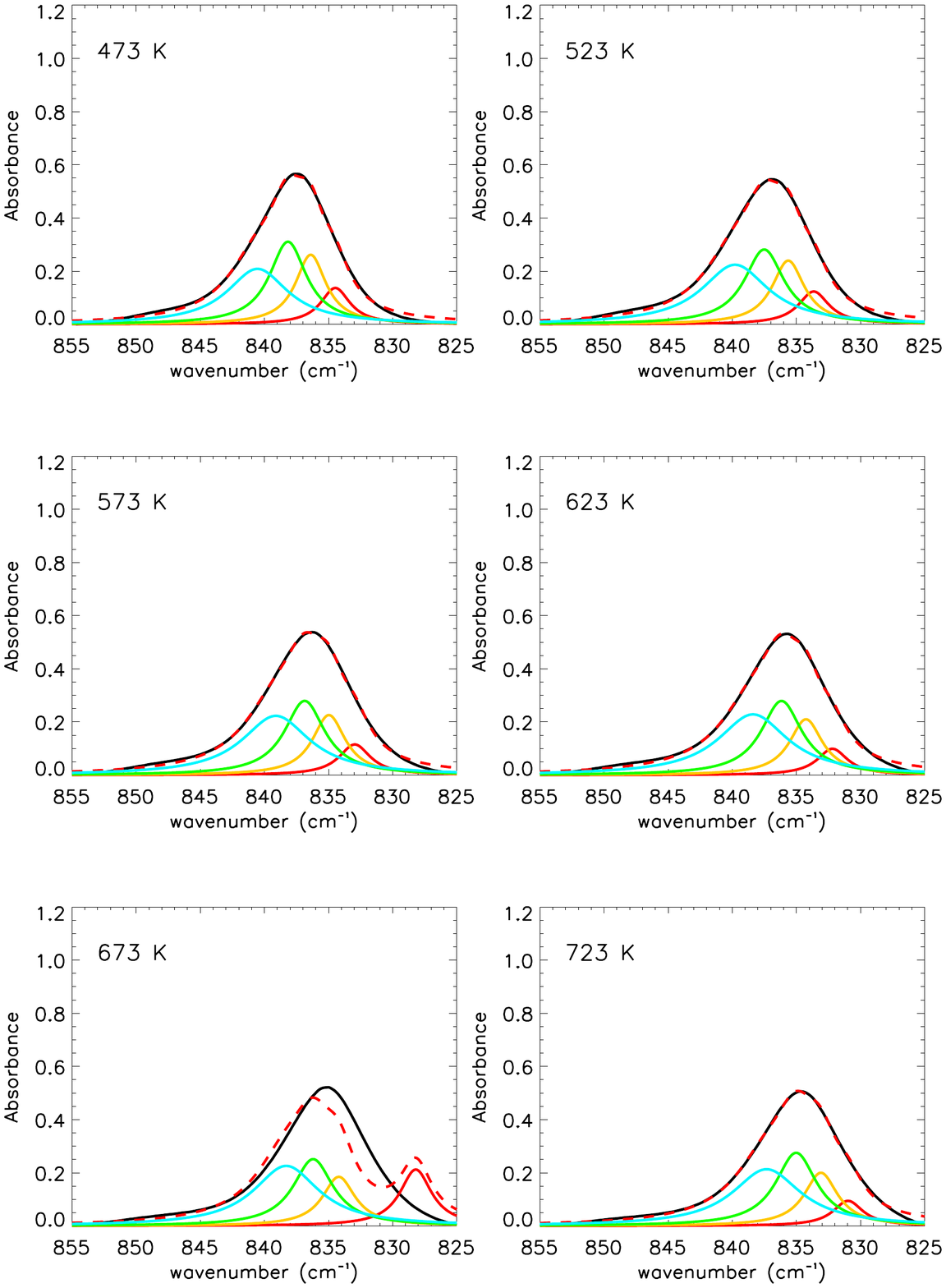} \\
\caption{Multi-component fit of the 839.9~\cm~band at 473, 523, 573, 623, 673, and 723~K (continued from the previous page)}
%\label{839band}
\end{figure*}

\begin{figure*}
\centering
\includegraphics[width=1.0\hsize]{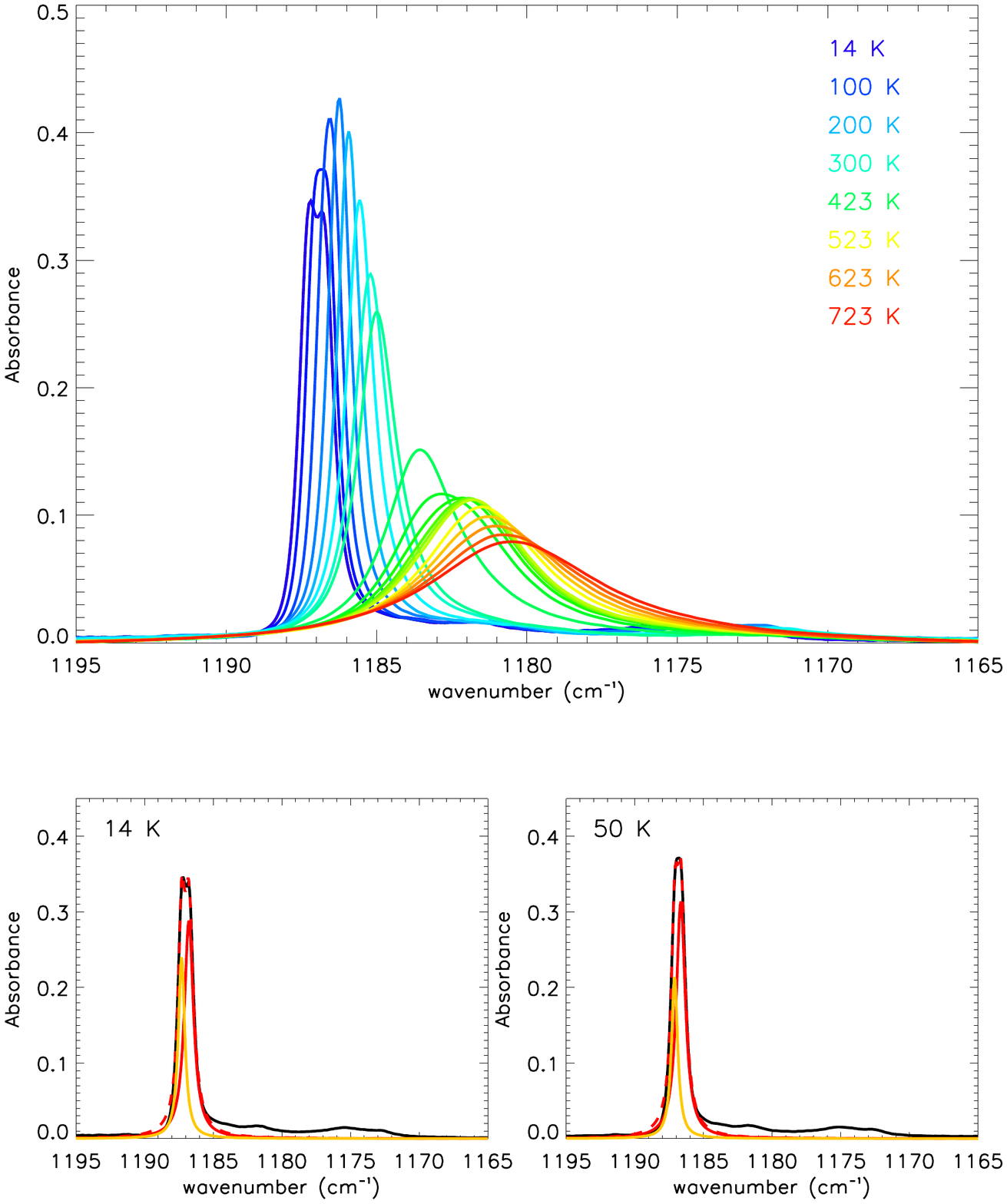} \\
\caption{Top: Evolution with temperature of the 1184.8~\cm~band of pyrene in KBr pellets. Bottom: Multi-component fit of the 1184.8~\cm~band at 14 and 50~K}
\end{figure*}

\begin{figure*}
\ContinuedFloat
\centering
\includegraphics[width=1.0\hsize]{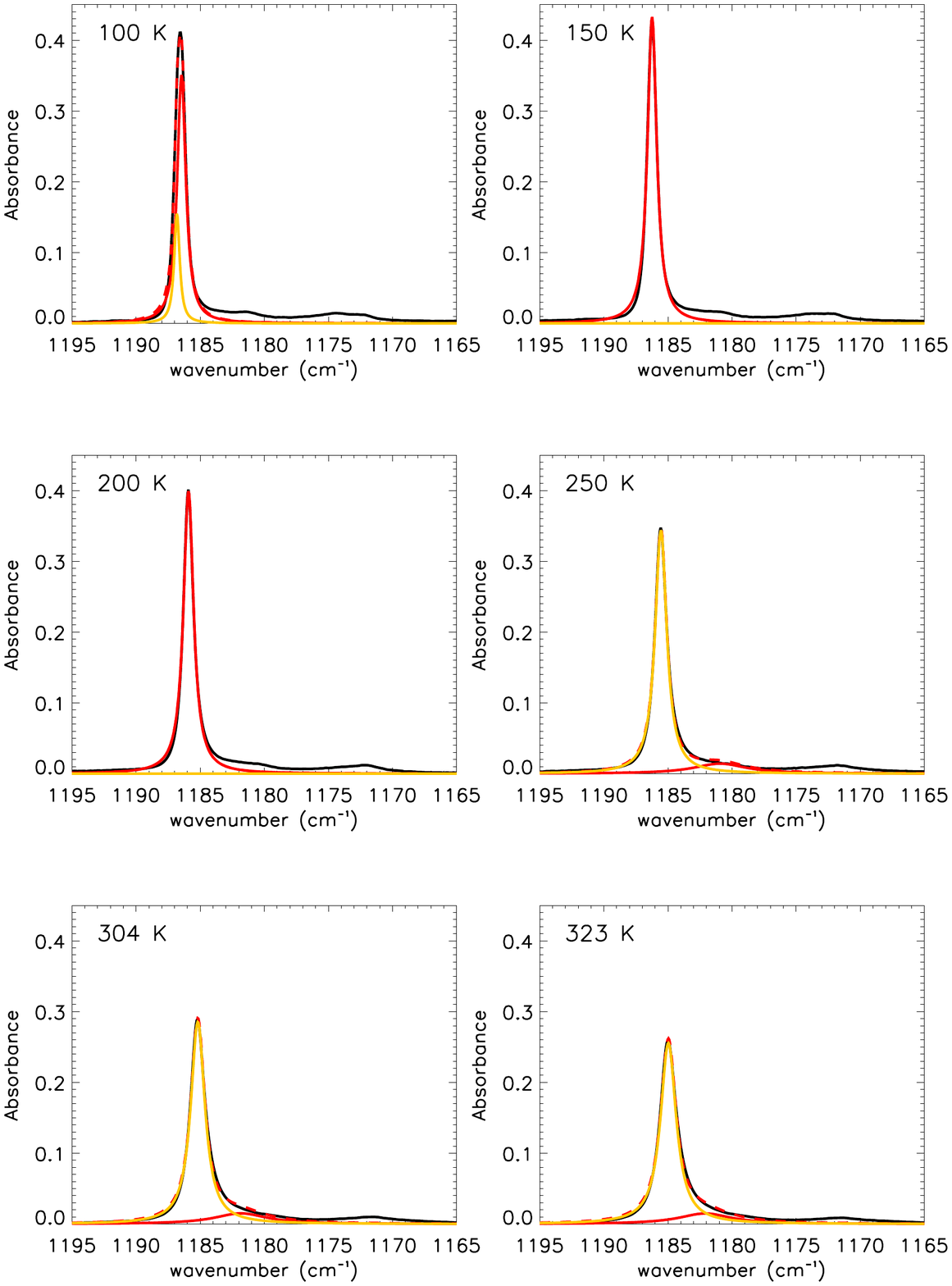} \\
\caption{Multi-component fit of the 1184.8~\cm~band at 100, 150, 200, 250, 304, and 323~K (continued from the previous page)}
\end{figure*}

\begin{figure*}
\ContinuedFloat
\centering
\includegraphics[width=1.0\hsize]{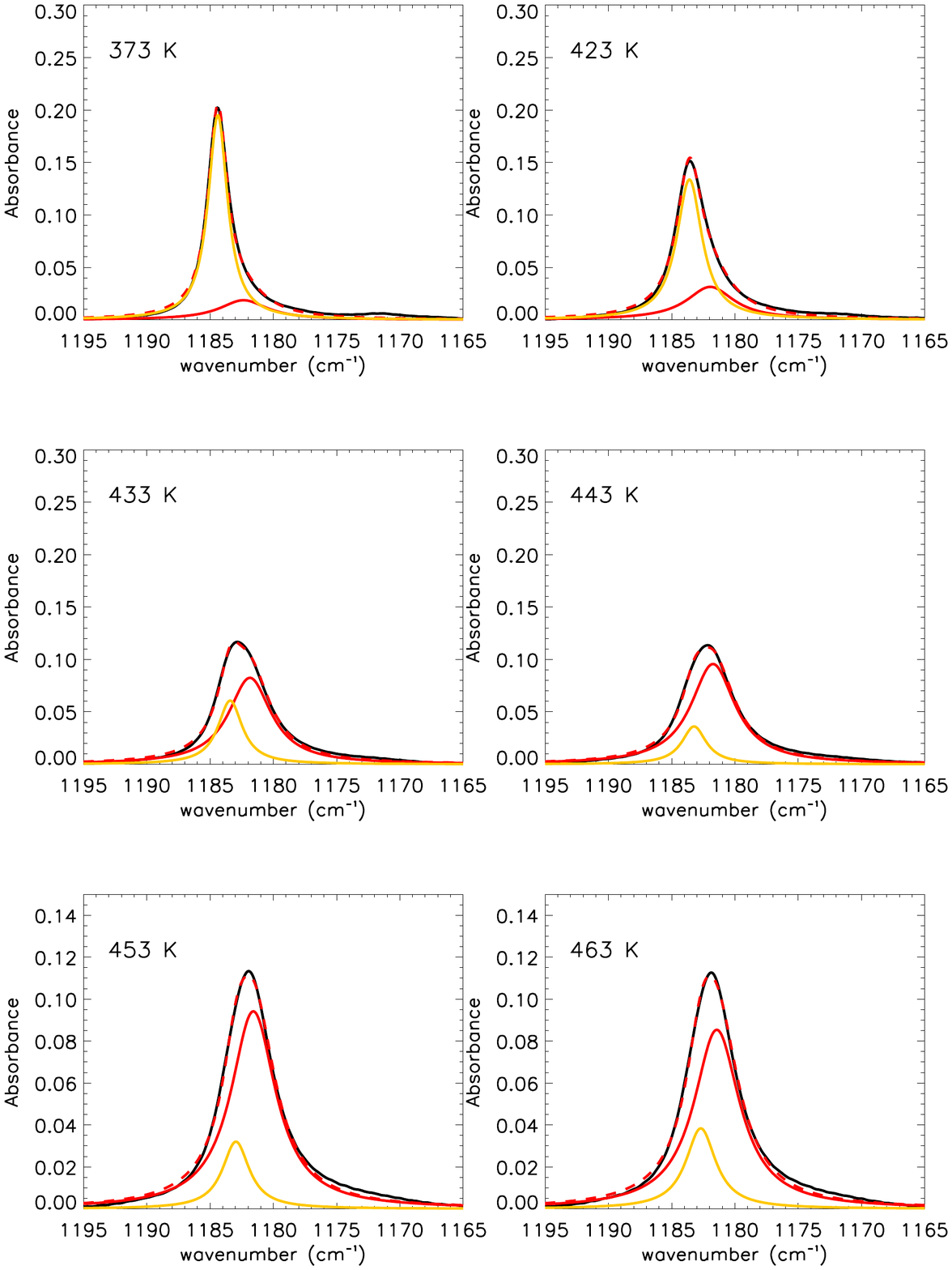} \\
\caption{Multi-component fit of the 1184.8~\cm~band at 373, 423, 433, 443, 453, and 463~K (continued from the previous page)}
\end{figure*}

\begin{figure*}
\ContinuedFloat
\centering
\includegraphics[width=1.0\hsize]{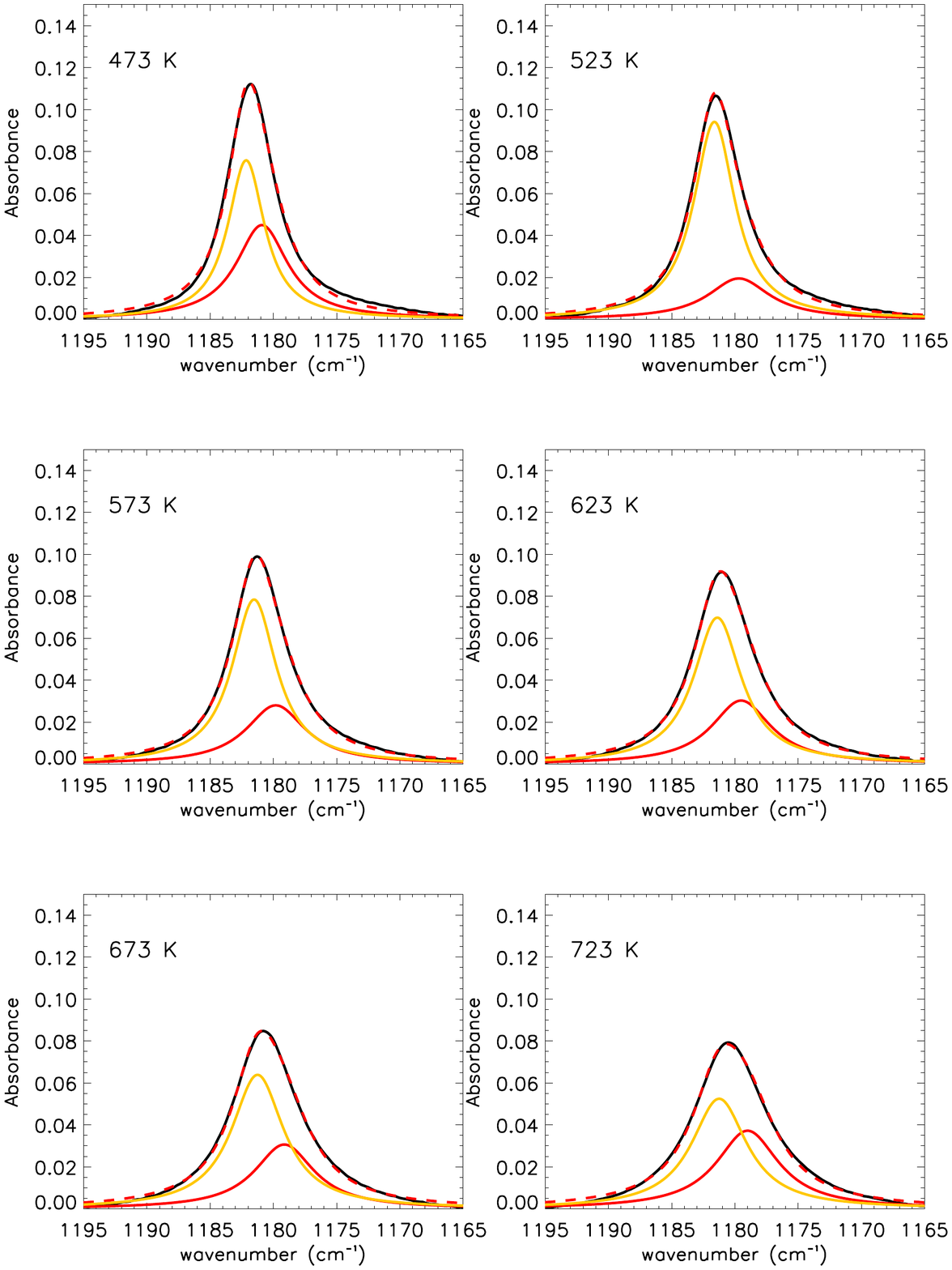} \\
\caption{Multi-component fit of the 1184.8~\cm~band at 473, 523, 573, 623, 673, and 723~K (continued from the previous page)}
\label{1185band}
\end{figure*}

\begin{figure*}
\centering
\begin{tabular}{c c}
\includegraphics[width=0.5\hsize]{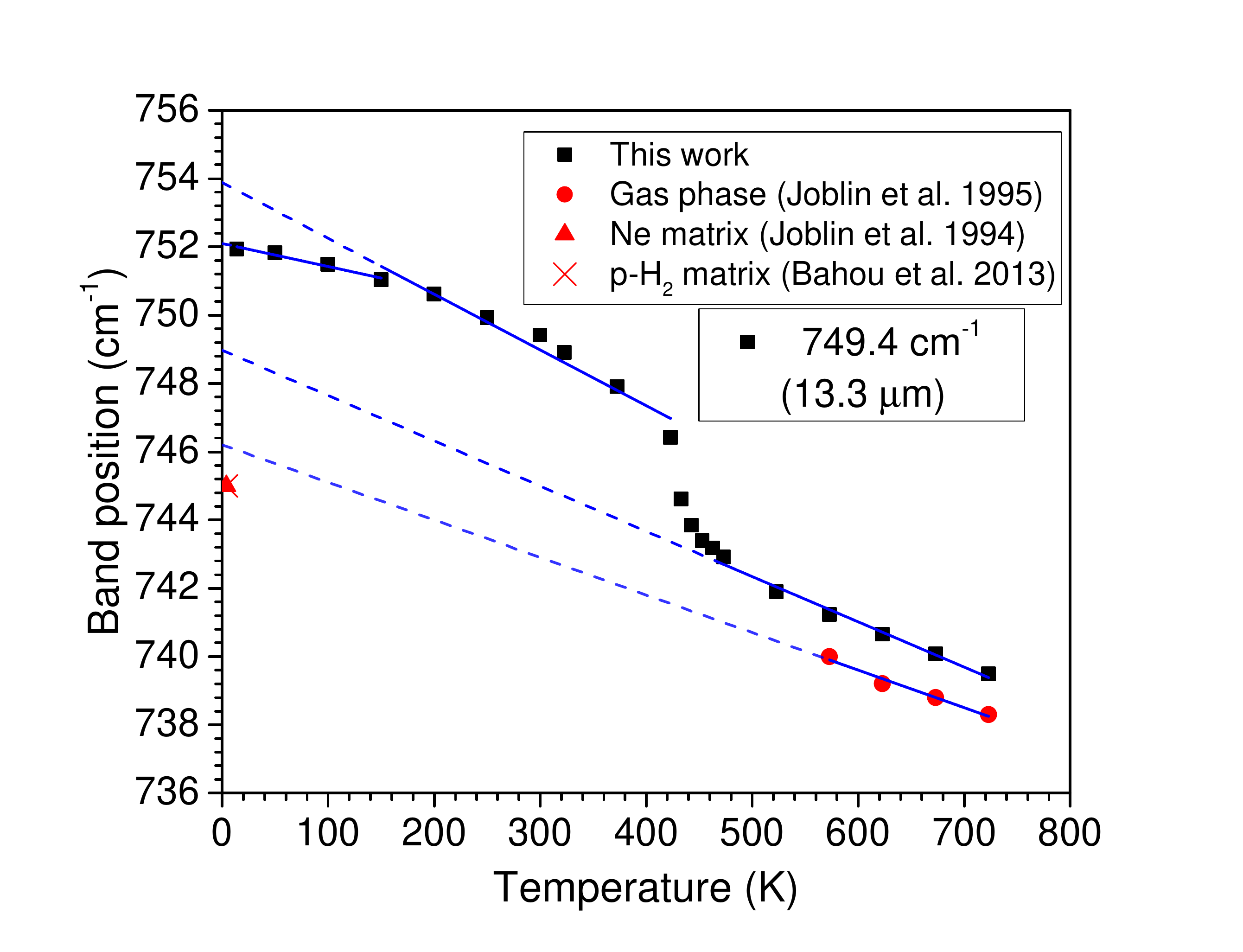}& \includegraphics[width=0.5\hsize]{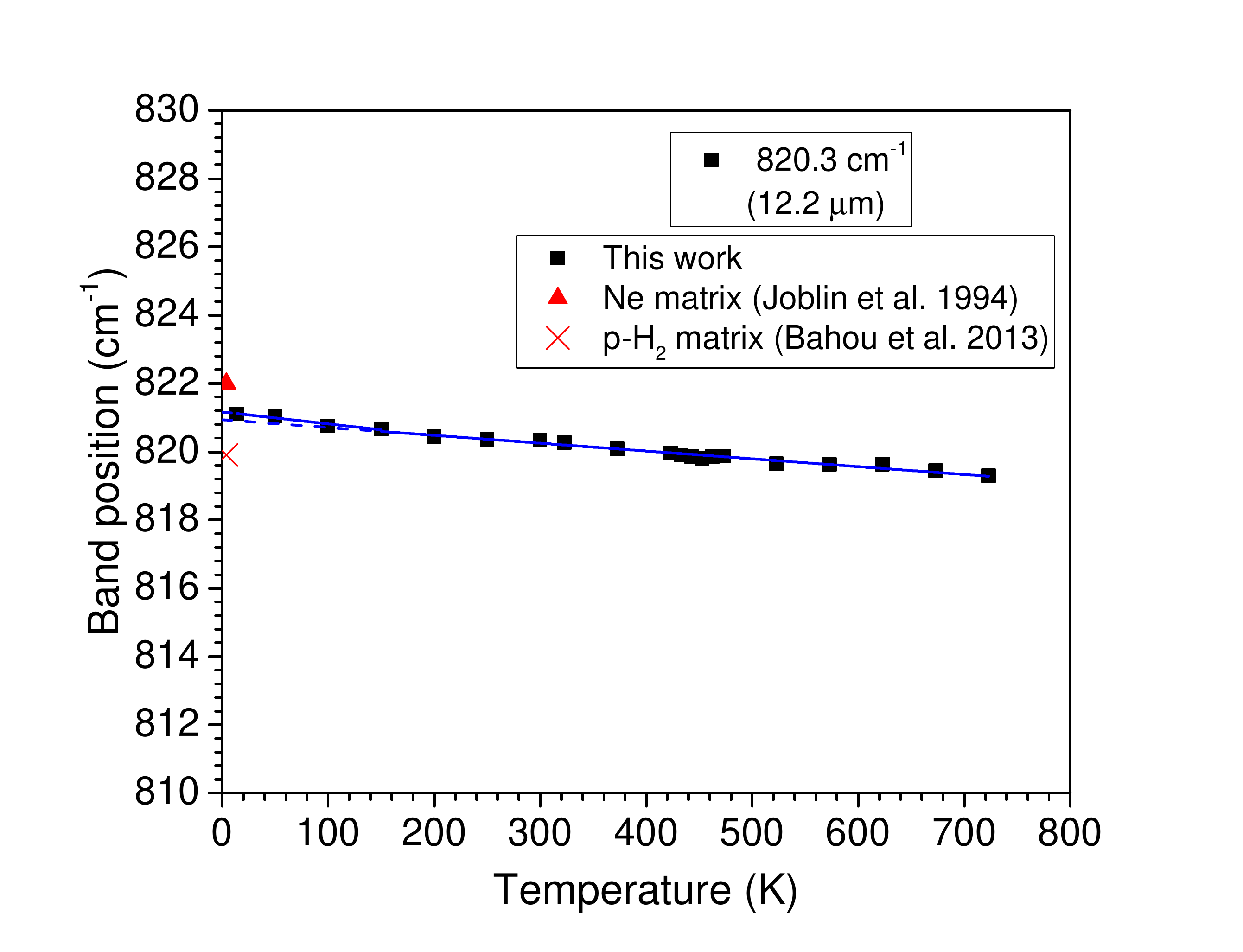} \\
(a) & (b)\\
\includegraphics[width=0.5\hsize]{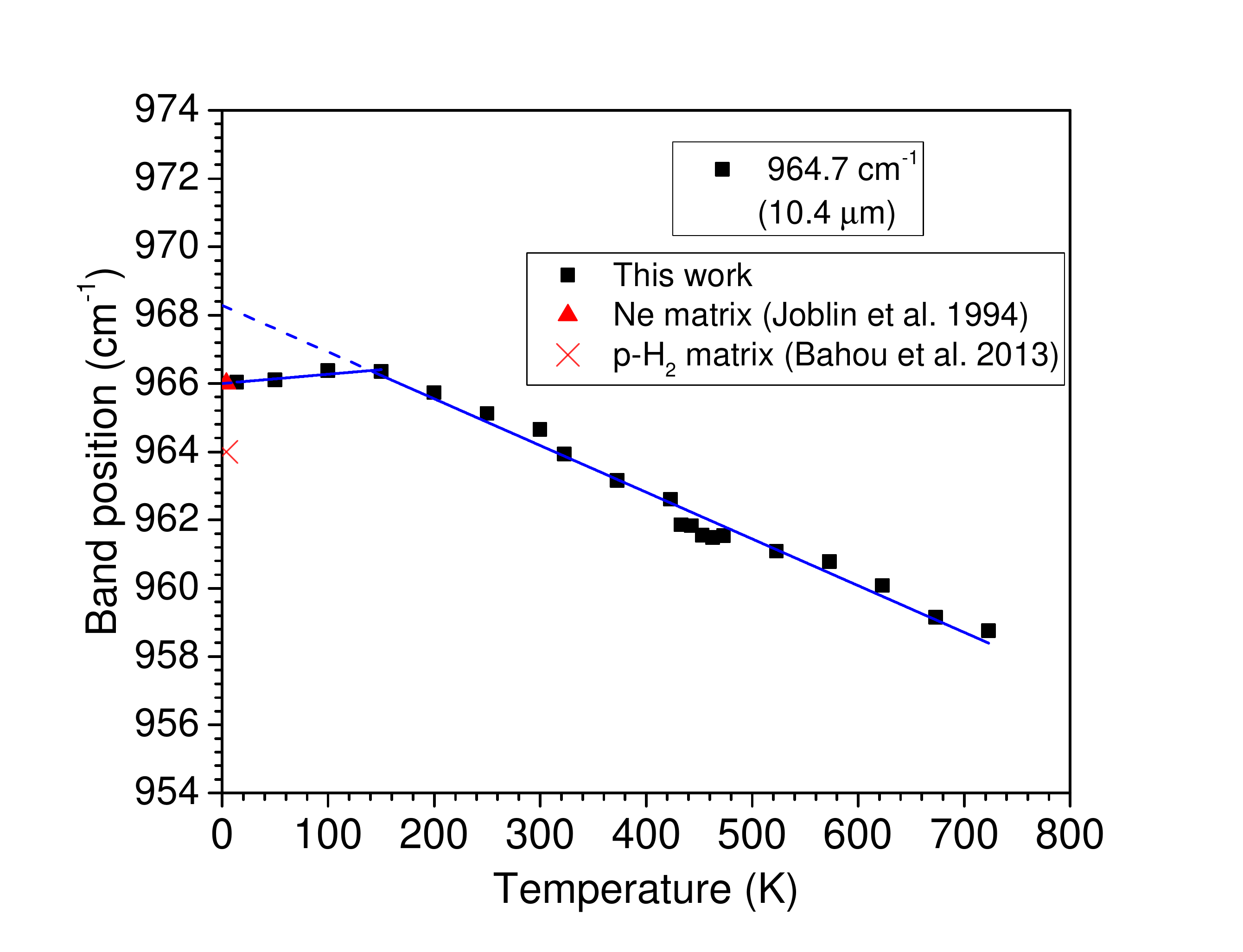}& \includegraphics[width=0.5\hsize]{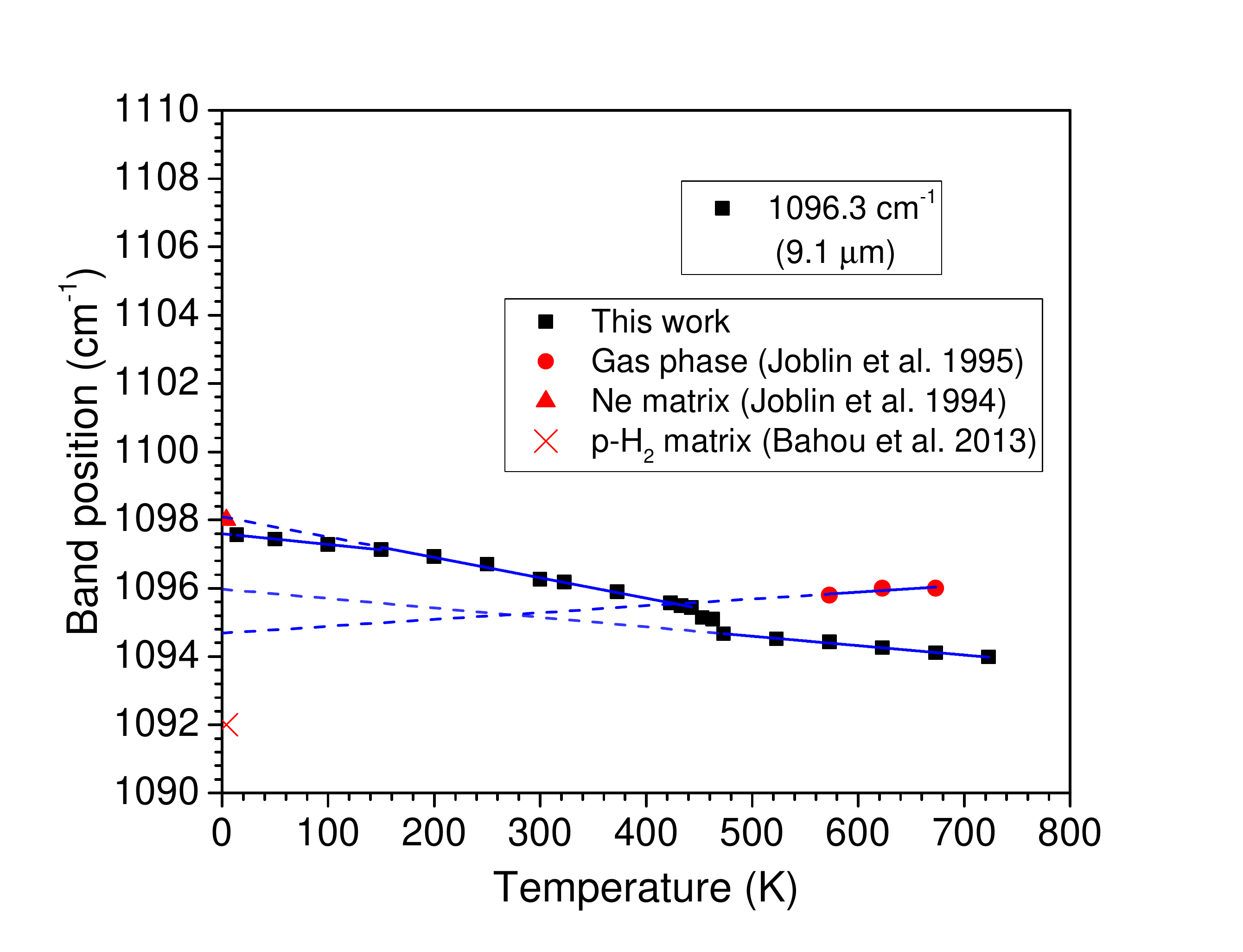}\\
(c) & (d)\\
\includegraphics[width=0.5\hsize]{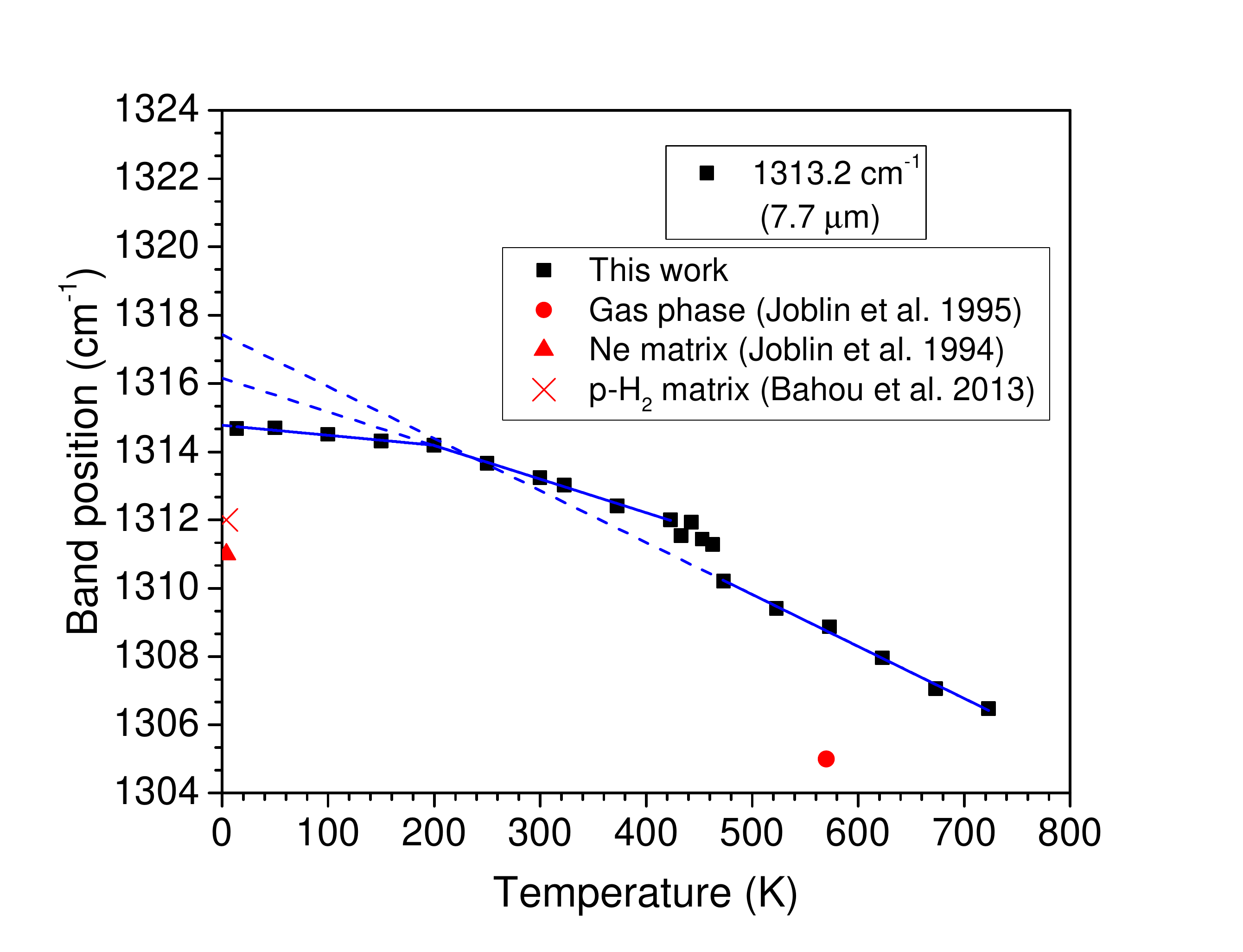}& \includegraphics[width=0.5\hsize]{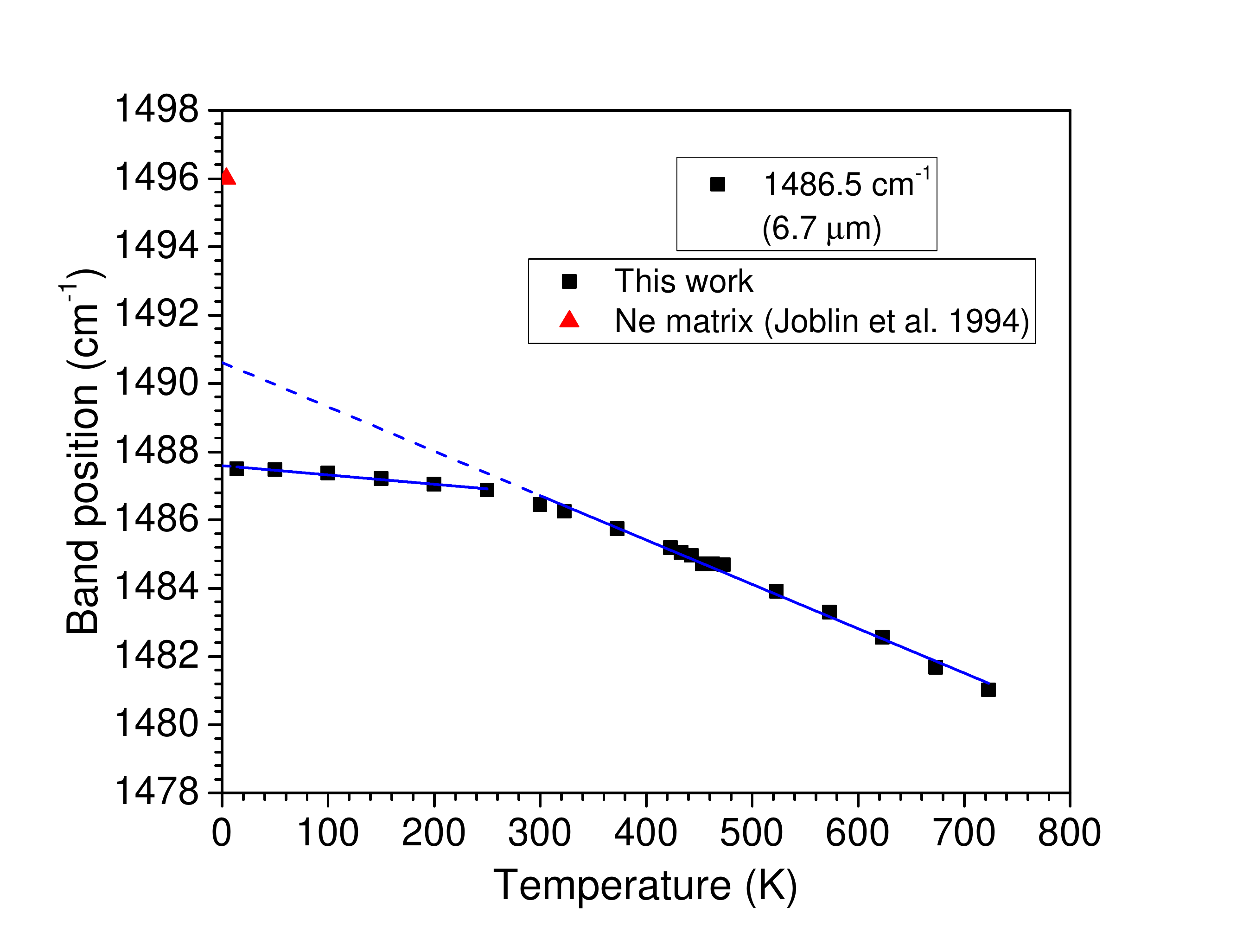}\\
(e) & (f)\\
\end{tabular}
\caption{Evolution of the IR band position with temperature for additional fundamental transitions of pyrene in KBr pellets. The linear fits are shown by blue lines. Also shown are published data at high temperature in the gas-phase\cite{Joblin1995} as well as in frozen Ne matrix (3-4\,K)\cite{Joblin1994} or p-H$_{2}$ matrix\cite{bahou2013}.}
\label{posn}
\end{figure*}

\begin{comment}
\begin{table}[H]
\centering
\footnotesize
\caption{Integration range of the major fundamental bands reported in Figures 2 and S2}
\begin{tabular}{c c c}
\hline
\multicolumn{2}{c}{Position} & \multicolumn{1}{c}{Integration range}\\
\multicolumn{2}{c}{(300K)} & \multicolumn{1}{c}{}\\
\multicolumn{1}{c}{$\mu$m} & \multicolumn{1}{c}{cm$^{-1}$} & \multicolumn{1}{c}{cm$^{-1}$}\\
\hline
14.1 & 709.7 & 716 - 690\\
13.3 & 749.4 & 760 - 720\\
12.2 & 820.3 & 830 - 810\\
11.9 & 839.9 & 855 - 825\\
10.4 & 964.7 & 985 - 930\\
 9.9 & 1000.9& 1008 - 988\\
 9.1 & 1096.3 & 1105 - 1185\\
 8.4 & 1184.8 & 1195 - 1165\\
 8.1 & 1241.7 & 1250 - 1230\\
 7.7 & 1313.2 & 1340 - 1280\\
 6.9 & 1433.6 & 1440 - 1410\\
 6.7 & 1486.5 & 1495 - 1470\\
 6.3 & 1599.8 & 1616 - 1558\\
 3.3 & 3048.3 & 3120 - 2920\\
\hline
\label{int_range}
\end{tabular}
\end{table}
\end{comment}

\begin{figure}
\centering
\includegraphics[width=1.15\columnwidth]{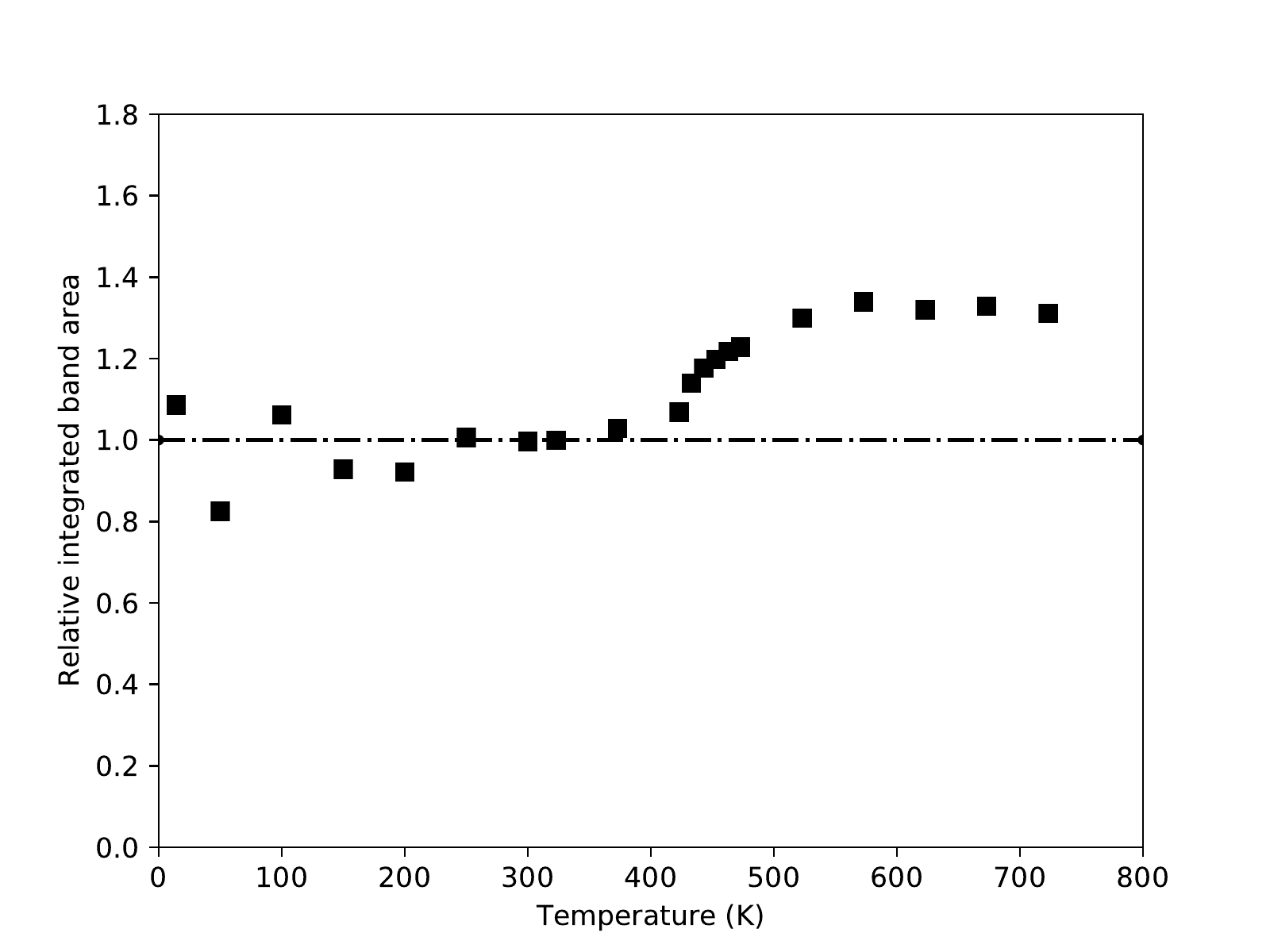}\\
\caption{Evolution of the integrated intensity of the pyrene spectrum from 3200 to 650~\cm, measured in KBr pellets, as a function of temperature. The values have been normalized with respect to the 300~K value.}
\label{tot_int}
\end{figure}

\begin{figure*}
\centering
\begin{tabular}{c c}
\includegraphics[width=0.5\hsize]{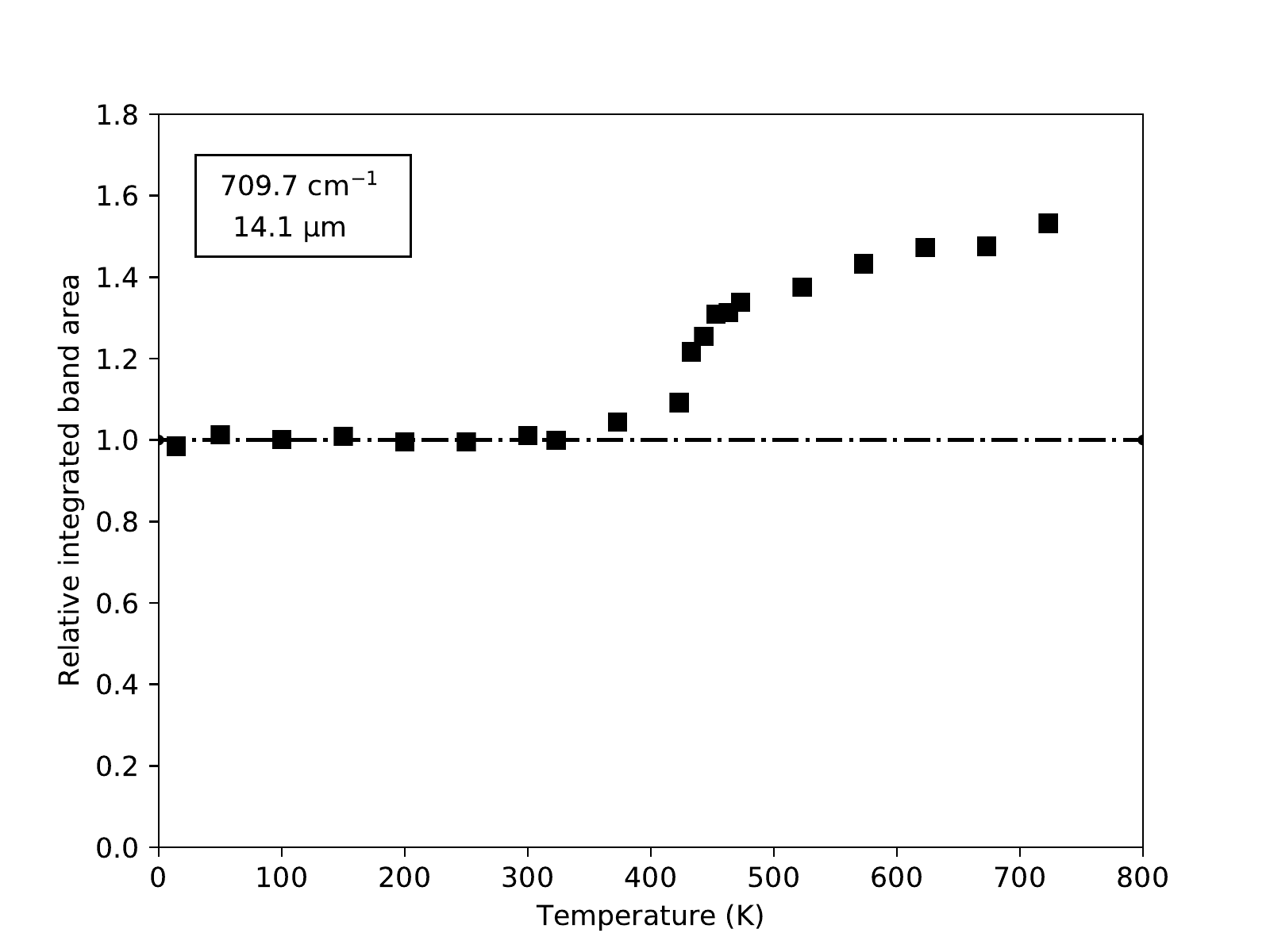} & \includegraphics[width=0.5\hsize]{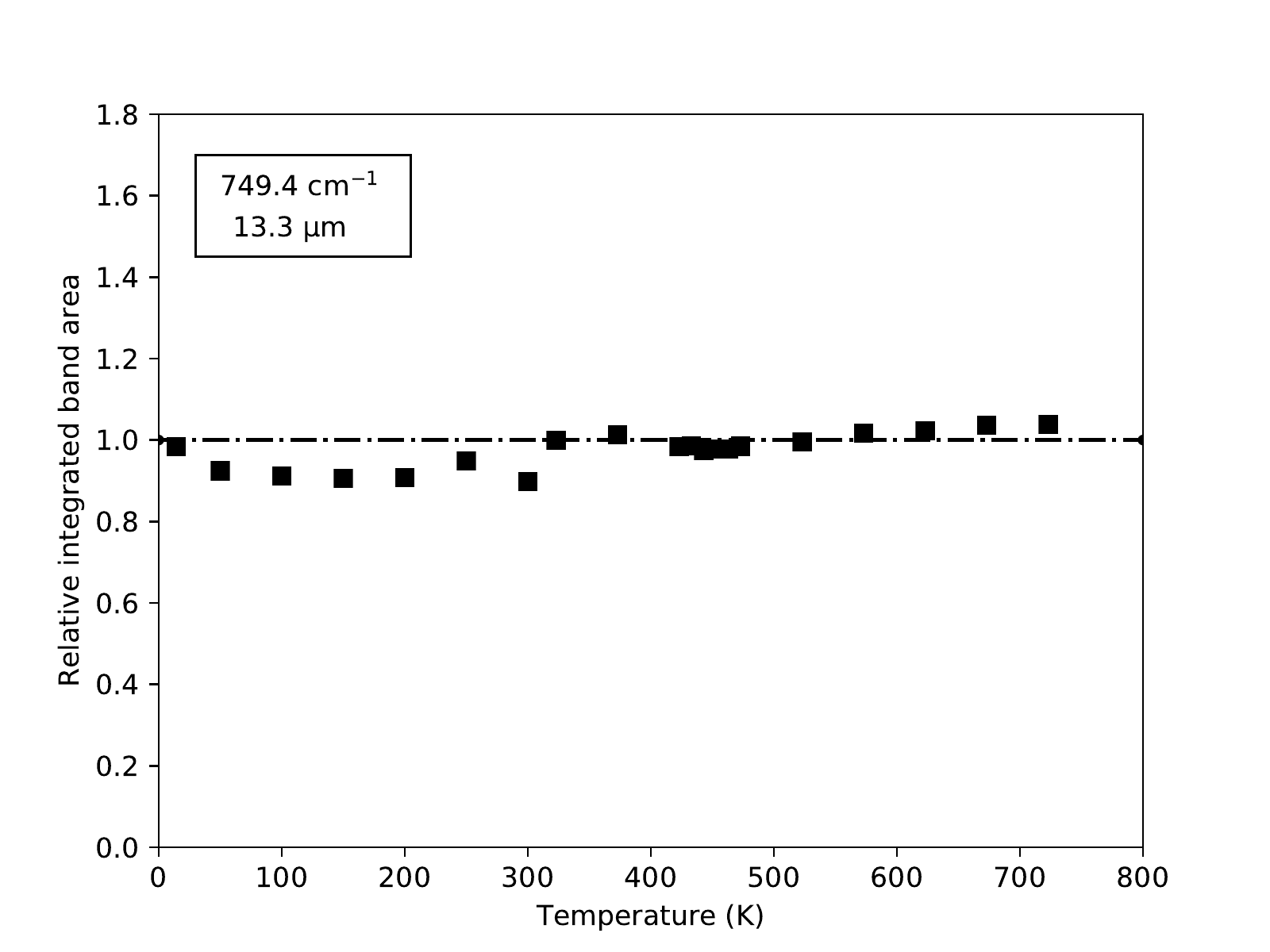}\\
(a) & (b)\\
\includegraphics[width=0.5\hsize]{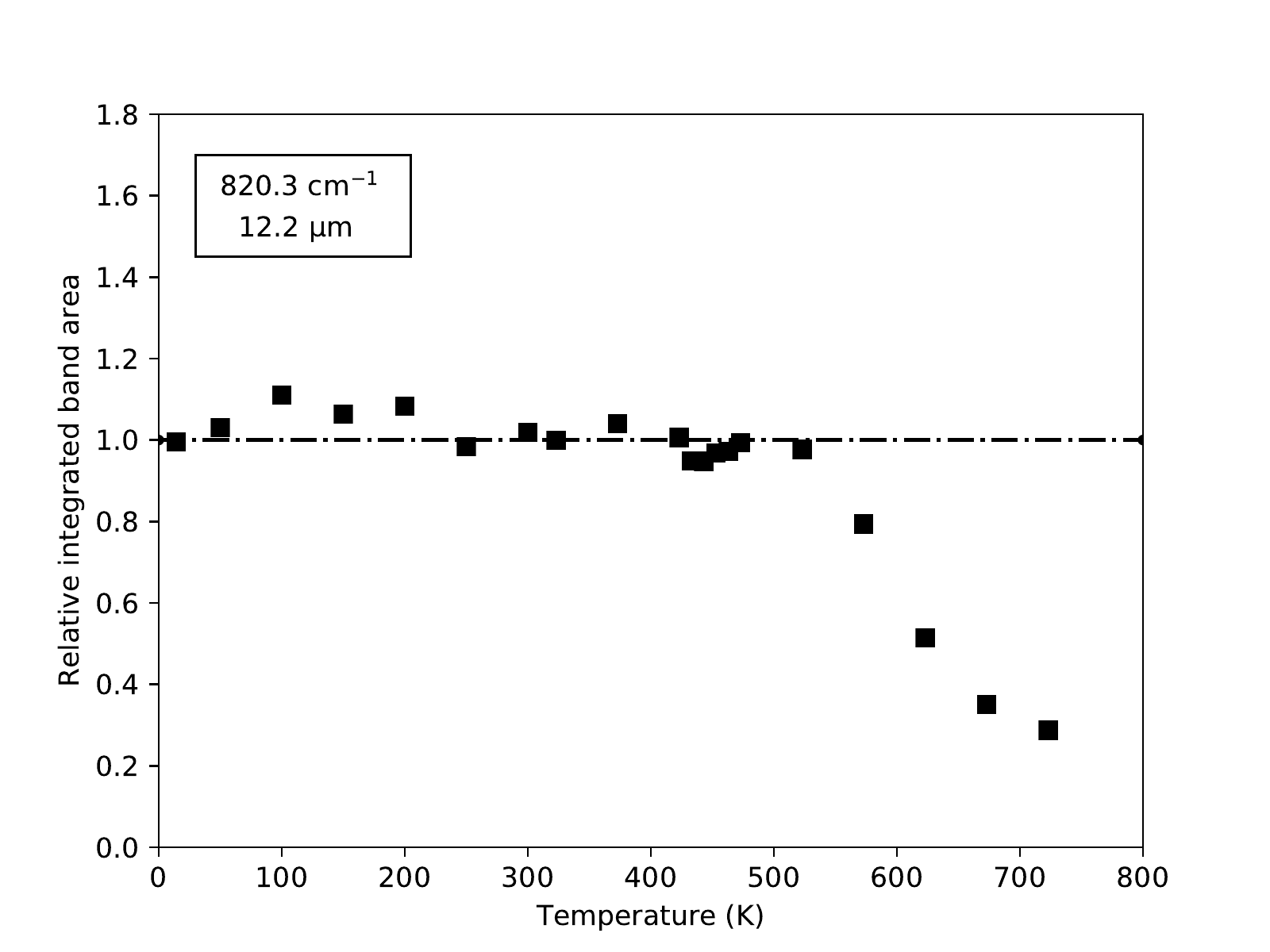} & \includegraphics[width=0.5\hsize]{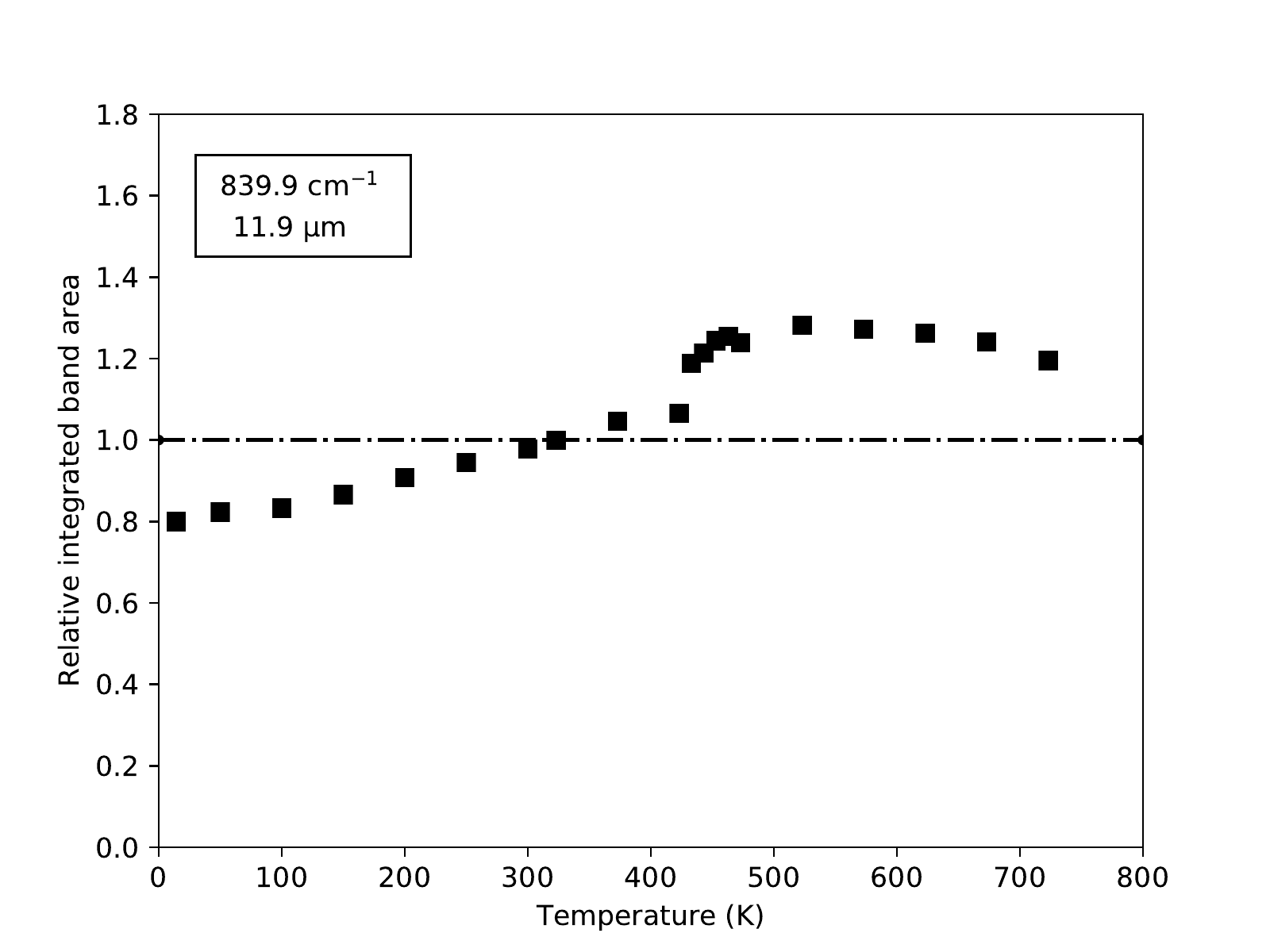}\\
(c) & (d)\\
\includegraphics[width=0.5\hsize]{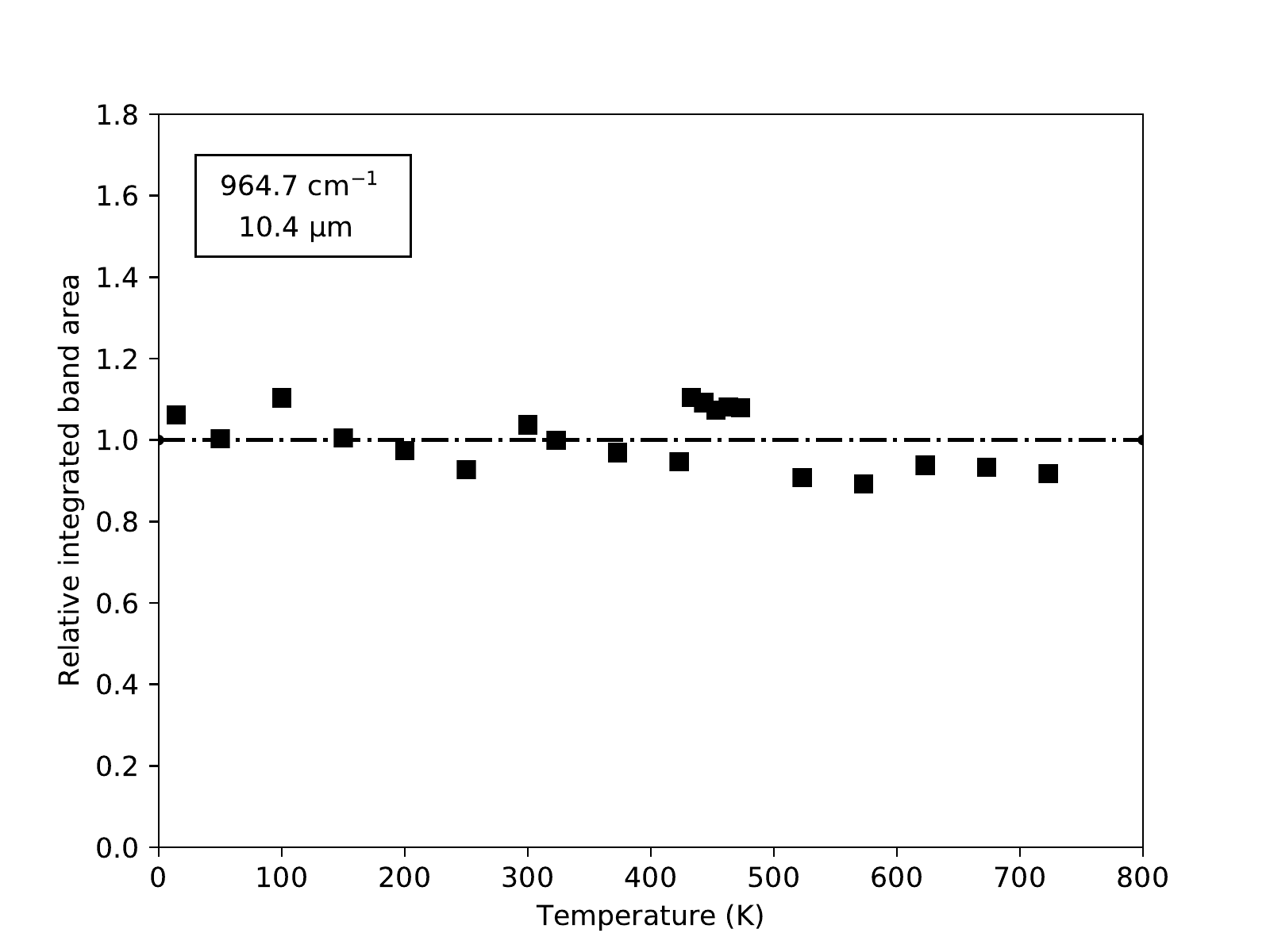}& \includegraphics[width=0.5\hsize]{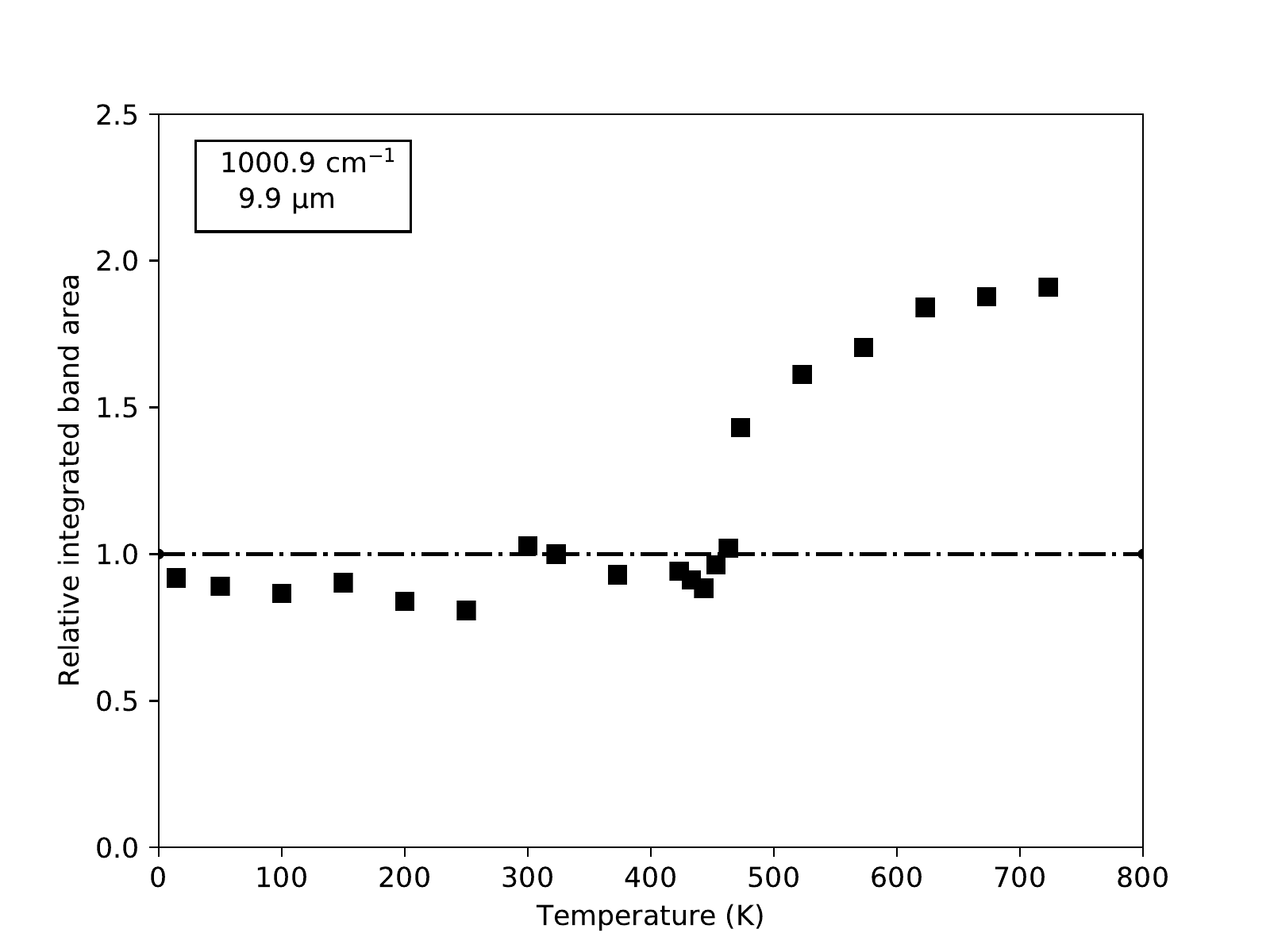}\\
(e) & (f)\\
\end{tabular}
\caption{Evolution of the integrated intensity of individual fundamental bands of pyrene, measured in KBr pellets, as a function of temperature. The values have been normalized with respect to the 300~K value.\label{band_int}}
\end{figure*}

\begin{figure*}
\ContinuedFloat
\centering
\begin{tabular}{c c}
\includegraphics[width=0.5\hsize]{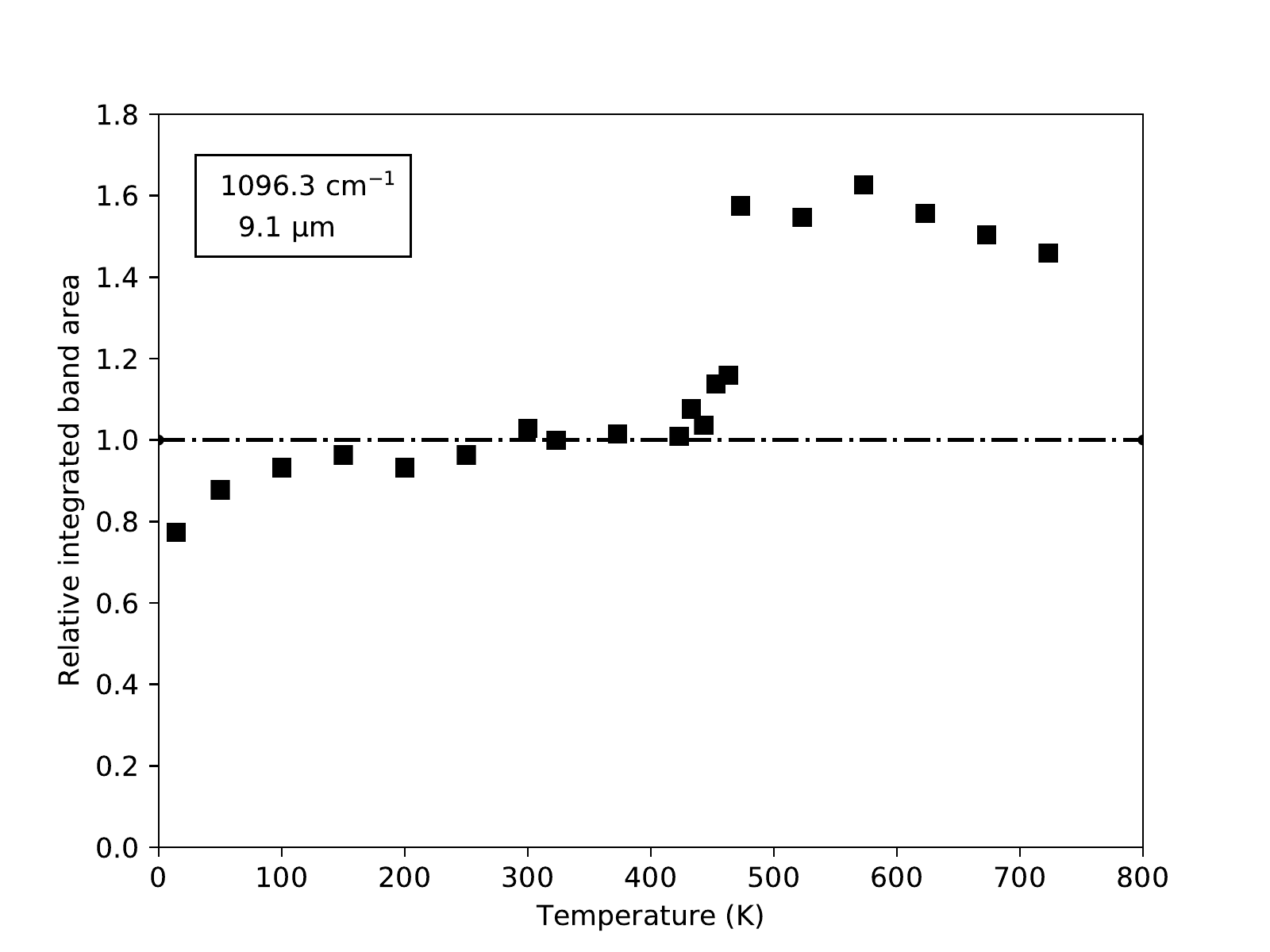} & \includegraphics[width=0.5\hsize]{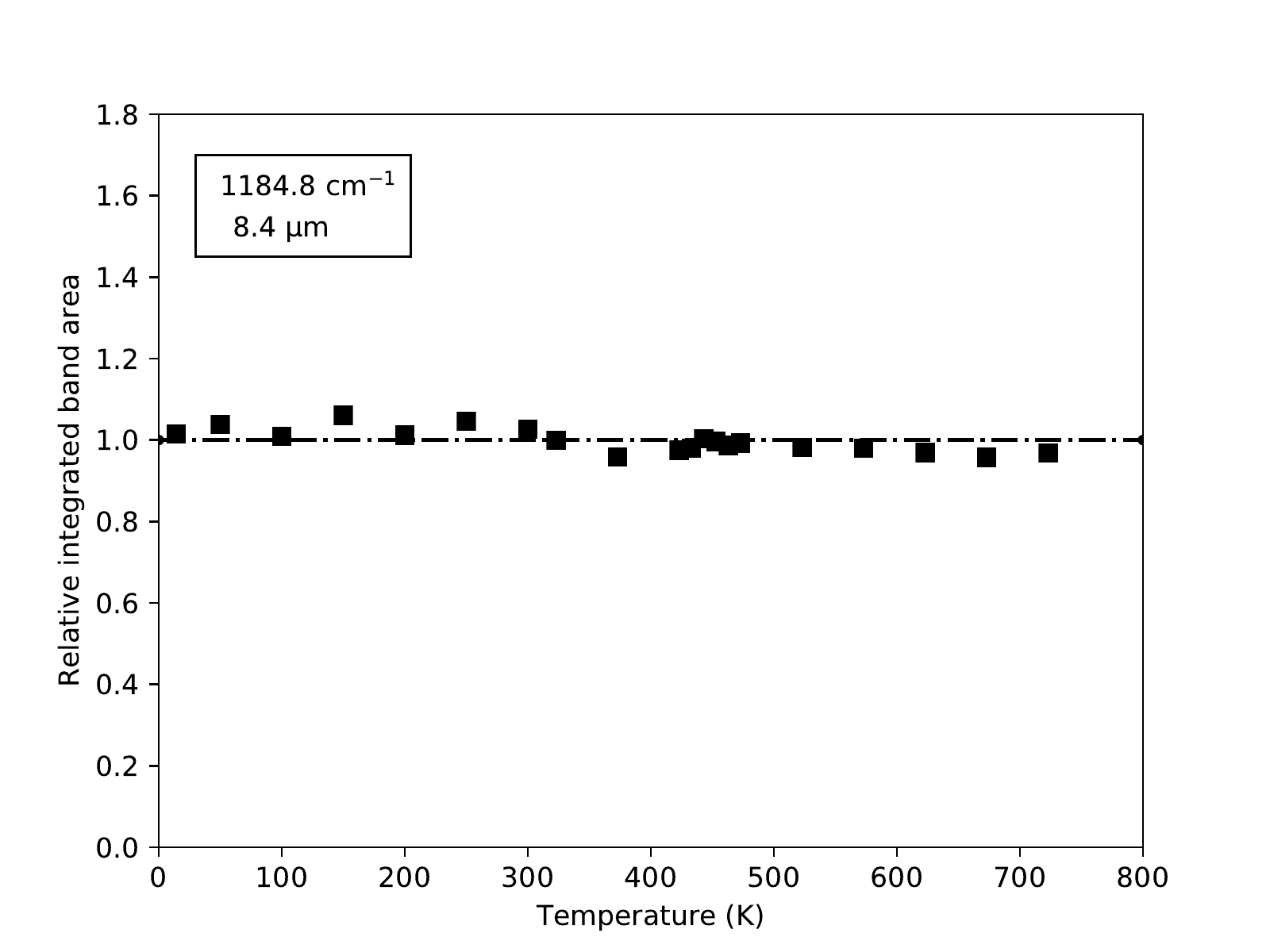}\\
(g) & (h)\\
\includegraphics[width=0.5\hsize]{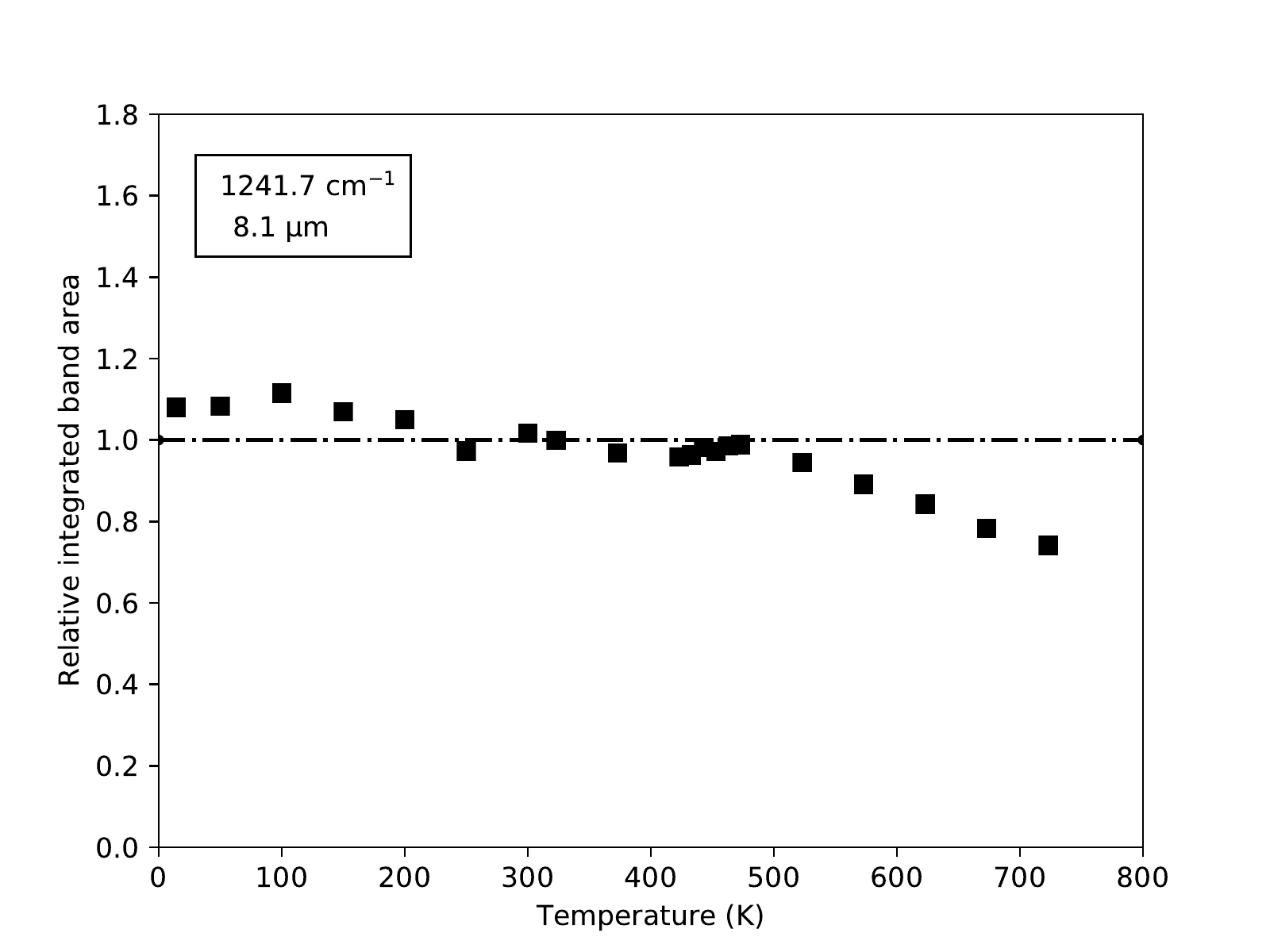} & \includegraphics[width=0.5\hsize]{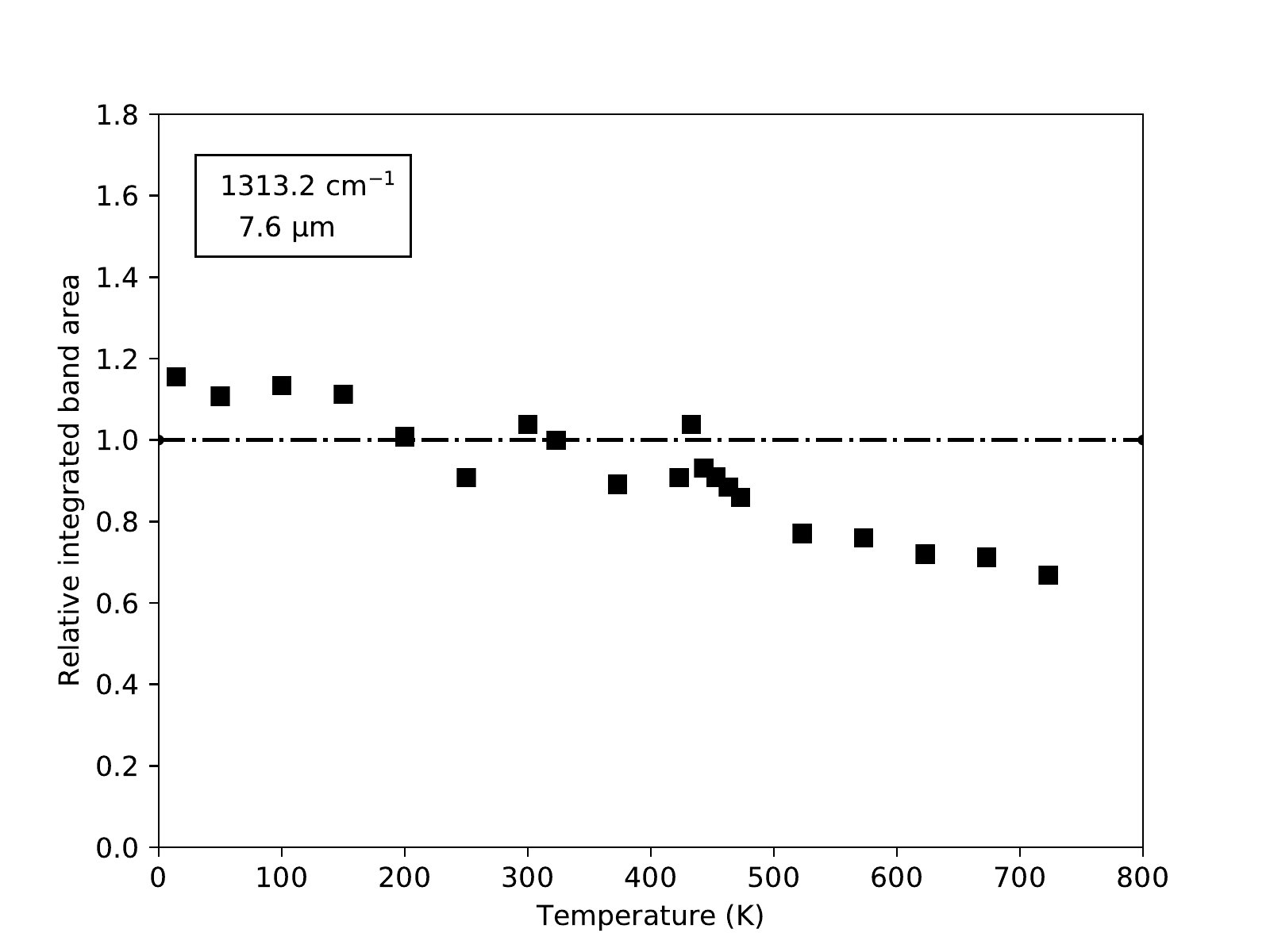}\\
(i) & (j)\\
\includegraphics[width=0.5\hsize]{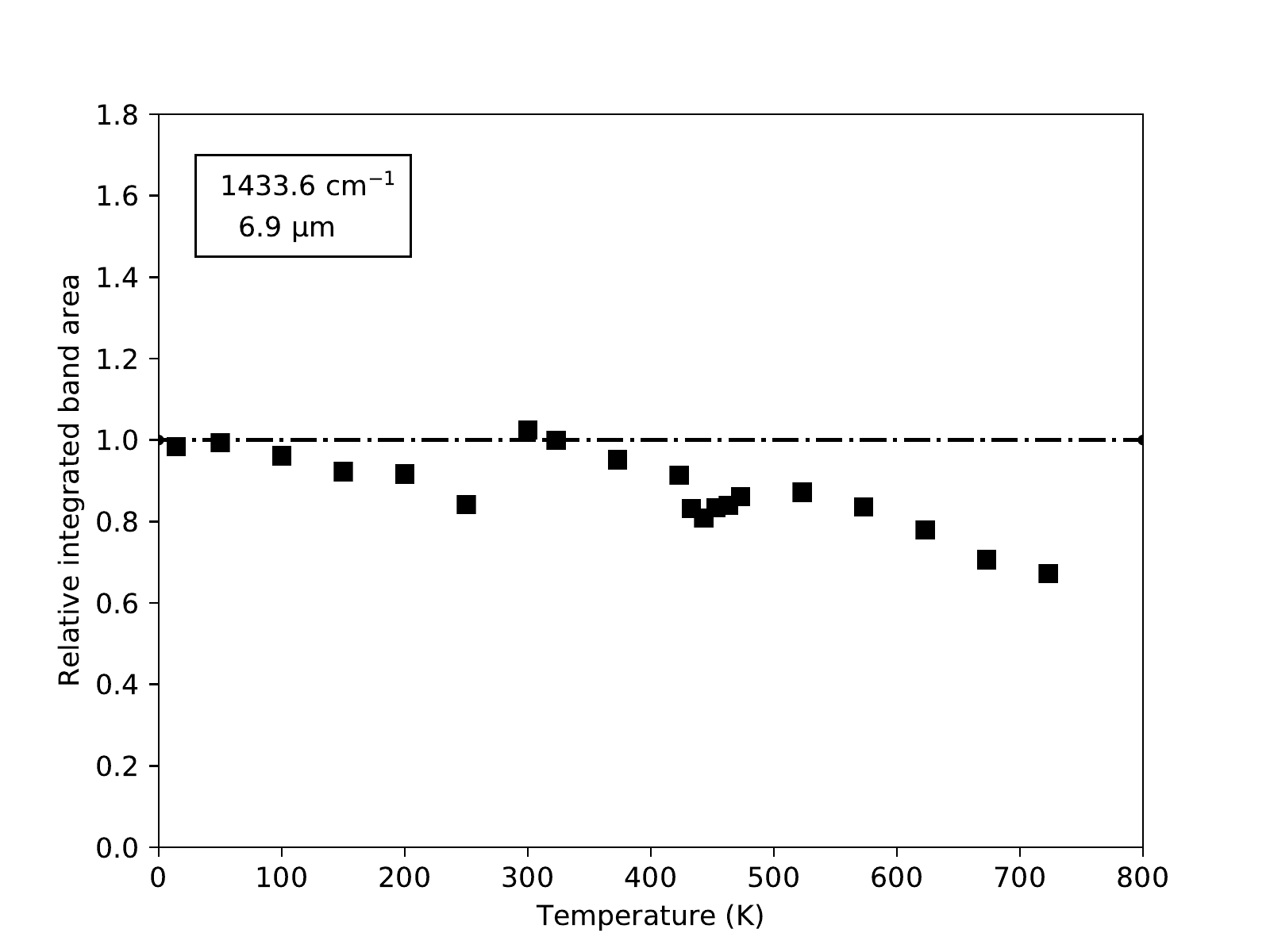}& \includegraphics[width=0.5\hsize]{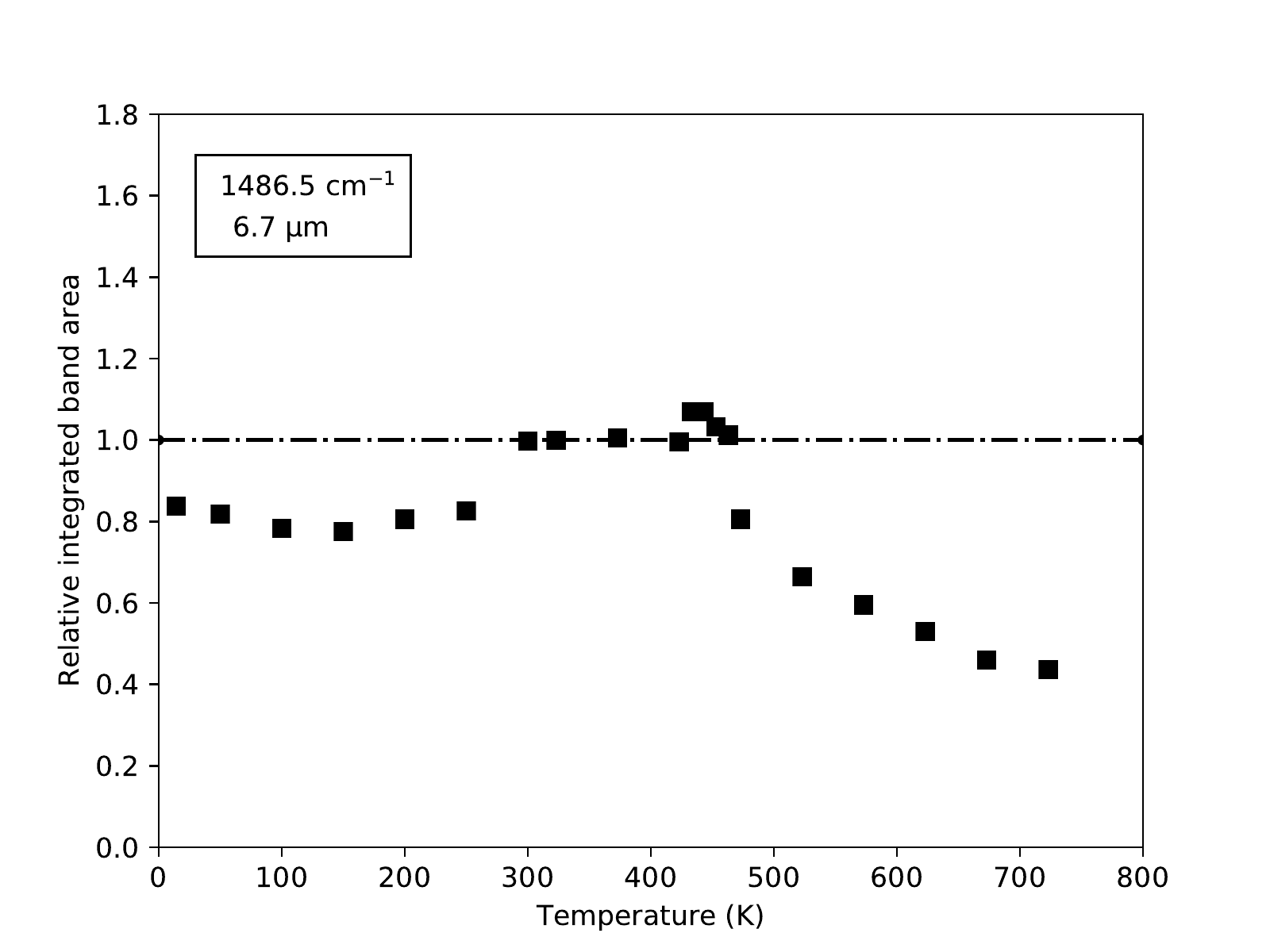}\\
(k) & (l)\\
\end{tabular}
%\label{band_int}
\caption{Evolution of the integrated intensity of individual fundamental bands of pyrene, measured in KBr pellets, as a function of temperature. The values have been normalized with respect to the 300~K value (continued from previous page). }
\end{figure*}

\begin{figure*}
\ContinuedFloat
\centering
\begin{tabular}{c c}
\includegraphics[width=0.5\hsize]{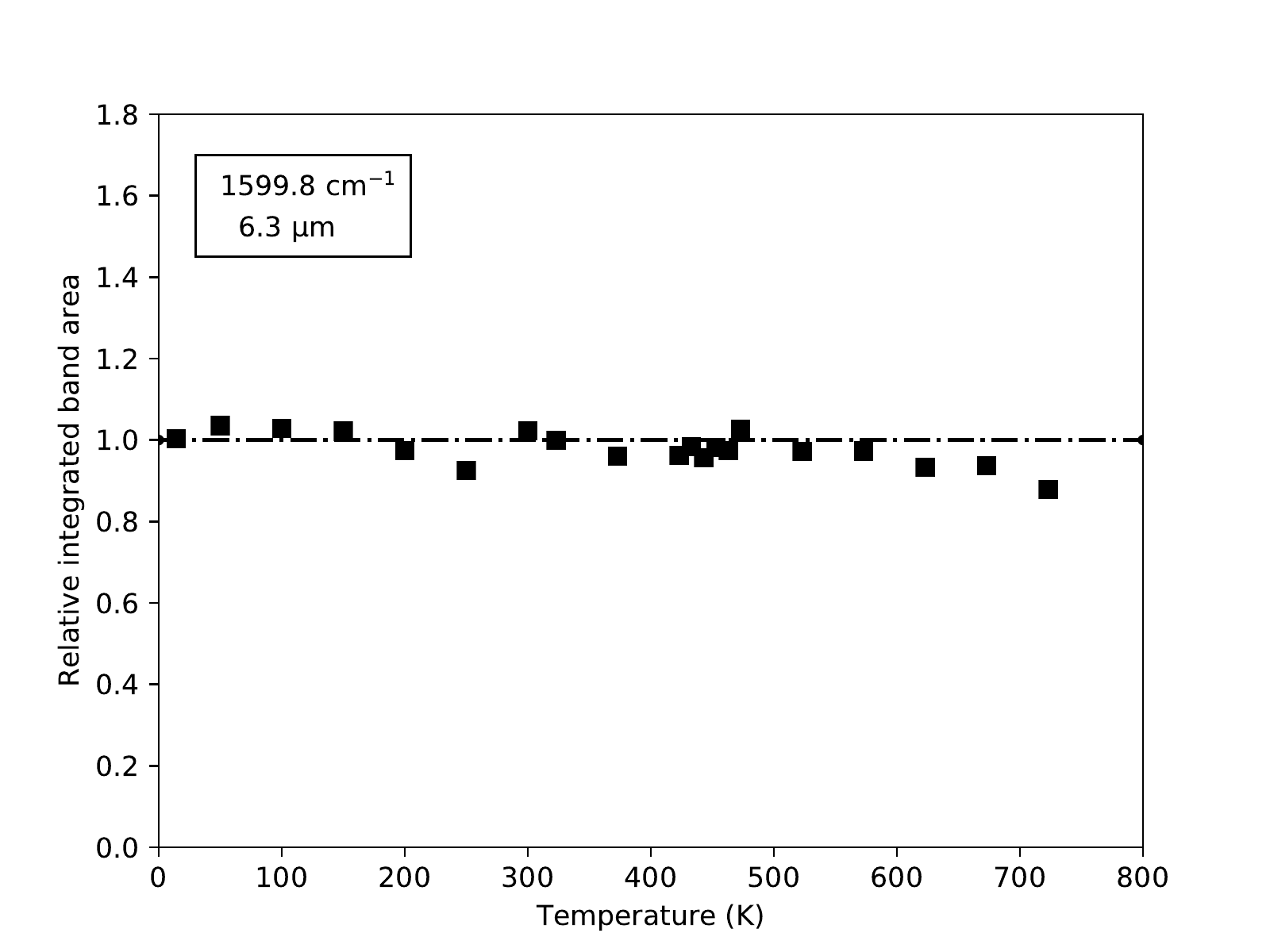} &\includegraphics[width=0.5\hsize]{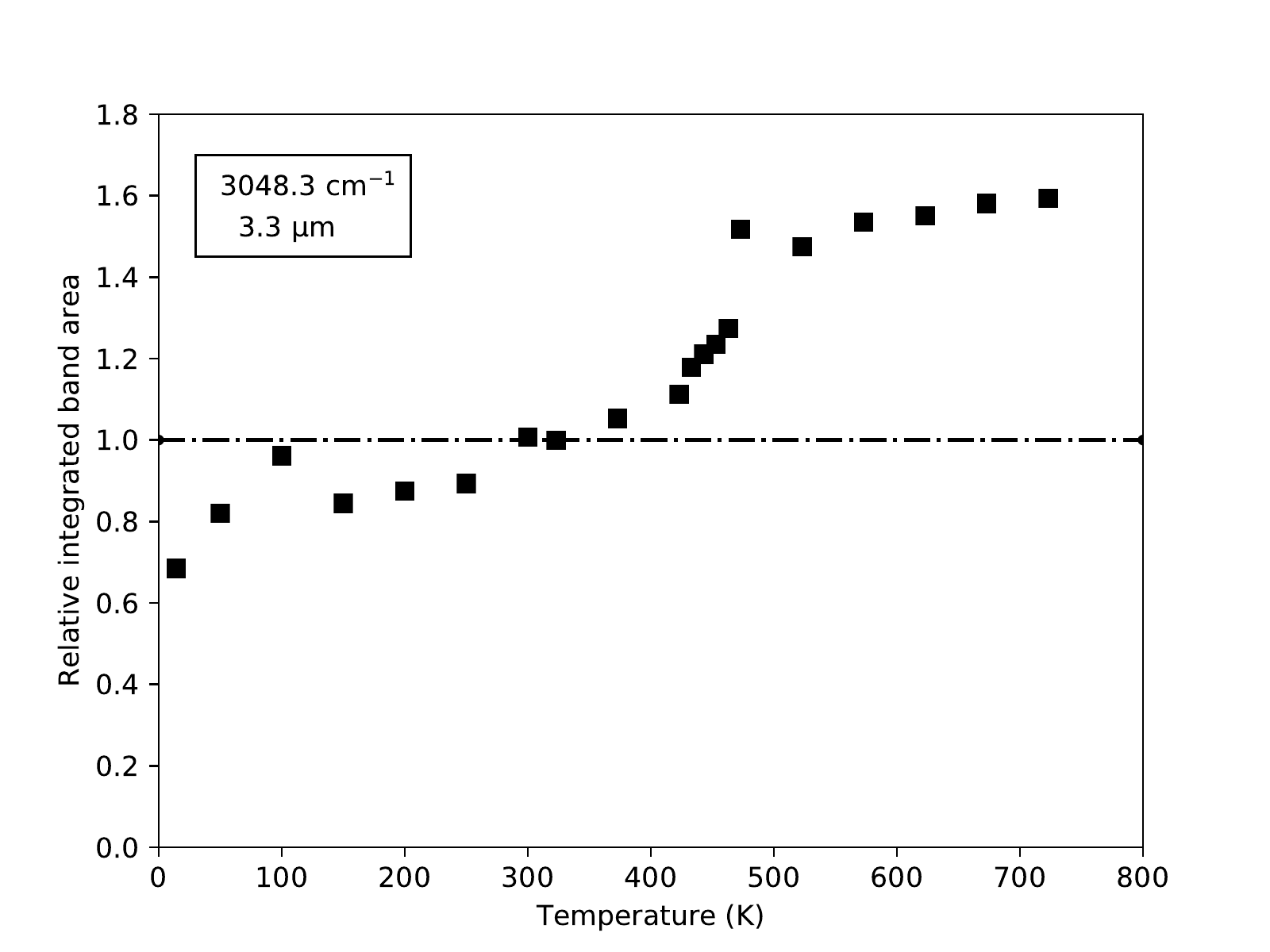} \\
(g) & (h)\\
\end{tabular}
\caption{Evolution of the integrated intensity of individual fundamental bands of pyrene, measured in KBr pellets, as a function of temperature. The values have been normalized with respect to the 300~K value (continued from previous page).} 
%\label{band_int}
\end{figure*}

\begin{figure*}
\centering
\includegraphics[width=0.8\hsize]{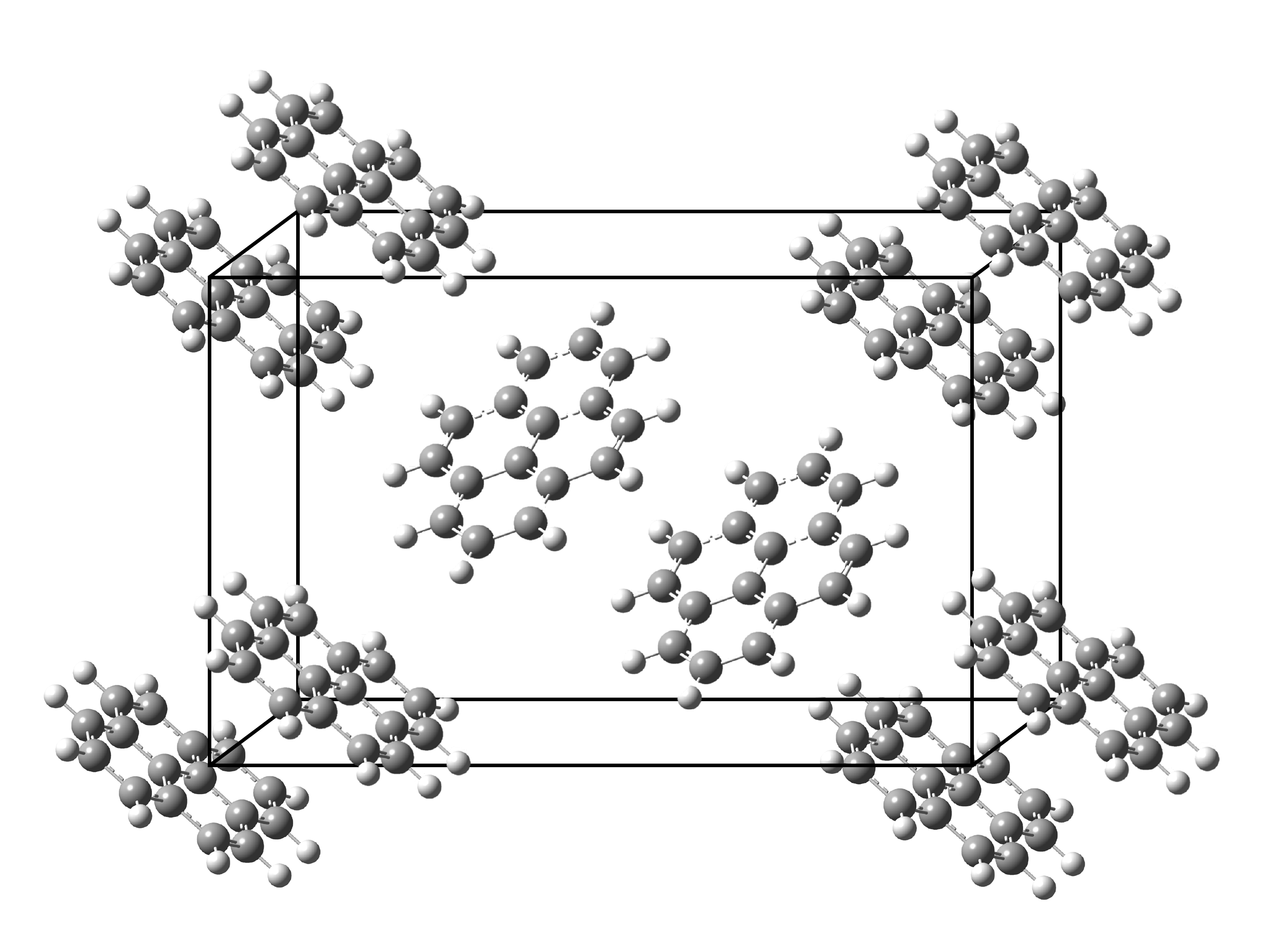} \\
 \caption{Schematic representation of the Phase I structure expected in pyrene microcrystals. The figure has been adapted from Robertson et. al. \cite{Robertson1947}}
\label{pyr_str}
\end{figure*}

\begin{figure*}
\centering
\begin{tabular}{c c}
\includegraphics[width=0.5\hsize]{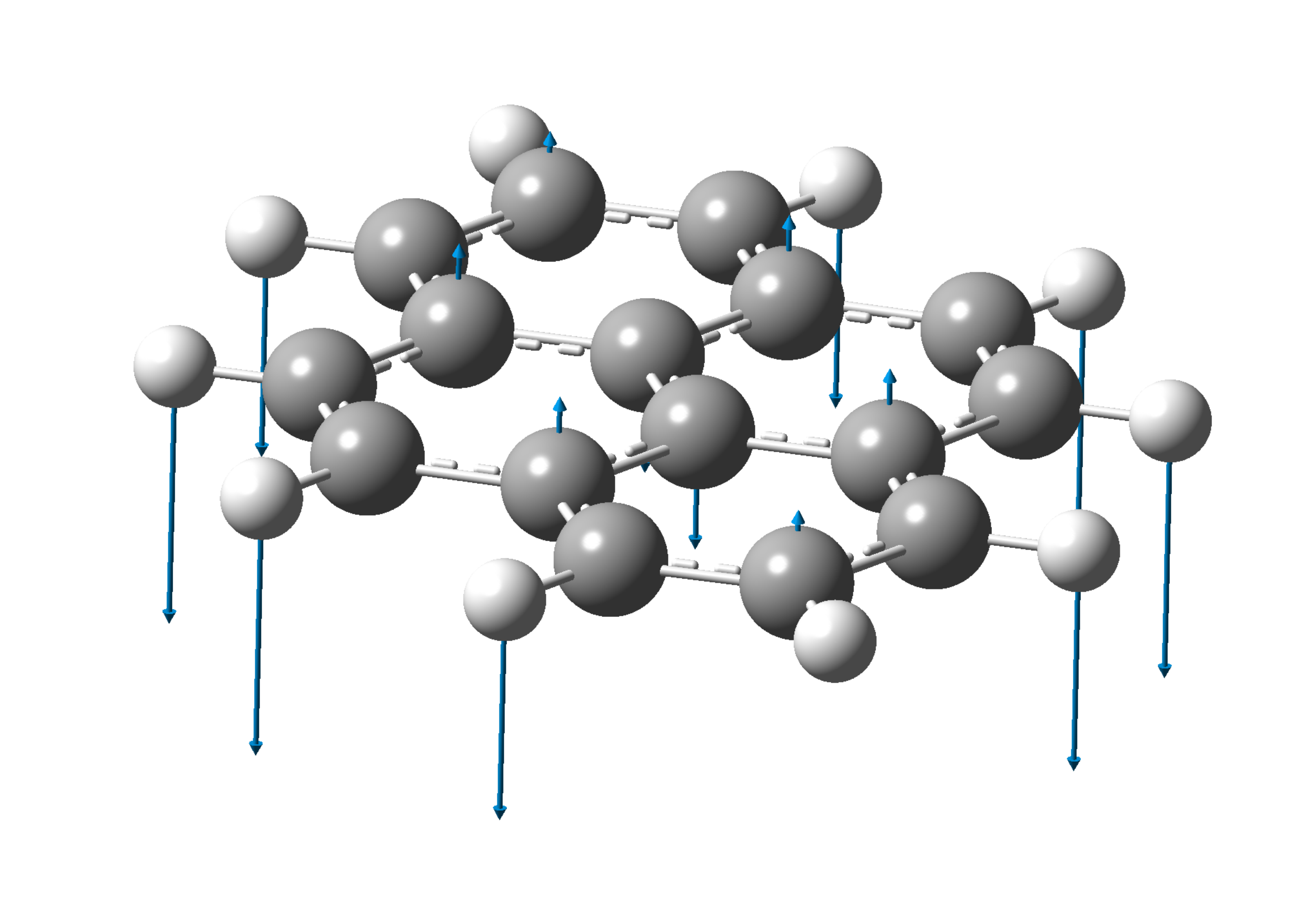} & \includegraphics[width=0.5\hsize]{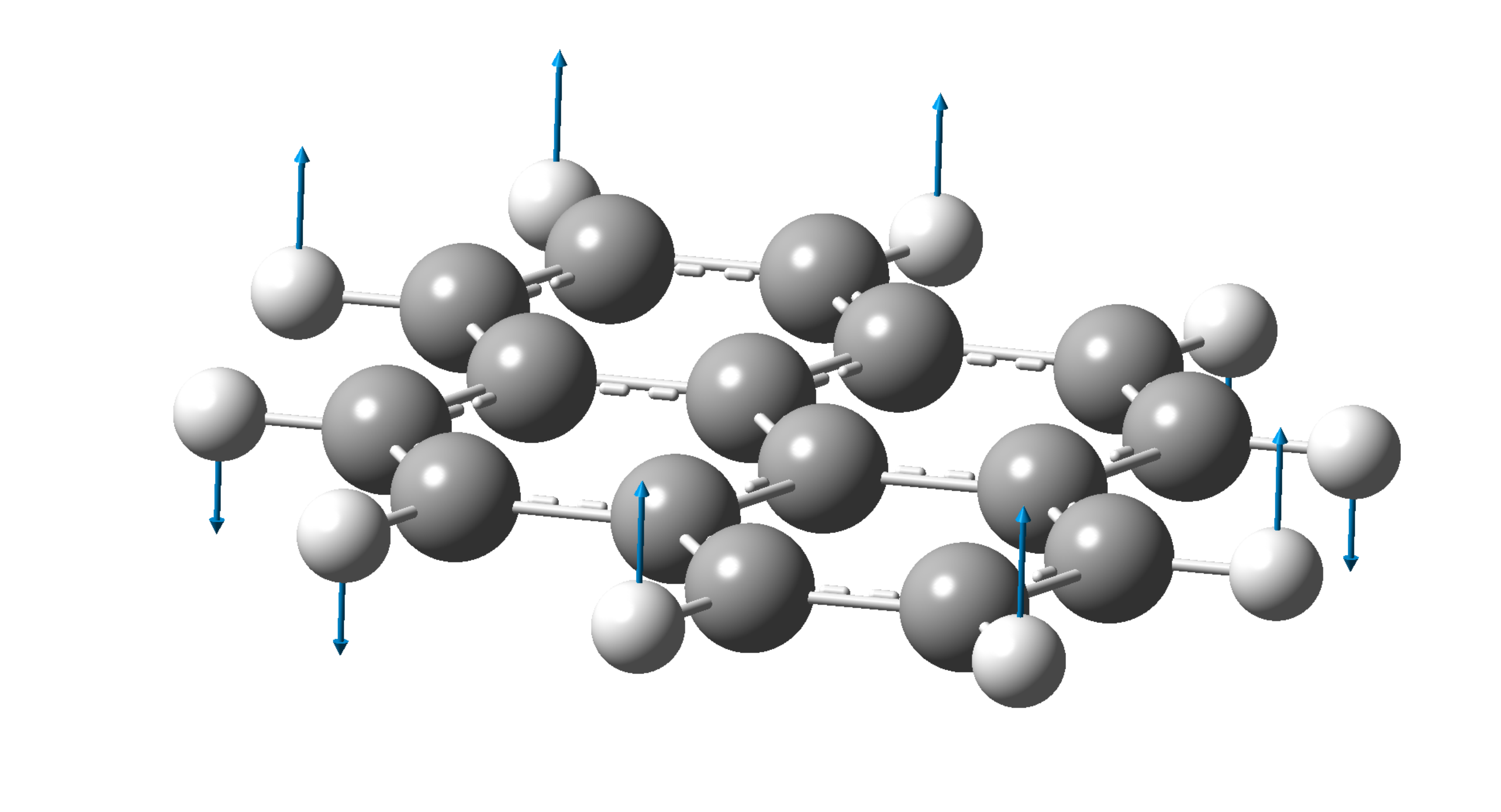}\\
(a) 709.7 cm$^{-1}$ & (b) 749.4 cm$^{-1}$ \\
\includegraphics[width=0.5\hsize]{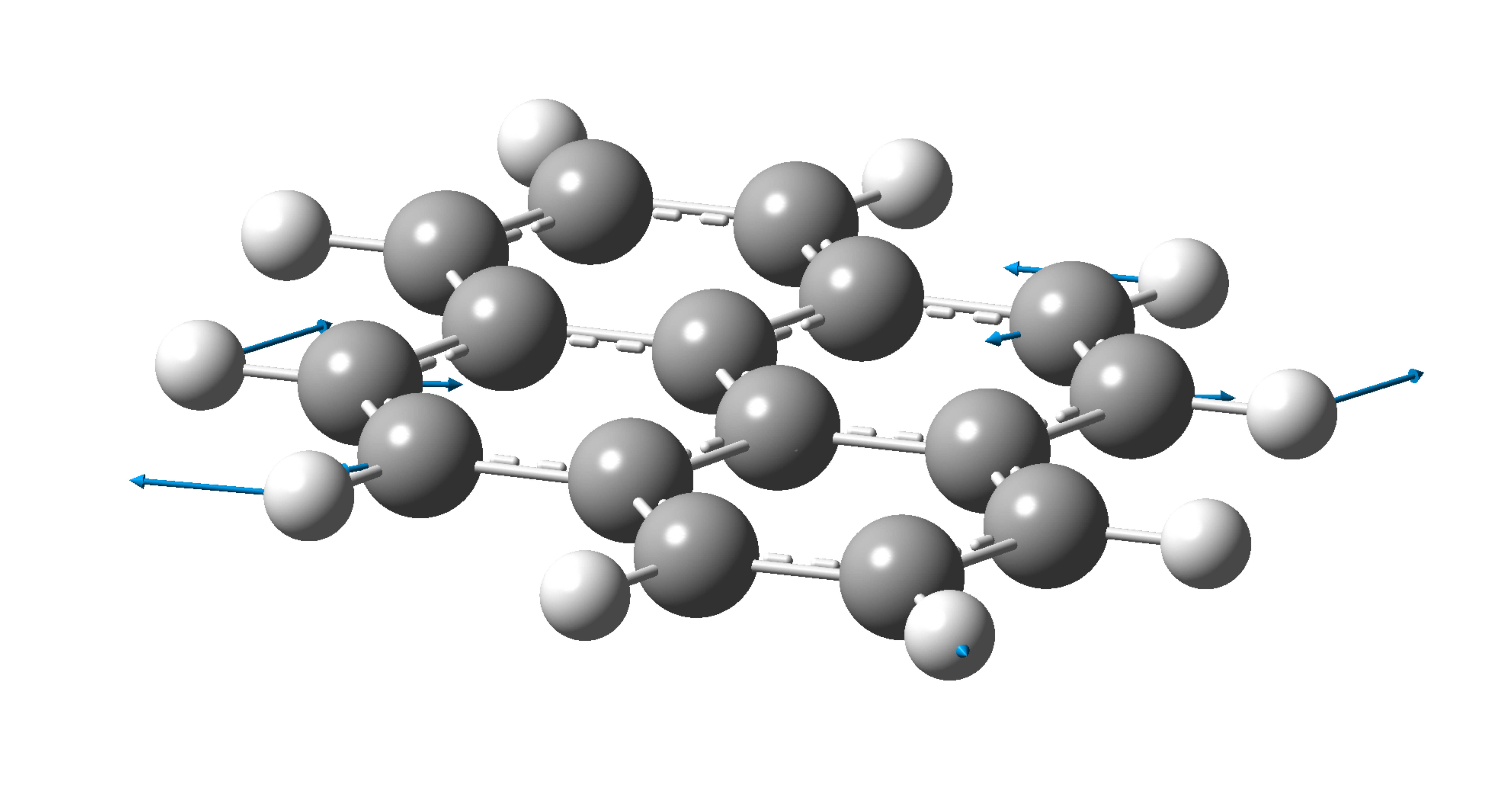} & \includegraphics[width=0.5\hsize]{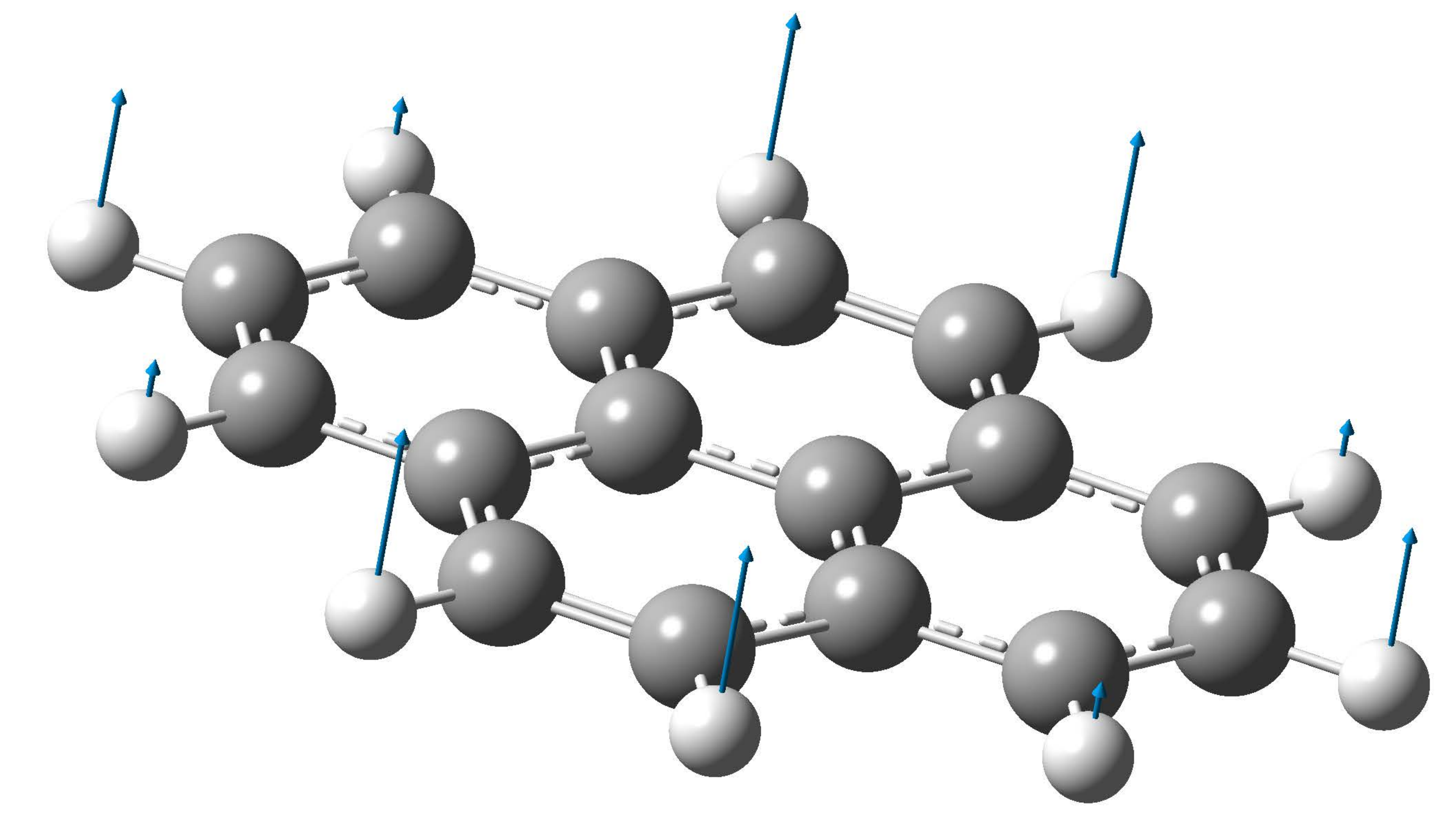}\\
(c) 820.3 cm$^{-1}$ & (d) 839.9 cm$^{-1}$\\
\includegraphics[width=0.5\hsize]{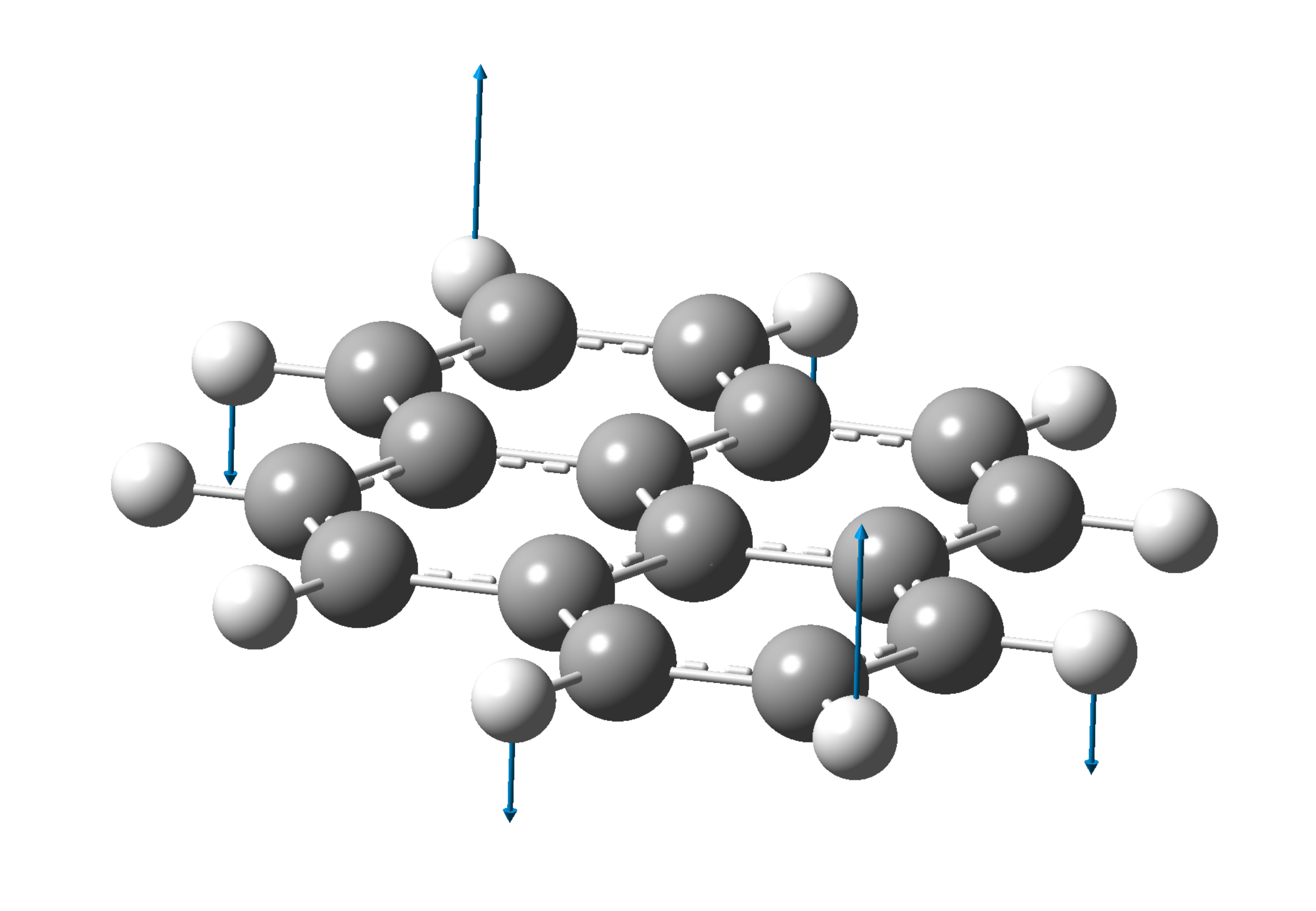} & \includegraphics[width=0.5\hsize]{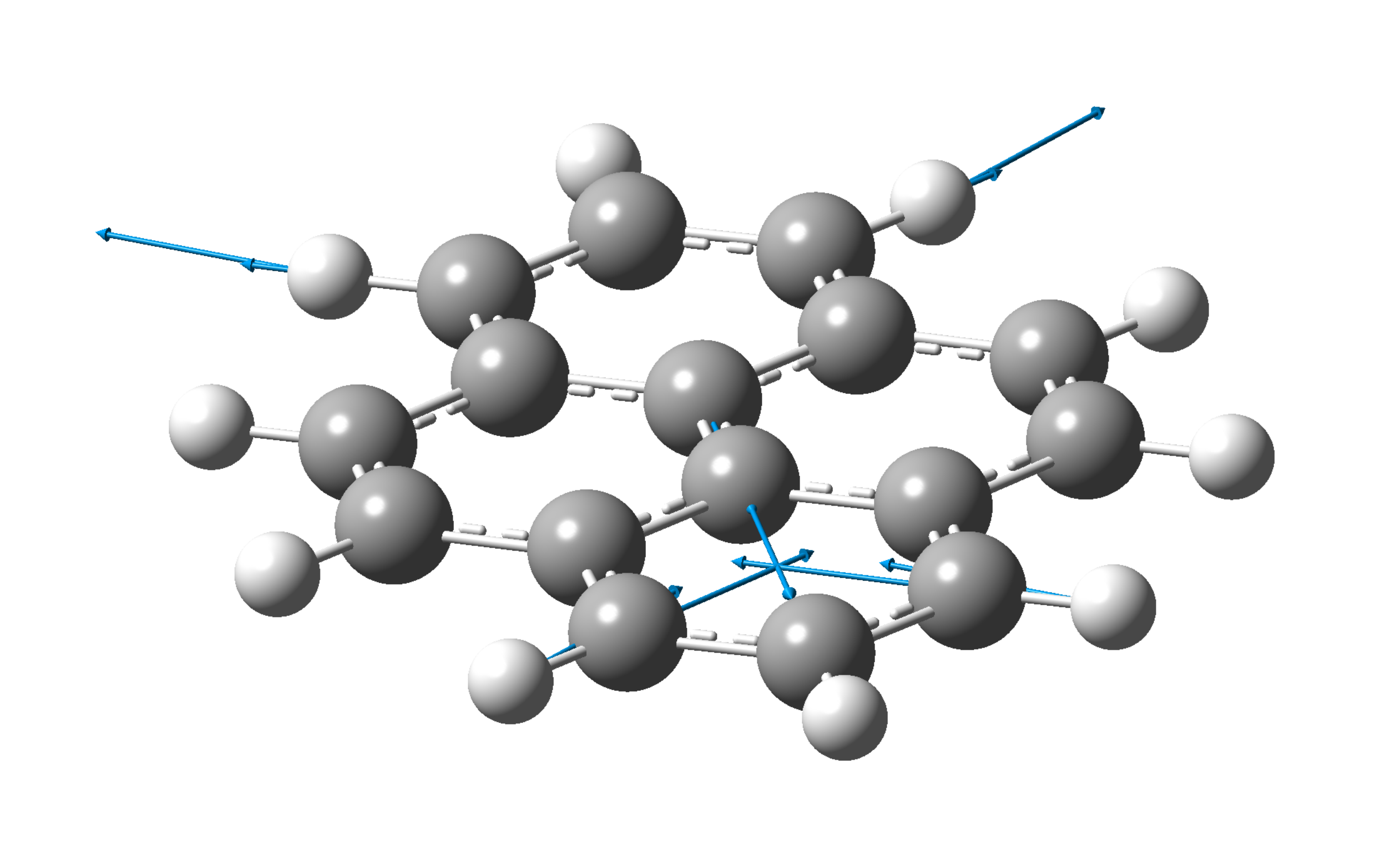}\\
(e) 964.7 cm$^{-1}$ & (f) 1000.9 cm$^{-1}$\\
\end{tabular}
\caption{Representation of the harmonic vibrational modes corresponding to the IR bands of pyrene studied in this manuscript. 
Assignments were made using the scaled harmonic frequencies of pyrene reported in Table~III of Mulas et al. \cite{mulas2018} The listed frequencies are those of the experimental bands at 300\,K reported in this work.}
\end{figure*}

\begin{figure*}
\ContinuedFloat
\centering
\begin{tabular}{c c}
\includegraphics[width=0.5\hsize]{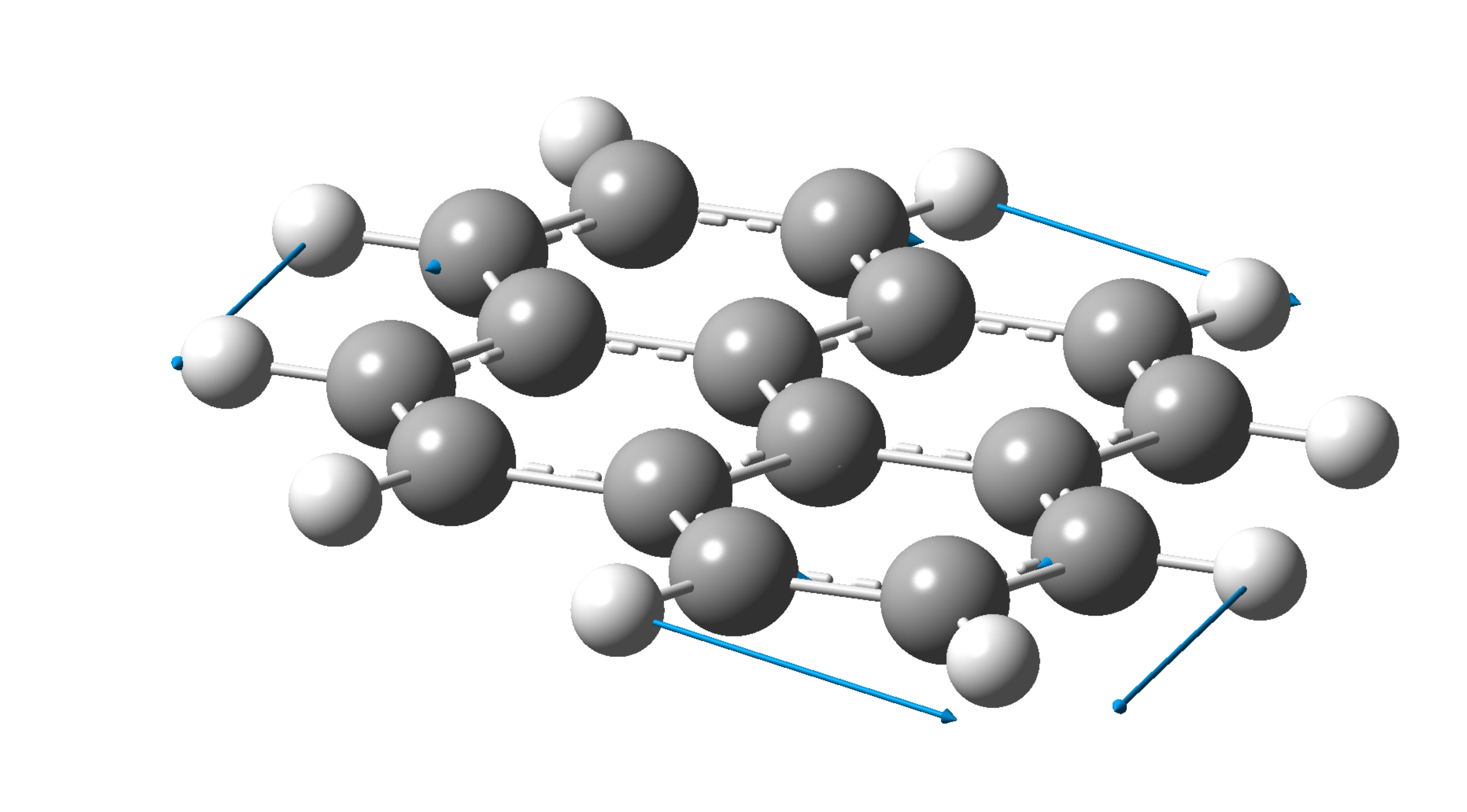} & \includegraphics[width=0.5\hsize]{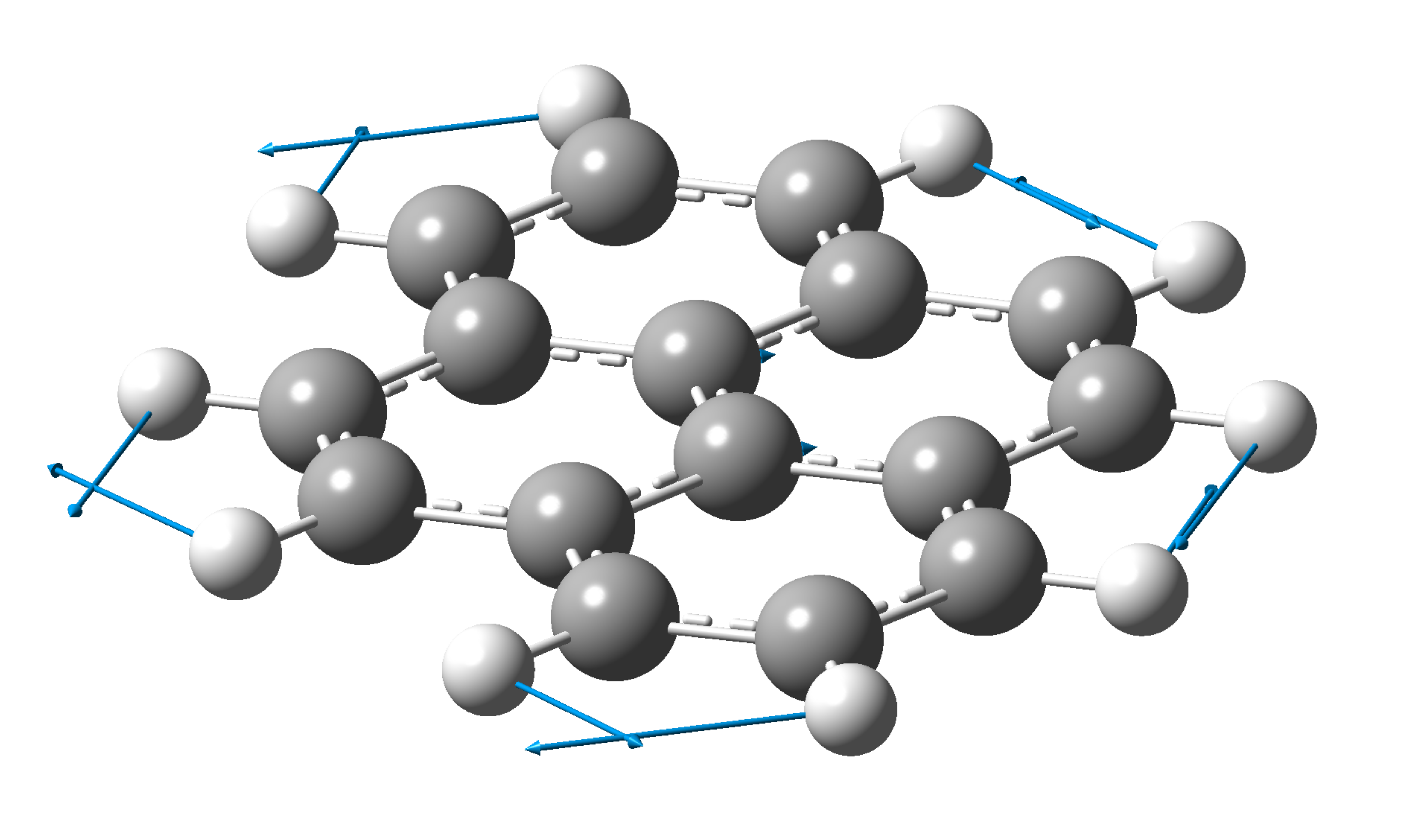}\\
(g) 1096.3 cm$^{-1}$ & (h) 1184.8 cm$^{-1}$\\
\includegraphics[width=0.5\hsize]{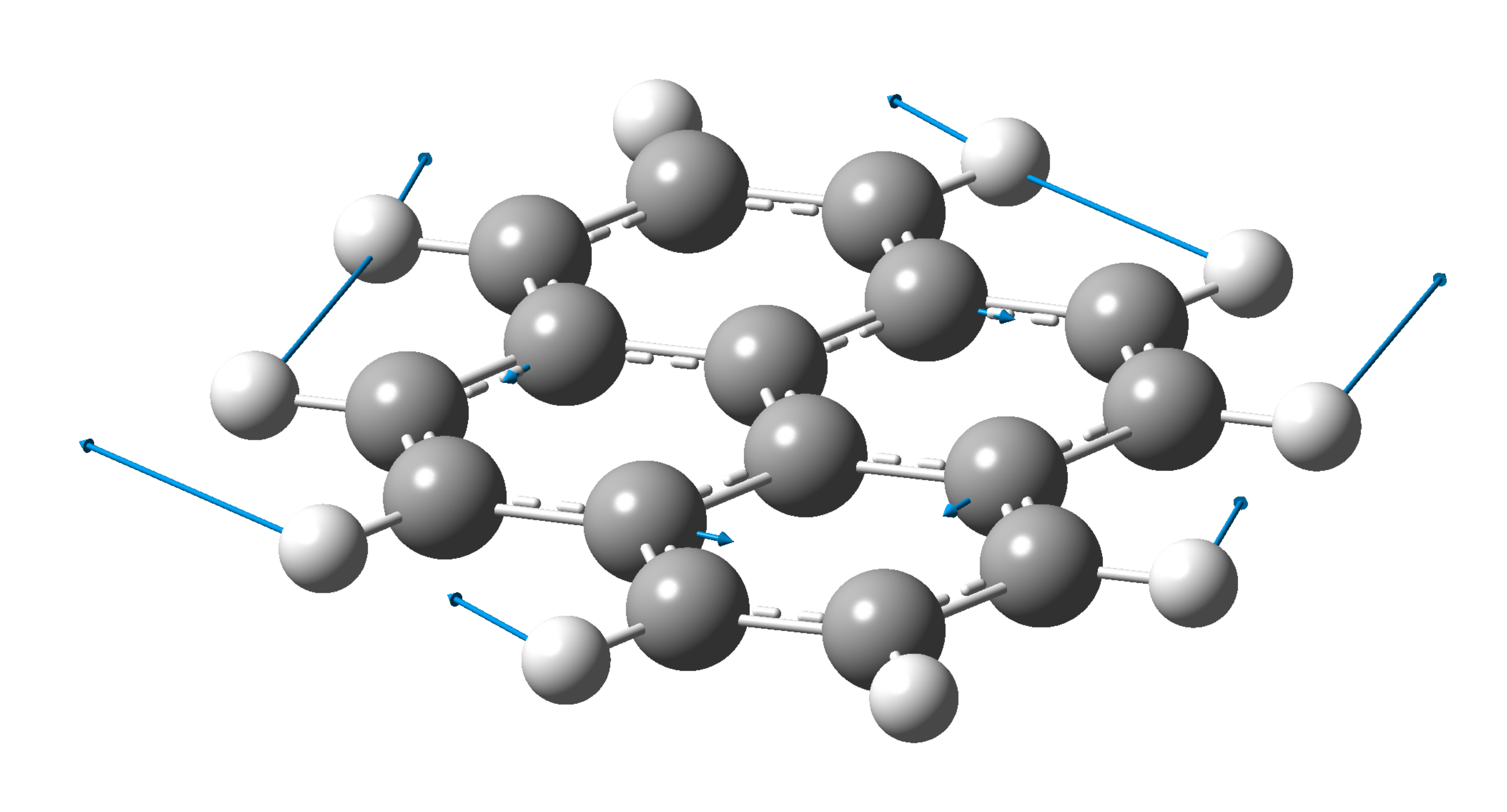} & \includegraphics[width=0.5\hsize]{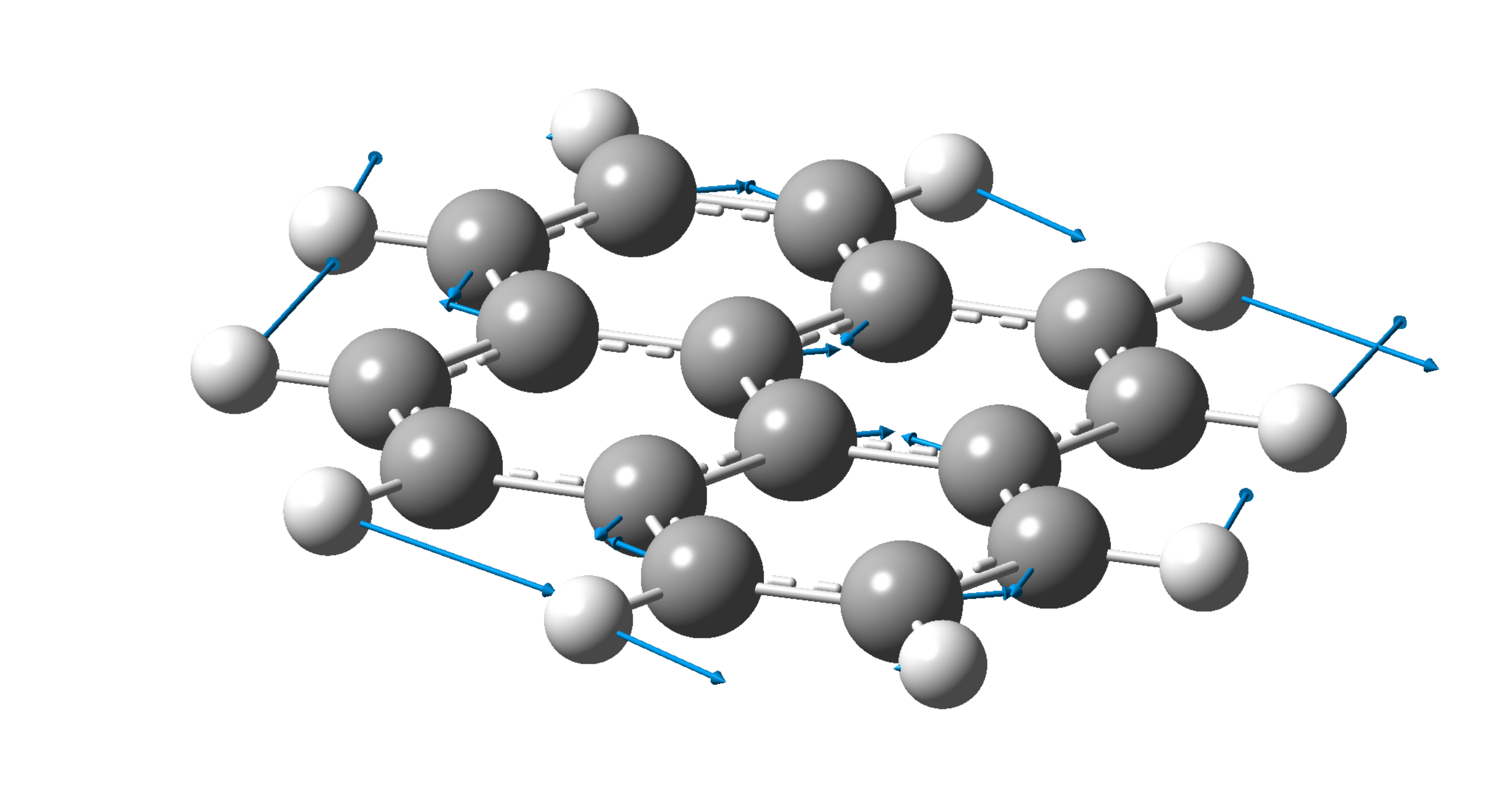}\\
(i) 1241.7 cm$^{-1}$ & (j) 1313.2 cm$^{-1}$\\
\includegraphics[width=0.5\hsize]{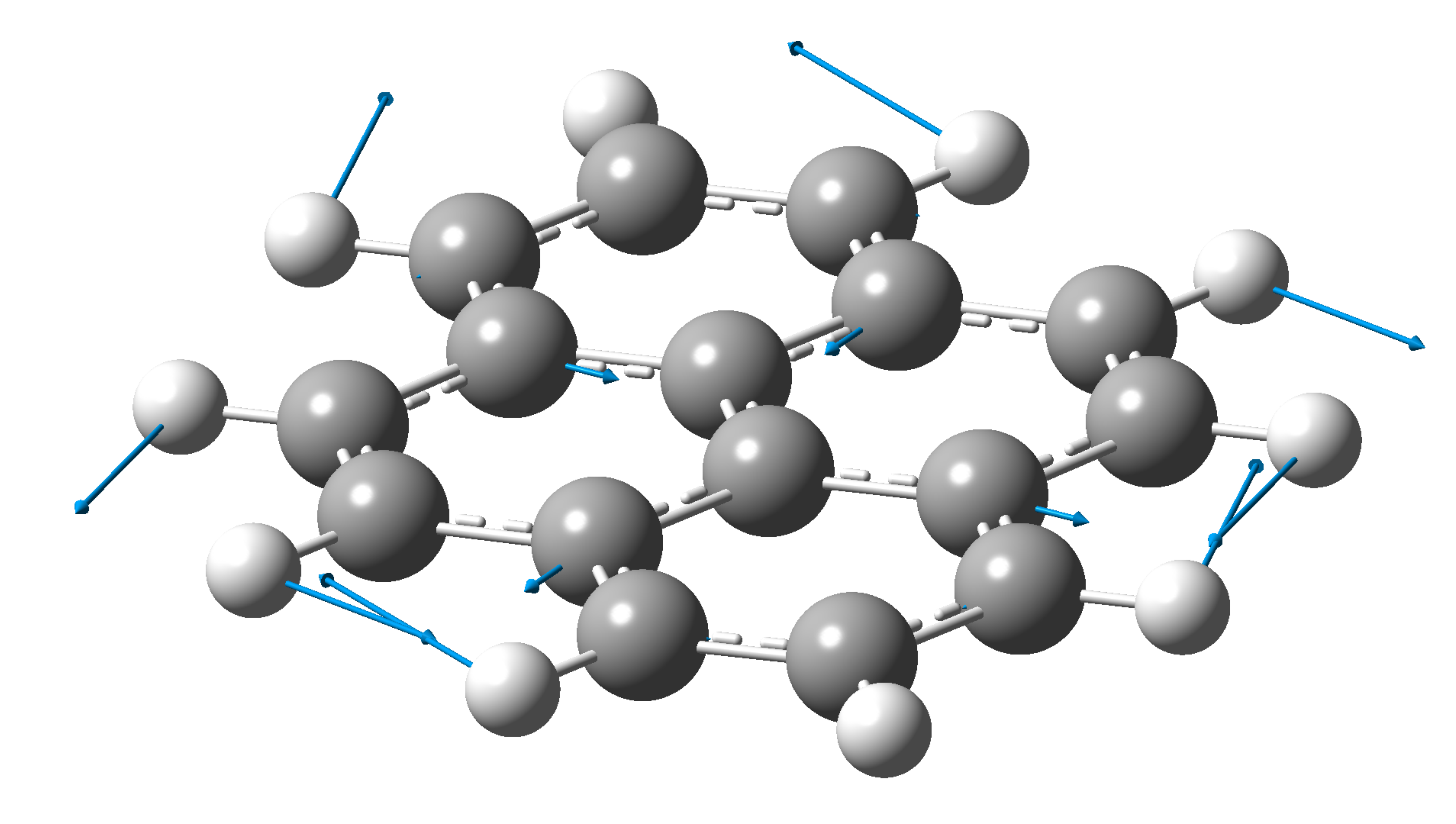} & \includegraphics[width=0.5\hsize]{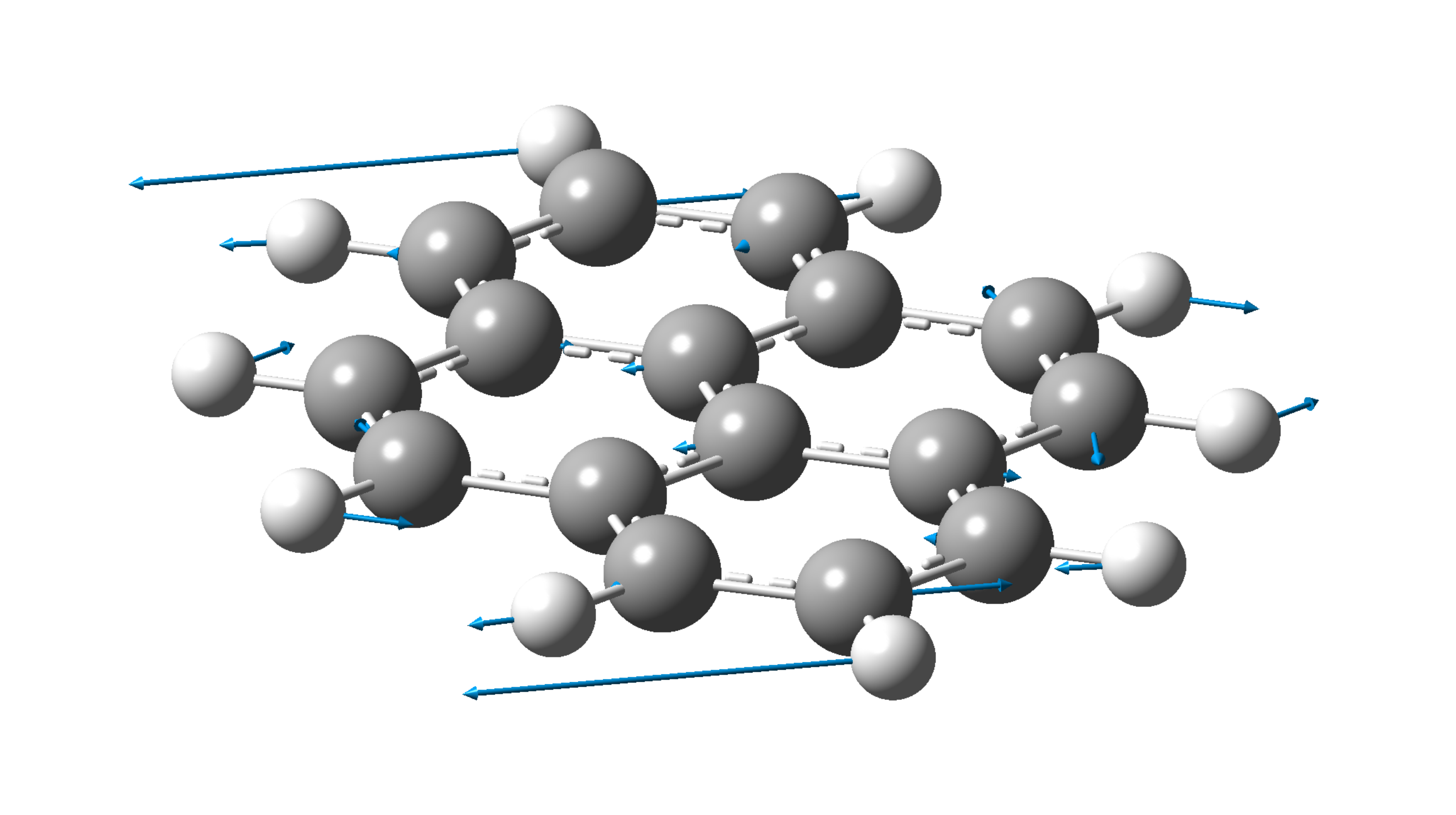}\\
(k) 1433.6 cm$^{-1}$ & (l) 1486.5 cm$^{-1}$\\
\end{tabular}
\caption{Representation of the harmonic vibrational modes corresponding to the IR bands of pyrene studied in this manuscript. 
Assignments were made using the scaled harmonic frequencies of pyrene reported in Table~III of Mulas et al. \cite{mulas2018} The listed frequencies are those of the experimental bands at 300\,K reported in this work (continued from previous page).}
\end{figure*}

\begin{figure*}
\ContinuedFloat
\centering
\begin{tabular}{c c}
\includegraphics[width=0.5\hsize]{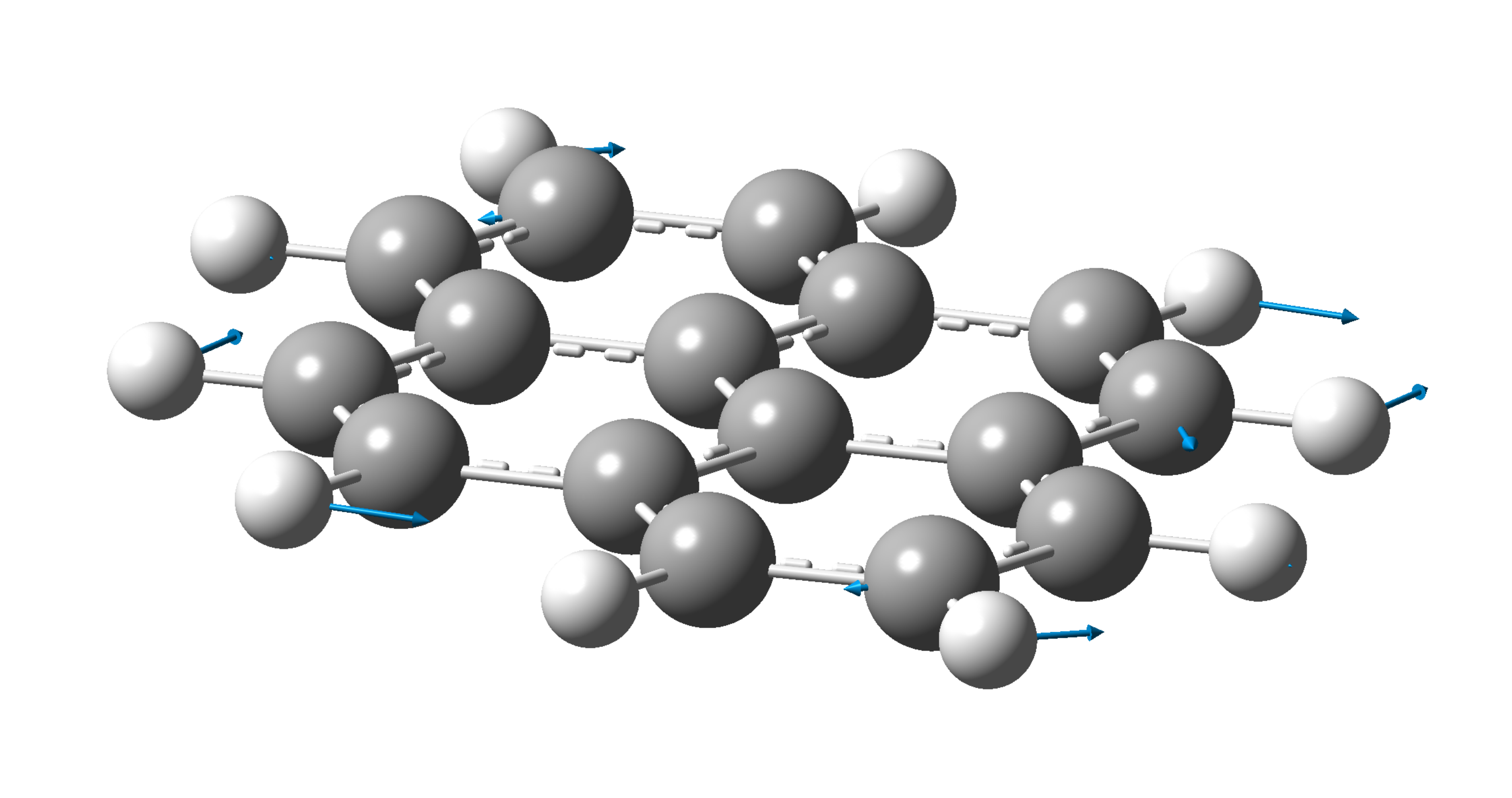} & \includegraphics[width=0.5\hsize]{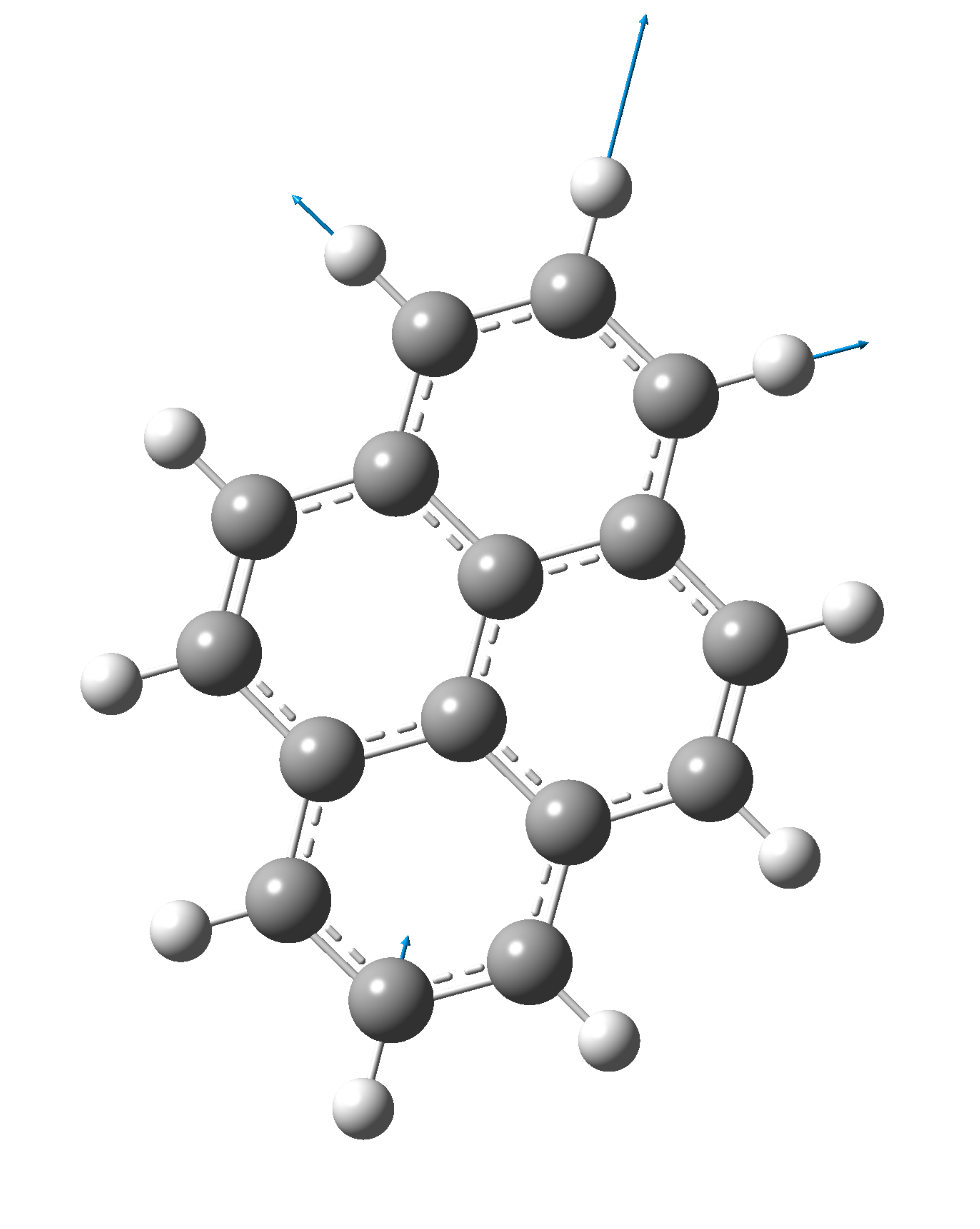}\\
(m) 1599.8 cm$^{-1}$ & (n) 3048.3 cm$^{-1}$\\
\includegraphics[width=0.5\hsize]{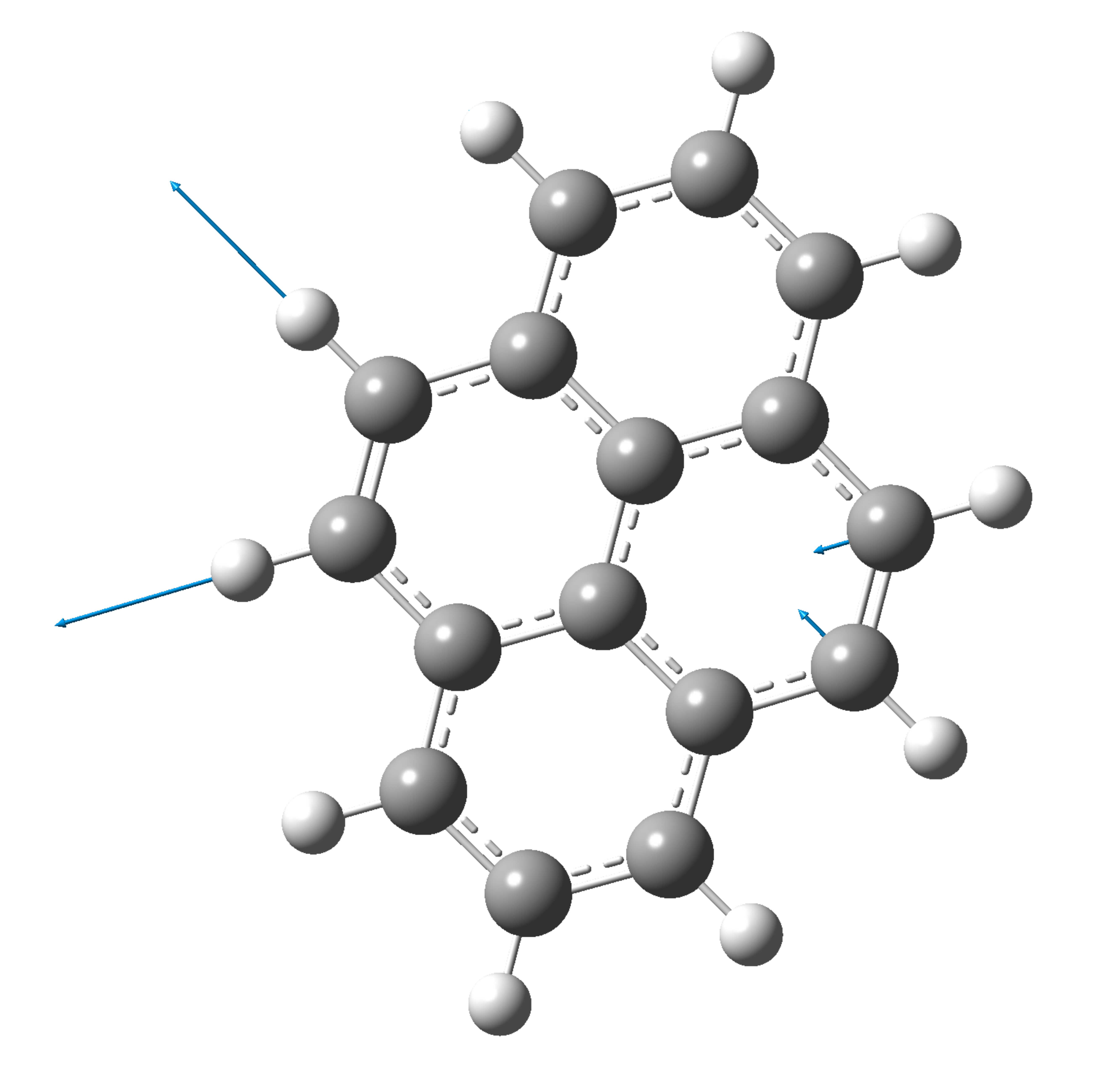} & \\
(o) 3043.6 cm$^{-1}$ & \\
\end{tabular}
\caption{Representation of the harmonic vibrational modes corresponding to the IR bands of pyrene studied in this manuscript. 
Assignments were made using the scaled harmonic frequencies of pyrene reported in Table~III of Mulas et al. \cite{mulas2018} The listed frequencies are those of the experimental bands at 300\,K reported in this work (continued from previous page).}
\label{vib_str}
\end{figure*}

\bibliography{pyrene_expt}